\documentclass[twocolumn]{aastex701}

\usepackage{graphicx}
\usepackage{subcaption}

\usepackage[percent]{overpic}
\usepackage{array}
\usepackage{booktabs}
\usepackage{multirow}
\usepackage{tabularx}
\usepackage{bm}
\usepackage{txfonts}

\begin{document}

\title{Evidence for CME--CME Interaction in a Magnetic-Cloud-Like Ejecta: Insights from Multipoint Observations and Polytropic Analysis}

\author[0000-0003-1713-119X]{Jyoti Sheoran}
\affiliation{Aryabhatta Research Institute of Observational Sciences, 
Beluwakhan, 263001, Uttarakhand, India } 
\affiliation{Department of Applied Physics, Mahatma Jyotiba Phule Rohilkhand University, Bareilly 243006, India } 
\email[show]{sheoran.j03@gmail.com} 
\author[0000-0002-6954-2276]{Vaibhav Pant}
\affiliation{Department of Physics, Indian Institute of Technology Delhi, Hauz Khas, New Delhi-110016, Delhi, India}
\email[show]{vpant@iitd.ac.in} 
\author[0009-0004-8761-3789]{Eva Weiler}
\affiliation{Austrian Space Weather Office, GeoSphere Austria, Reininghausstraße 3, Graz, 8020, Austria}
\affiliation{Institute of Physics, University of Graz, Universit\"atsplatz 5, Graz, 8010, Austria}
\email{eva.weiler@geosphere.at}

\author[0009-0007-8208-3960]{Vivek Menon}
\affiliation{Bachelor of Science (Research) Programme, Indian Institute of Science, Bengaluru, 560012, India}
\email[]{vivekmenon@iisc.ac.in}
\author[0000-0001-9992-8471]{Emma E. Davies}
\affiliation{Austrian Space Weather Office, GeoSphere Austria, Reininghausstraße 3, Graz, 8020, Austria}
\email[]{emma.davies@geosphere.at}
\author[0000-0001-6868-4152]{Christian M\"ostl}
\affiliation{Austrian Space Weather Office, GeoSphere Austria, Reininghausstraße 3, Graz, 8020, Austria}
\email[]{chris.moestl@outlook.com}
\author[0000-0003-4653-6823]{Dipankar Banerjee}
\affiliation{Indian Institute of Space Science and technology, Valiamala, Thiruvananthapuram - 695 547,Kerala, India}
\affiliation{Center of Excellence in Space Sciences India, IISER Kolkata, Mohanpur 741246, West Bengal, India}
\email[]{dipu@iiap.res.in}
\author{M. Saleem Khan}
\affiliation{Department of Applied Physics, Mahatma Jyotiba Phule Rohilkhand University, Bareilly 243006, India } 
\email{saleem.hepru@gmail.com}

\begin{abstract}

Using in-situ observations from Solar Orbiter, STEREO-A, and Wind, we investigate the heliospheric evolution of an interplanetary coronal mass ejection (ICME). The magnetic ejecta (ME) shows a magnetic-cloud-like (MCL) configuration: a front region in which the magnetic field rotates and its magnitude declines, followed by a weakly rotating ``back region'' of nearly constant field magnitude. Near-Sun EUV and white-light observations reveal two fast CMEs launched in rapid succession and interacting at low coronal heights, providing direct evidence that an MCL ejecta can arise from interaction between two closely spaced CMEs sampled near their apex. Using near-radially aligned Solar Orbiter and STEREO-A observations, we find that the sheath expands more rapidly than the ME, consistent with the ``snow-plow'' effect, while the ME properties broadly follow trends reported in previous ICME studies. Comparison of STEREO-A and Wind observations, separated by only $9.2^{\circ}$ in longitude, reveals measurable mesoscale variability within both the sheath and the ME. We identify dual polytropic behavior within both substructures, consistent with the merging of two interacting CMEs to form the observed MCL ejecta. This dual behavior in the ME is most pronounced in the inner heliosphere and weakens with heliocentric distance, indicating thermodynamic homogenization during radial expansion, whereas the sheath maintains thermodynamic contrasts from the inner heliosphere to 1~au. Our findings suggest that polytropic diagnostics can help reveal thermodynamically distinct plasma populations associated with interacting CMEs, providing additional evidence for CME--CME interaction in MCL ejecta where conventional in-situ signatures show no clear evidence of such interactions.

\end{abstract}


\section{Introduction}\label{sec:intro}

Coronal mass ejections (CMEs) are large-scale eruptions of magnetized plasma from the solar corona into interplanetary space \citep{2012_Webb}. The interplanetary counterparts of CMEs are referred to as interplanetary coronal mass ejections \citep[ICMEs;][]{2006_Zurbuchen}. An ICME typically consists of three distinct regions: a shock, a sheath, and a magnetic ejecta (ME). An ICME traveling faster than the ambient solar wind drives a forward shock, compressing and accumulating plasma and magnetic fields in front of the ejecta, forming a sheath region characterized by enhanced density, elevated proton temperatures, and strongly fluctuating magnetic fields \citep{2017_Kilpua}. The ME represents the core magnetic structure of the ICME and is often interpreted as a magnetic flux rope (MFR), though some events contain multiple MFRs or lack clear flux rope signatures \citep{2006_Jian, 2019_nieves}. The MFR has a coherent magnetic field structure in which helical magnetic field lines wrap around a central axis \citep{1981_Burlaga}, and is commonly described by analytical force-free flux-rope models, such as the Lundquist \citep{1950_Lundquist} and Gold--Hoyle \citep{1960_Gold} solutions. A subset of MEs containing a single MFR that exhibits a smooth magnetic field rotation through a large angle (up to $\sim180^{\circ}$), low plasma beta, and reduced proton temperatures and densities is classified as a magnetic cloud \citep[MC;][]{1981_Burlaga, MARUBASHI1986335, 1988_Burlaga}. However, such idealized MC (or MFR) signatures occur in only a subset of ICMEs \citep{2002_Cane}, and many events exhibit more irregular or complex magnetic configurations \citep{2006_Zurbuchen, 2010_Richardson, 2025_Al-haddad}. Furthermore, a subset of magnetic flux ropes that cannot be fitted by force-free flux-rope models \citep{1990_Lepping} or reconstructed using the Grad--Shafranov technique \citep{2002_Hu} is classified as magnetic-cloud-like \citep[MCL;][]{2006_Lepping} ejecta.  A commonly reported signature of MCL ejecta is the presence of a ``back region'', which follows the main magnetic-field rotation and exhibits little or no additional field rotation while maintaining a relatively elevated and nearly constant magnetic-field strength \citep{2010_Mostl, 2024_Rudisser}.

To date, two main approaches have been employed to investigate the heliospheric evolution of ICMEs: (i) statistical studies, which analyze ICMEs observed by different spacecraft to derive empirical relationships between ICME parameters and heliocentric distance \citep{KumarRust1996, Bothmer1998, Liu2005, Wang2005, Leitner2007, Gulisano2010, Gulisano2012, 2021_Davies, 2026_Mostl}, and (ii) multi-point observational studies, which examine the same ICME sampled by radially aligned spacecraft to track the radial evolution of its properties \citep{1981_Burlaga, Burlaga1982, 1993a_Osherovich, Bothmer1998, Mulligan2001, 2019_Vrvsnak, 2020_Salman, 2022_Davies, 2022_Mostl, 2025b_Zhang}. Using these approaches, previous work has investigated the radial evolution of ICME width, magnetic field strength, density, and temperature, and has derived power-law dependencies of these parameters on heliocentric distance \citep{Liu2005, Wang2005, Gulisano2010, Gulisano2012, 2015_Good, 2019_Vrvsnak, 2020_Salman, 2021_Davies, 2022_Davies, 2023_Zhuang, 2025b_Zhang, 2026_Mostl}. Nevertheless, evolution of the thermodynamic properties of ICME plasma, especially within their distinct substructures, remains poorly understood \citep{2006_Forsyth, 2012_Webb}. 

Thermodynamic properties of ICMEs can be characterized using a polytropic description, which links plasma density and temperature through a power-law index ($\Gamma$) and describes the evolution of the thermodynamic state of the ICME plasma. For an ideal proton population, the polytropic relation can be expressed as \citep{1933_Chandrasekhar,2018b_Livadiotis,2018a_Livadiotis}
\begin{equation}
P_{\mathrm{th}} \propto N_p^{\Gamma_p} \quad or \quad
T_p \propto N_p^{\Gamma_p - 1},
\label{eq:polytropic_relation}
\end{equation}
where \(P_{\mathrm{th}} = N_p k_{\mathrm B} T_p\) is the proton thermal pressure and \(\Gamma_p\) is the proton polytropic index.
The polytropic approximation is widely used in heliospheric plasma modeling \citep[e.g.][]{1993b_Osherovich,2014_Nicolaou,2020_Nicolaou,2023_Nicolaou,2019_Livadiotis}, and several CME models adopt fixed indices, such as $\Gamma = 4/3$ in the analytical 3D CME model of \citet{1998_Gibson} and $\Gamma \approx 1.2$ in theoretical models describing ICME initiation and propagation \citep{1996_Chen,2000_Krall}. Several efforts have aimed to determine the polytropic index of ME plasma using both observational and theoretical approaches. Using the remote-sensing Flux Rope Internal State (FRIS) model, \citet{2018_Mishra, Mishra_2023} found that, for a slow CME, $\Gamma_{\mathrm{p}}$ decreases from $1.32$ to $0.98$ between $5.8$ and $13.6~R_{\odot}$, consistent with a heat-absorption state throughout its propagation in the corona, whereas a fast CME exhibits $\Gamma_{\mathrm{p}}$ values ranging from $1.7$ to $1.9$ out to $1~\mathrm{au}$, suggesting continuous heat release during propagation \citep{2020_Mishra}. More recent studies suggest that both fast and slow CMEs evolve toward nearly isothermal conditions ($\Gamma_{\mathrm{p}} \approx 1$) at heights of $\sim3$–$7 R_{\odot}$, with fast CMEs transitioning from heat release to heat absorption at lower heights than slower events \citep{2023_Khuntia,2024_Khuntia}. Remote-sensing observations by \citet{2023_Sheoran} further show that a slow CME maintained a nearly isothermal state much closer to the Sun ($\sim1.05$--$1.35 R_{\odot}$). 

Using in situ measurements of ICME plasma, \citet{2006_Liu} found $\Gamma_{\mathrm{p}} \approx 1.3$ inside expanding ejecta between $0.3$ and $20$~au, while proton and electron indices near 1~au were reported as $\approx 1.15$ and $\approx 0.73$, respectively \citep{Liu2005}. A multi-spacecraft analysis by \citet{2024_Ghag} indicated nearly isothermal expansion in several ICME flux ropes, and a recent statistical study at 1~au found a wide range of $\Gamma_{\mathrm{p}}$ within ME, spanning isothermal ($\Gamma_{\mathrm{p}} \approx 1$), adiabatic ($\Gamma_{\mathrm{p}} = 5/3$), and super-adiabatic ($\Gamma_{\mathrm{p}} \gtrsim 2$) regimes \citep{2025_Shaikh}. However, thermodynamic characterization across ICME substructures, particularly the sheath region, remains limited \citep{2022_Dayeh}. Recent statistical studies have further highlighted the diversity of thermodynamic states within ICMEs near 1~au. \citet{katsavrias2025polytropic} found that sheaths are generally sub-adiabatic, with $\Gamma_{\mathrm{p}}$ depending on the presence of a shock, whereas MEs span a broad range from sub-adiabatic to super-adiabatic states depending on the flux-rope magnetic orientation. Examining ICMEs across multiple solar cycles, \citet{khuntia2025evolution} showed that MEs can exist in both heating ($\Gamma_{\mathrm{p}}<5/3$) and cooling ($\Gamma_{\mathrm{p}}>5/3$) states, with heating MEs dominating during solar maxima and exhibiting strong solar-cycle modulation, while cooling MEs maintain nearly constant $\Gamma_{\mathrm{p}}\sim2$ across cycles. More recently, \citet{khuntia2026thermal} investigated a series of interacting CMEs associated with the intense geomagnetic storm of 10 May 2024 and reported diverse thermal states within the resulting complex ejecta at 1~au.

Despite these advances, the continuous evolution of the thermodynamic state of an individual ICME throughout its heliospheric propagation is still not well constrained. Multi-viewpoint observations from radially aligned spacecraft provide a powerful means to track how plasma heating and cooling evolve as an ICME expands through the heliosphere. In particular, examining thermodynamic properties as a function of heliocentric distance offers key insights into the physical state of the ejecta and the processes governing plasma heating. In this study, we investigate the thermodynamic evolution of an ICME observed sequentially by Solar Orbiter (SolO), STEREO-A (STA), and Wind on 2024 March 23--24, enabling near-radial sampling from the inner heliosphere to 1~au. We examine the radial evolution of plasma properties and proton polytropic indices within ICME substructures (sheath and ME). We present evidence for dual polytropic indices within both the ICME sheath and ME. The paper is organized as follows. Section~\ref{sec:data} describes the data sets and event observations. Section~\ref{sec:analysis} presents the analysis and results of the radial evolution of ICME properties. The main conclusions are summarized in Section~\ref{sec:conclusions}.

\section{Data and Observations}\label{sec:data}

We analyze in situ observations of an ICME listed in the \texttt{HELIO4CAST} LineupCAT\footnote{\url{https://helioforecast.space/lineups}} catalog \citep{2022_Mostl} and detected sequentially by SolO, STA, and Wind during 2024 March 23--24 \citep{2025arXiv250813892D}. On 2024 March 23, the spacecraft were located at heliocentric distances of 0.37~au (SolO), 0.96~au (STA), and 1.00~au (Wind), with heliographic coordinates (longitude, latitude) of SolO (9.6$^\circ$, $-8.0^\circ$), STA (9.2$^\circ$, $-6.5^\circ$), and Wind (0.0$^\circ$, $-6.9^\circ$), respectively. This configuration provides a favorable multi-point sampling of the ICME at three distinct radial distances. For SolO, we use 1-min magnetic field measurements from
magnetometer \citep[MAG;][]{2020_Horbury} and proton moments from Solar Wind Analyser Proton-Alpha Sensor \citep[SWA-PAS;][]{2020_Owen} at a typical cadence of 2~s. For STA, proton plasma parameters are obtained from Plasma and Suprathermal Ion Composition \citep[PLASTIC;][]{Galvin2008} and magnetic field data from magnetometer \citep[MAG;][]{Acuna2008} within In situ Measurements of Particles and CME Transients (IMPACT) Investigation \citep{Luhmann2008}, both at 1-min cadence. For Wind, we use proton plasma measurements from Solar Wind Experiment \citep[SWE;][]{Ogilvie1995} at $\sim$100~s cadence and magnetic field measurements from Magnetic Field Investigation \citep[MFI;][]{Lepping1995} at 60~s cadence. The plasma quantities analyzed include proton bulk speed $V_{p}$, proton number density $N_{p}$, and proton temperature $T_{p}$. All in situ data used in this study are in the radial-tangential-normal (RTN) coordinate system.

To identify the near-Sun CME counterparts of this ICME, we examine extreme ultraviolet (EUV) and white-light observations from two viewpoints. For Earth’s perspective, we use EUV observations from the Atmospheric Imaging Assembly (AIA) onboard the Solar Dynamics Observatory \citep[SDO;][]{Lemen2012, Pesnell2012} and coronagraph observations from the C2 camera of the Large Angle and Spectrometric Coronagraph \citep[LASCO;][]{Brueckner1995} onboard the Solar and Heliospheric Observatory \citep[SOHO;][]{1995_Domingo}. From the STA viewpoint, we use EUV observations from the Extreme Ultraviolet Imager (EUVI) and white-light coronagraph observations from the COR2 instrument of the Sun Earth Connection Coronal and Heliospheric Investigation \citep[SECCHI;][]{Howard2008} suite.

\subsection{In situ observations at SolO, STA and WIND}

\begin{figure*}[htbp]
\centering


\begin{minipage}[t]{0.495\textwidth}
\centering
\begin{overpic}[width=\linewidth]{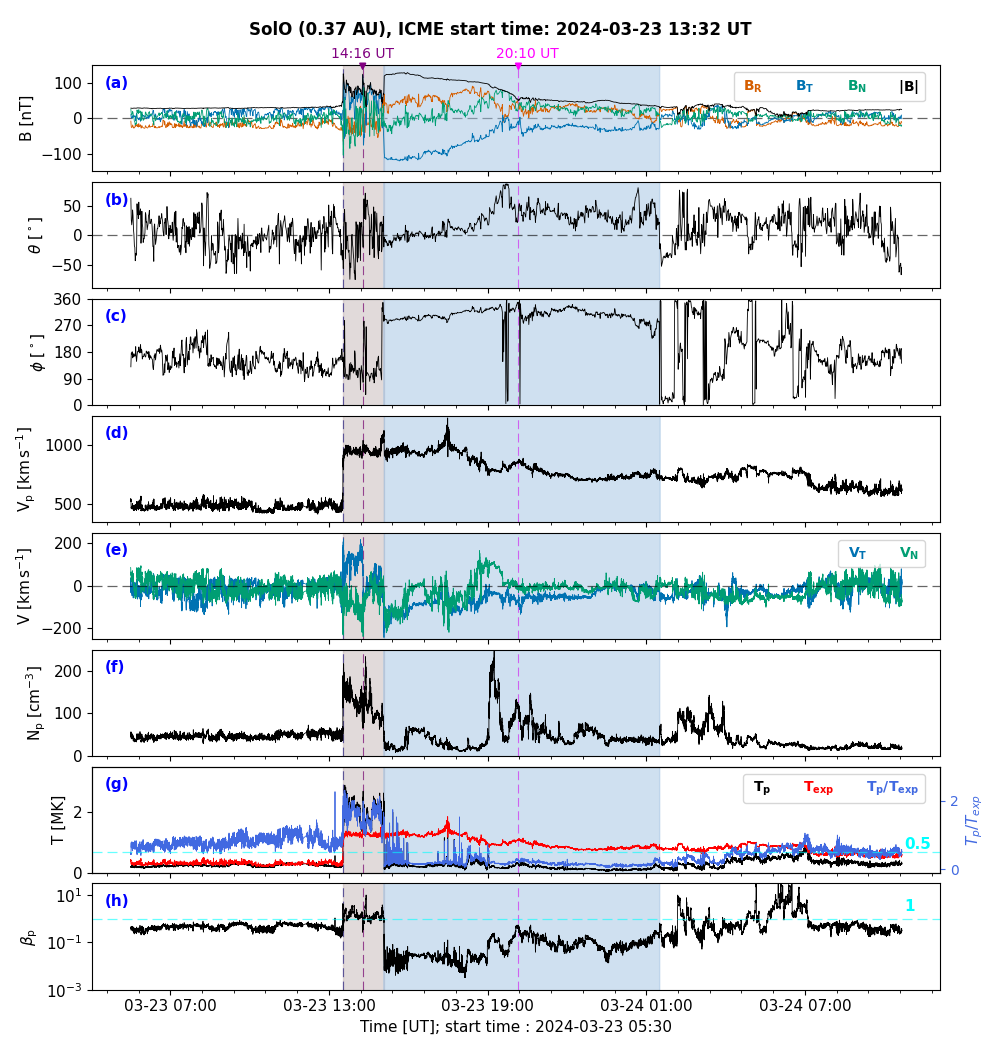}
\put(50,-5){\large (i)}
\end{overpic}
\end{minipage}
\hfill
\begin{minipage}[t]{0.495\textwidth}
\centering
\begin{overpic}[width=\linewidth]{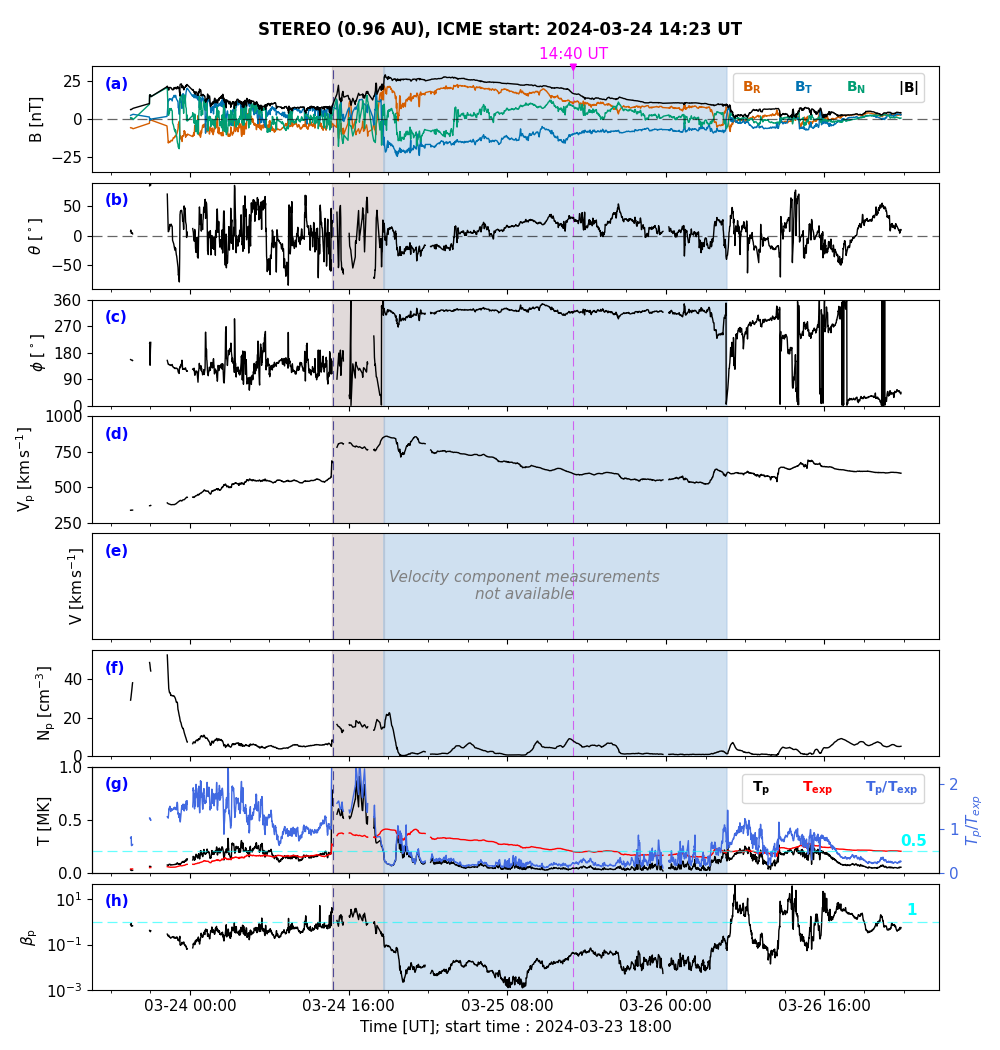}
\put(50,-5){\large (ii)}
\end{overpic}
\end{minipage}

\vspace{0.8cm}

\begin{minipage}[t]{0.495\textwidth}
\centering
\begin{overpic}[width=\linewidth]{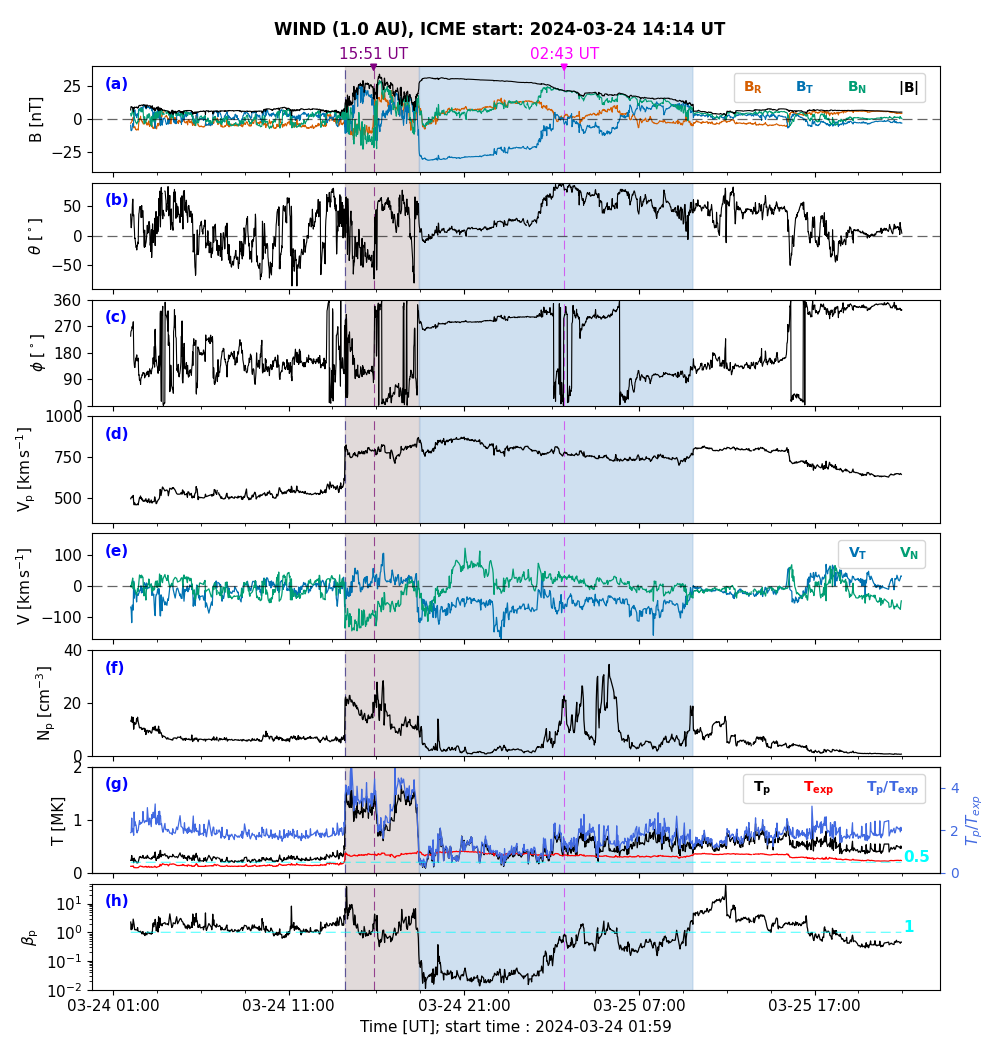}
\put(50,-5){\large (iii)}
\end{overpic}
\end{minipage}
\hfill
\begin{minipage}[t]{0.247\textwidth}
\centering
\begin{overpic}[width=\linewidth]{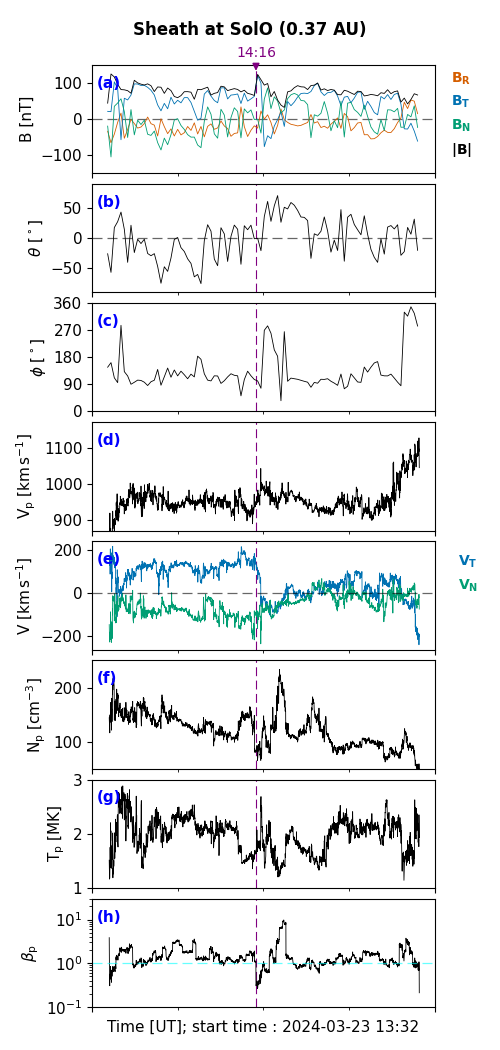}
\put(22,-5){\large (iv)}
\end{overpic}
\end{minipage}
\hfill
\begin{minipage}[t]{0.247\textwidth}
\centering
\begin{overpic}[width=\linewidth]{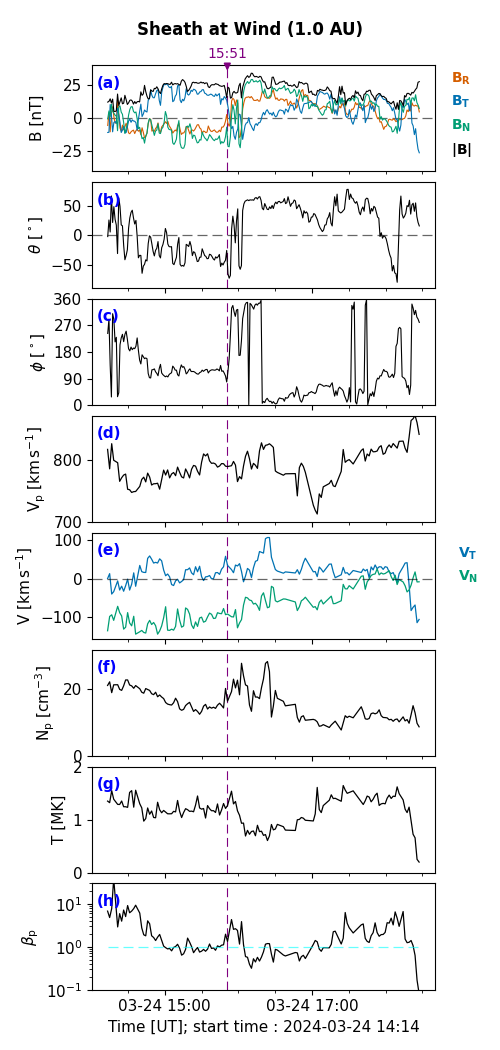}
\put(22,-5){\large (v)}
\end{overpic}
\end{minipage}

\vspace{0.4cm}

\caption{Time series of magnetic and plasma parameters observed by (i) SolO, (ii) STA, and (iii) Wind during the ICME event of 2024 March 23--24. From top to bottom, the panels show: (a) the magnetic field components ($B_R$, $B_T$, and $B_N$) and magnitude ($|B|$); (b) the latitude ($\theta$) of the magnetic field vector; (c) the longitude ($\phi$) of the magnetic field vector; (d) the proton bulk speed ($V_p$); (e) the proton velocity components ($V_T$ and $V_N$); (f) the proton number density ($N_p$); (g) the proton temperature ($T_p$, black), the expected proton temperature ($T_{\mathrm{exp}}$, red) derived from the empirical $V_p$--$T_p$ relation, and the temperature ratio $T_p/T_{\mathrm{exp}}$ (blue, right axis), where the horizontal dashed cyan line marks $T_p/T_{\mathrm{exp}}=0.5$; and (h) the proton plasma beta ($\beta_p$), where the horizontal dashed cyan line indicates $\beta_p=1$. The ICME sheath and ME are shaded in light gray and light blue, respectively. The blue and purple dashed lines mark the arrival of the shock and a secondary discontinuity within the sheath, while the magenta dashed line indicates the onset of the MCL ejecta "back region" (see text). Subfigures (iv) and (v) show these measurements for the sheath region observed by SolO and Wind, respectively, with panel descriptions identical to those in subfigures (i)--(iii).}

\label{fig:raw_data_all}
\end{figure*}

Figure~\ref{fig:raw_data_all} presents in situ magnetic field and plasma observations of the 2024 March 23--24 ICME from SolO, STA, and Wind. The panels show the time series of the magnetic field components ($B_R$, $B_T$, and $B_N$) and magnitude ($|B|$), the latitude ($\theta$) and longitude ($\phi$) of the magnetic field vector, the proton bulk speed ($V_p$), the proton velocity components ($V_T$ and $V_N$), the proton number density ($N_p$), the proton temperature ($T_p$), the expected proton temperature ($T_{\mathrm{exp}}$), the temperature ratio $T_p/T_{\mathrm{exp}}$, and the proton plasma beta ($\beta_p$). Here, $T_{\mathrm{exp}}$ denotes the expected temperature derived from the empirical $V_{p}$--$T_{p}$ relationship \citep{1986_Lopez,Liu2005}. The panel format is identical for all three spacecraft, with the ICME sheath region highlighted in light gray and the ME region in light blue. At SolO (Figure~\ref{fig:raw_data_all}(i)), an interplanetary shock associated with this ICME arrived at 13:32~UT on 23 March, followed by a turbulent sheath characterized by strongly fluctuating magnetic fields, enhanced density, elevated proton temperature $(T_{p}/T_{\mathrm{exp}}>1)$, and \(\beta_p > 1\). The subsequent ME (23 March 15:05~UT--24 March 01:30~UT) exhibits enhanced \(|B|\), reduced proton temperature $(T_{p}/T_{\mathrm{exp}}<0.5)$, and low $\beta_{p}$, consistent with MC signatures \citep{1981_Burlaga, 2006_Zurbuchen}. However, magnetic rotation is evident only in the leading portion of the ejecta, while the trailing portion shows minimal rotation and nearly constant \(|B|\). A transition at 20:10~UT (magenta dashed line) marks the boundary between the rotating front and the non-rotating ``back region'', identifying the structure as MCL \citep{2006_Lepping} ejecta. STA (Figure~\ref{fig:raw_data_all}(ii)) observed a shock at 24 March 14:23~UT, followed by a sheath and a ME interval spanning 24 March 19:32--26 March 06:11~UT. The ME again displays a rotating front portion and a trailing, nearly non-rotating ``back region'', with a transition at 25 March 14:40~UT, consistent with a MCL structure. Wind observations (Figure~\ref{fig:raw_data_all}(iii)) show the arrival of the shock at 14:14~UT on 24 March, followed by a sheath; the ME is detected from 18:28~UT on 24 March to 10:04~UT on 25 March. A transition to a flat, non-rotating ``back region" at 25 March 02:43~UT confirms that Wind also encountered an MCL configuration during this event. We note that the purple dashed lines within the sheath at SolO and Wind indicate a discontinuity within the sheath region.

To examine the sheath in greater detail, Figure~\ref{fig:raw_data_all}(iv) and Figure~\ref{fig:raw_data_all}(v) show the in situ time series of the sheath intervals observed at SolO and Wind, respectively. At SolO, around 14:16~UT on March 23, the magnetic field undergoes an abrupt reconfiguration characterized by enhanced fluctuations and a reversal in polarity. In particular, the $B_T$ component reverses its polarity and decreases to values close to zero shortly afterward. Correspondingly, the $\phi$ changes rapidly from approximately $90^\circ$ to $270^\circ$, before returning to lower values and subsequently rotating back toward $\sim270^\circ$, indicating a substantial change in the magnetic-field orientation. At the same time,  the plasma parameters also exhibit pronounced changes. The $V_p$ increases rapidly, $N_p$ decreases sharply, and $T_p$ shows a slight enhancement. A pronounced flow shear is also observed, with $V_T$ changing from approximately $140$ to $-50$~km~s$^{-1}$. Meanwhile, $\beta_p$ decreases abruptly to values below unity, subsequently increases to values greater than one, and then gradually returns to values close to unity. The simultaneous changes in the magnetic-field and plasma parameters indicate a localized transition within the sheath, possibly associated with a discontinuity or a distinct compression region. A similar transition is also observed at Wind around 15:51~UT on March 24, where the magnetic field undergoes an abrupt reconfiguration, with rapid changes in all three RTN components. In contrast to SolO, only a weak flow shear is observed across this interval. Correspondingly, $\phi$ changes abruptly across this interval and subsequently evolves more gradually, indicating comparatively smoother magnetic-field rotation in the trailing portion of the sheath. The plasma parameters exhibit comparatively modest changes. The $N_p$ shows no obvious systematic variation across the transition, whereas $T_p$ decreases significantly. The $\beta_p$ initially increases gradually, decreases to values below unity across the transition, and subsequently increases again to values close to unity. Overall, compared to the SolO observations, the plasma parameters at Wind evolve more gradually across this discontinuity.

\begin{figure*}
\centering
\includegraphics[width=0.855\linewidth]{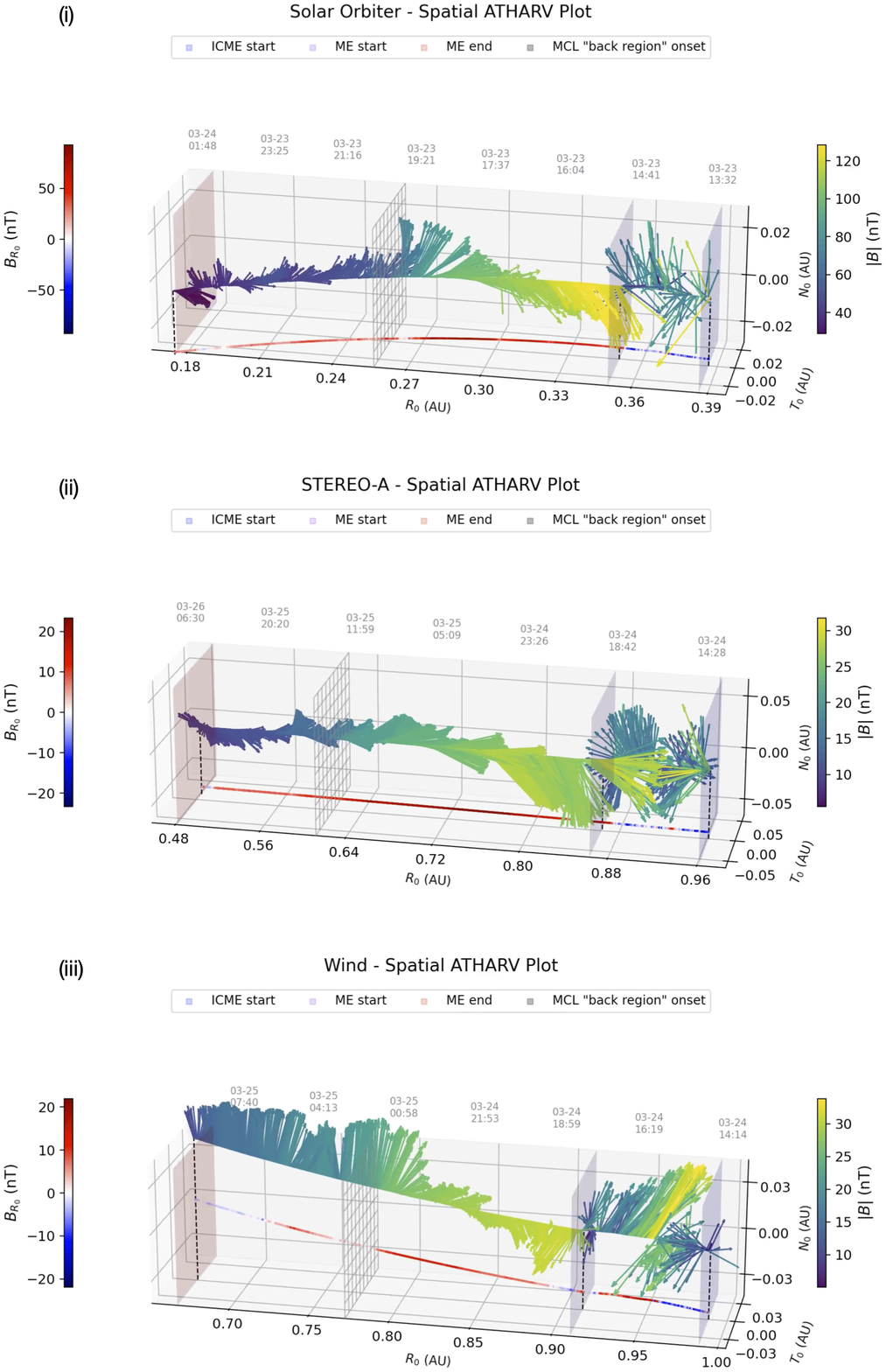}
\caption{ 3D visualization of the magnetic-field vectors measured along the spacecraft trajectories during the ICME event of 2024 March 23--24 for (i) SolO, (ii) STA, and (iii) Wind, generated using the ATHARV tool (Menon et al. 2026, under revision). Arrows indicate the magnetic field direction, with length and color representing $|B|$ in the $R_{0}T_{0}N_{0}$ frame at the initial ICME encounter time $t_{0}$. The arrow tails trace the remapped spacecraft trajectory in the $R_{0}T_{0}N_{0}$ frame, illustrating the path sampled through the ICME. A parallel red--white--blue strip shows the magnitude and sign of the $B_{R_{0}}$ component along the same trajectory. The ICME propagates radially outward (to the right); the blue, indigo, and red planes mark the ICME start, ME start, and ME end, respectively. The gray meshed plane marks the onset of the MCL ejecta ``back region". An animation of the ATHARV visualizations for each spacecraft is available online.}
\label{fig:vector_plot_all}
\end{figure*}

\subsubsection{Three-dimensional (3D) Representation of the ICME Magnetic Field}

Figure~\ref{fig:vector_plot_all} shows a 3D visualization of the magnetic field vectors within the ICME at each spacecraft, generated using the ATHARV tool (Menon et al. 2026, under revision). ATHARV maps the time-series magnetic-field observations into spatial coordinates along the spacecraft trajectory using the measured plasma velocity, assuming that the ICME magnetic structure is approximately frozen into the solar-wind flow. In deriving these spatial coordinates, the reconstruction accounts for the changing spacecraft position during the encounter and corrects for ICME expansion using an isotropic self-similar approximation, thereby mapping the measurements to their corresponding spatial locations at a chosen reference time. The resulting visualization complements conventional time-series analysis and facilitates assessment and comparison of the spatial scales of large-scale magnetic-field rotations, coherent structures, and boundaries within the ejecta across spacecraft. The $R_{0}T_{0}N_{0}$ coordinate system corresponds to the spacecraft radial, tangential, and normal directions at the initial ICME encounter time $t_{0}$ (e.g., $t_{0}=13{:}32$~UT for SolO). Magnetic field vectors are displayed as arrows whose orientation represents the local field direction, while their length and color encode the field magnitude ($|B|$). The arrow tails are anchored along the remapped spacecraft trajectory in the $R_{0}T_{0}N_{0}$ frame, illustrating the path sampled through the ICME. A parallel red--white--blue strip indicates the magnitude and sign of the $B_{R_{0}}$ component of the magnetic field along the same trajectory. The ICME propagates radially outward (to the right in Figure~\ref{fig:vector_plot_all}), with the blue, indigo, and red planes marking the start of the sheath, the onset of the ME, and the end of the ME, respectively.

At all spacecraft, the sheath is characterized by highly disordered magnetic field vectors, consistent with strong turbulence. In contrast, the ME exhibits clear magnetic field rotation and a gradual decrease in field strength in its leading portion, followed by a trailing region with little or no rotation and nearly constant $|B|$. The gray meshed plane in Figure~\ref{fig:vector_plot_all} marks the transition between these regions, corresponding to the onset of the ``back region'' identified in the time-series observations and further supporting the MCL classification of the ejecta. To aid visualization of the three-dimensional magnetic field structure, animations of the ATHARV plots have been generated for each spacecraft, showing snapshots of the 3D vector fields from multiple viewing angles. These animations are available for all three spacecraft at \url{https://doi.org/10.6084/m9.figshare.32043171}.

\begin{figure*}[t]
\centering

\begin{subfigure}{0.33\textwidth}
    \centering
    \begin{overpic}[width=\linewidth]{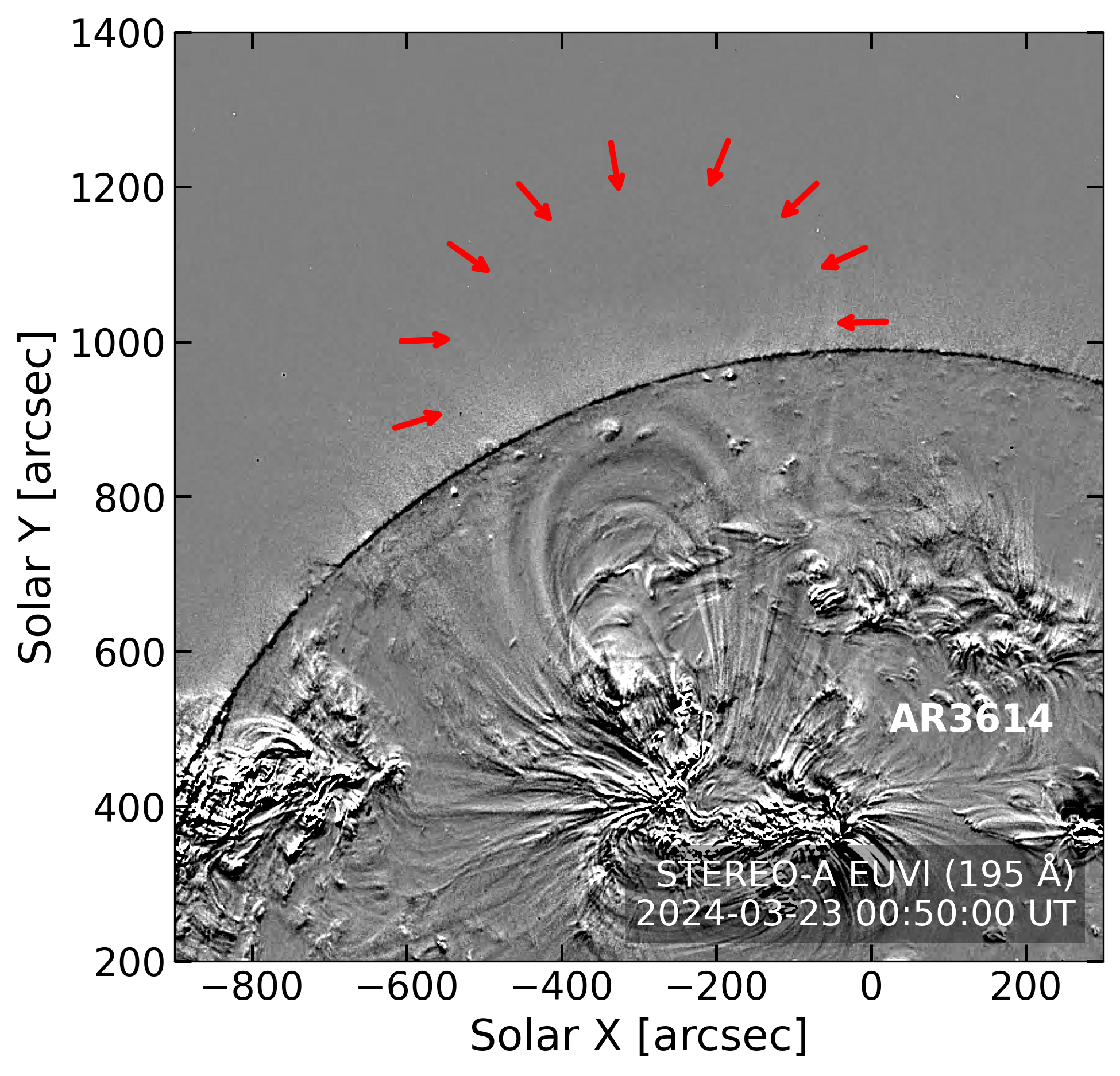}
        \put(18,85){\large\bfseries (a)}
    \end{overpic}
\end{subfigure}
\hfill
\begin{subfigure}{0.33\textwidth}
    \centering
    \begin{overpic}[width=\linewidth]{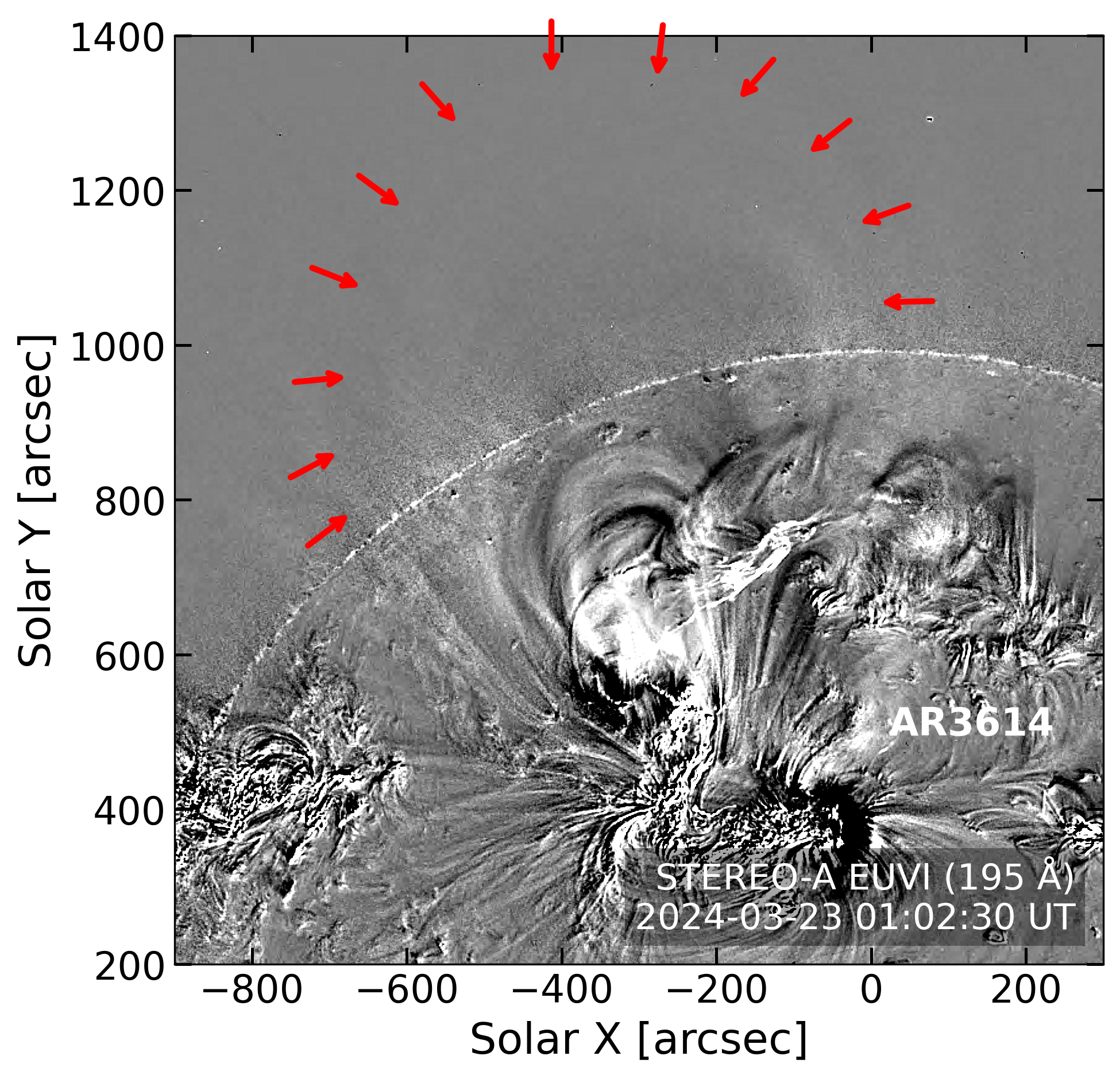}
        \put(18,85){\large\bfseries (b)}
    \end{overpic}
\end{subfigure}
\hfill
\begin{subfigure}{0.33\textwidth}
    \centering
    \begin{overpic}[width=\linewidth]{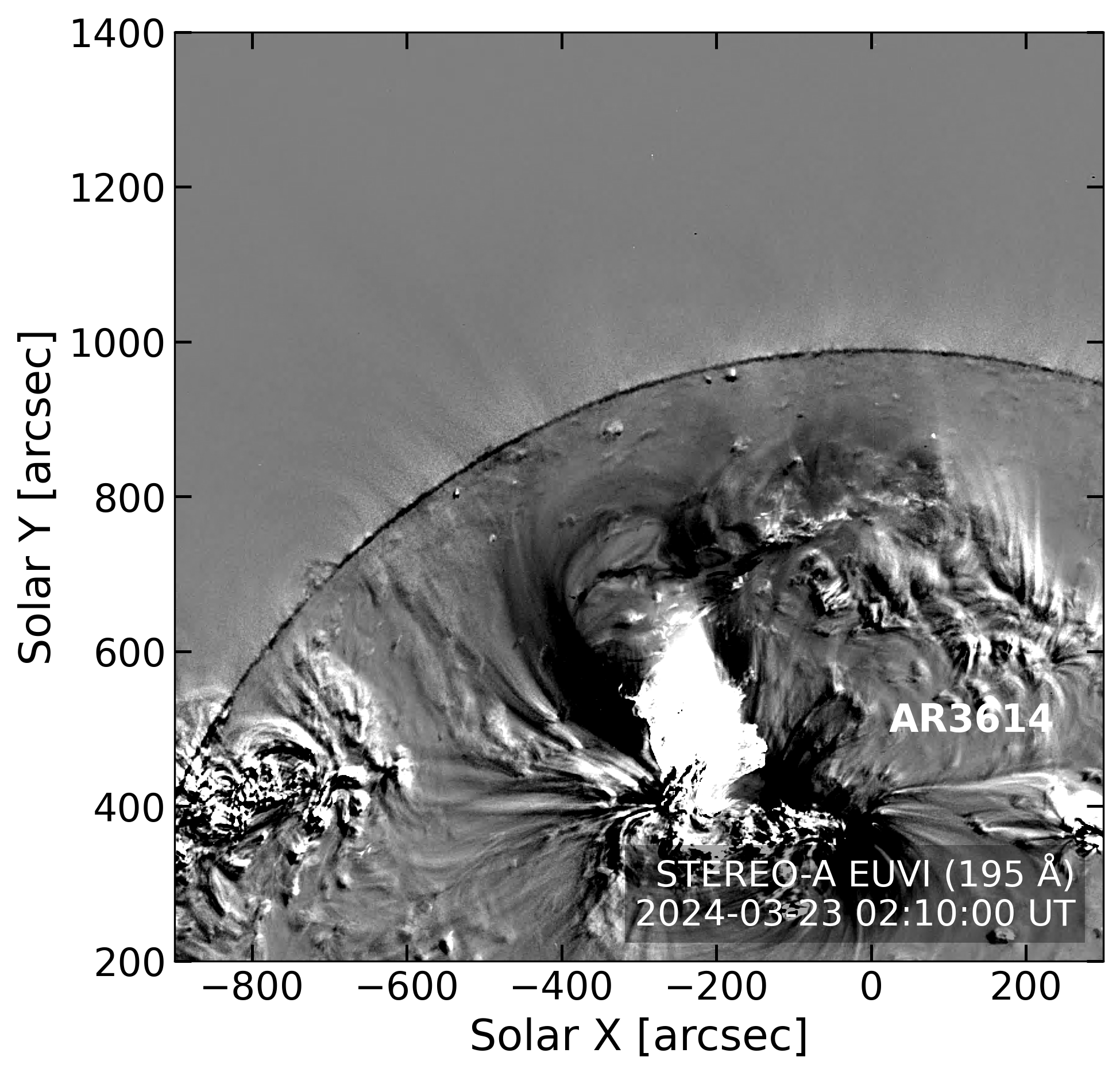}
        \put(18,85){\large\bfseries (c)}
    \end{overpic}
\end{subfigure}

\vspace{0.3cm}

\begin{subfigure}{0.33\textwidth}
    \centering
    \begin{overpic}[width=\linewidth]{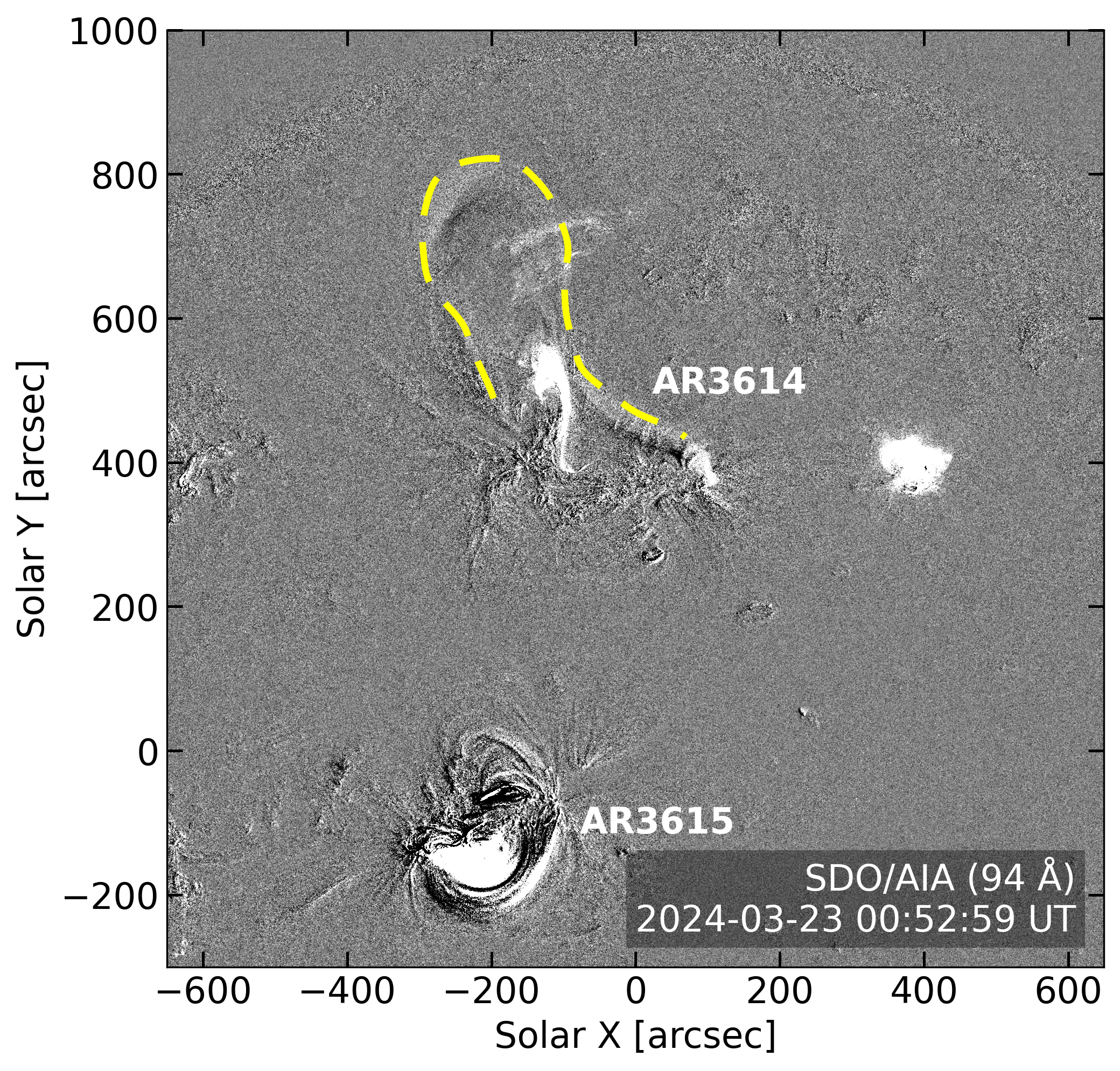}
        \put(18,85){\large\bfseries (d)}
    \end{overpic}
\end{subfigure}
\hfill
\begin{subfigure}{0.33\textwidth}
    \centering
    \begin{overpic}[width=\linewidth]{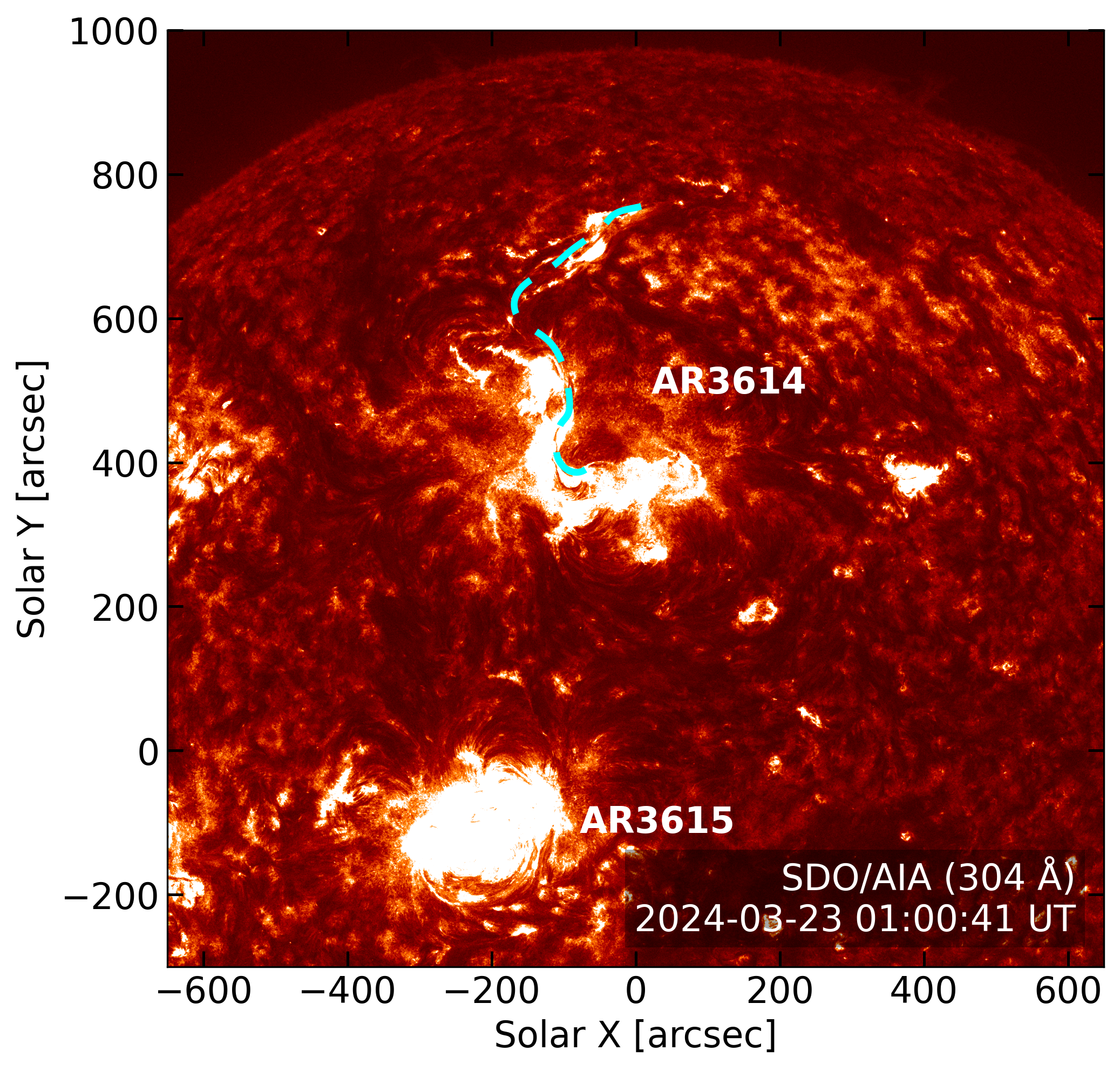}
        \put(18,85){\large\bfseries (e)}
    \end{overpic}
\end{subfigure}
\hfill
\begin{subfigure}{0.33\textwidth}
    \centering
    \begin{overpic}[width=\linewidth]{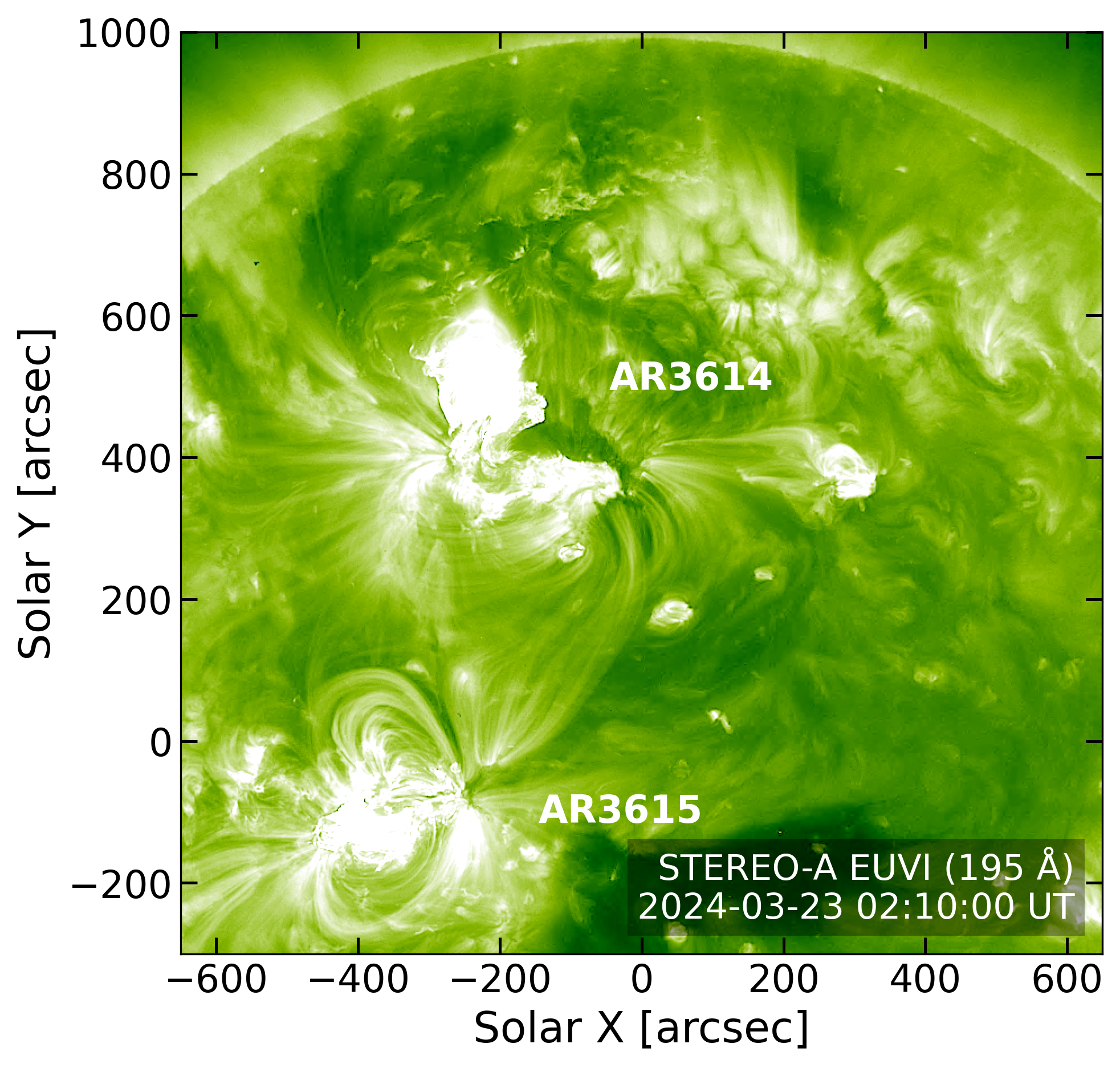}
        \put(18,85){\large\bfseries (f)}
    \end{overpic}
\end{subfigure}
\caption{Overview of the EUV observations of the 2024 March 23 solar eruption. (a)–(c) Running-difference EUVI 195~\AA\ images. Panels~(a) and (b): expansion and rise of a faint coronal loop system above AR~3614, with red arrows marking its outer boundary. Panel~(c): erupted loop system and post-eruptive arcade above AR~3614. (d) Base-difference AIA 94~\AA\ image showing the hot-channel MFR, the source of CME1, outlined by a yellow dashed line. (e) AIA 304~\AA\ image depicting the erupting filament, the source of CME2, highlighted by a green dashed line. (f) EUVI 195~\AA\ image showing the post-eruptive arcade above AR~3614, the associated coronal dimming, and a sympathetic eruption from AR~3615.}
\label{fig:euv_obs}
\end{figure*}

\begin{figure*}[t]
\centering


\begin{subfigure}{0.33\textwidth}
    \centering
    \begin{overpic}[width=\linewidth]{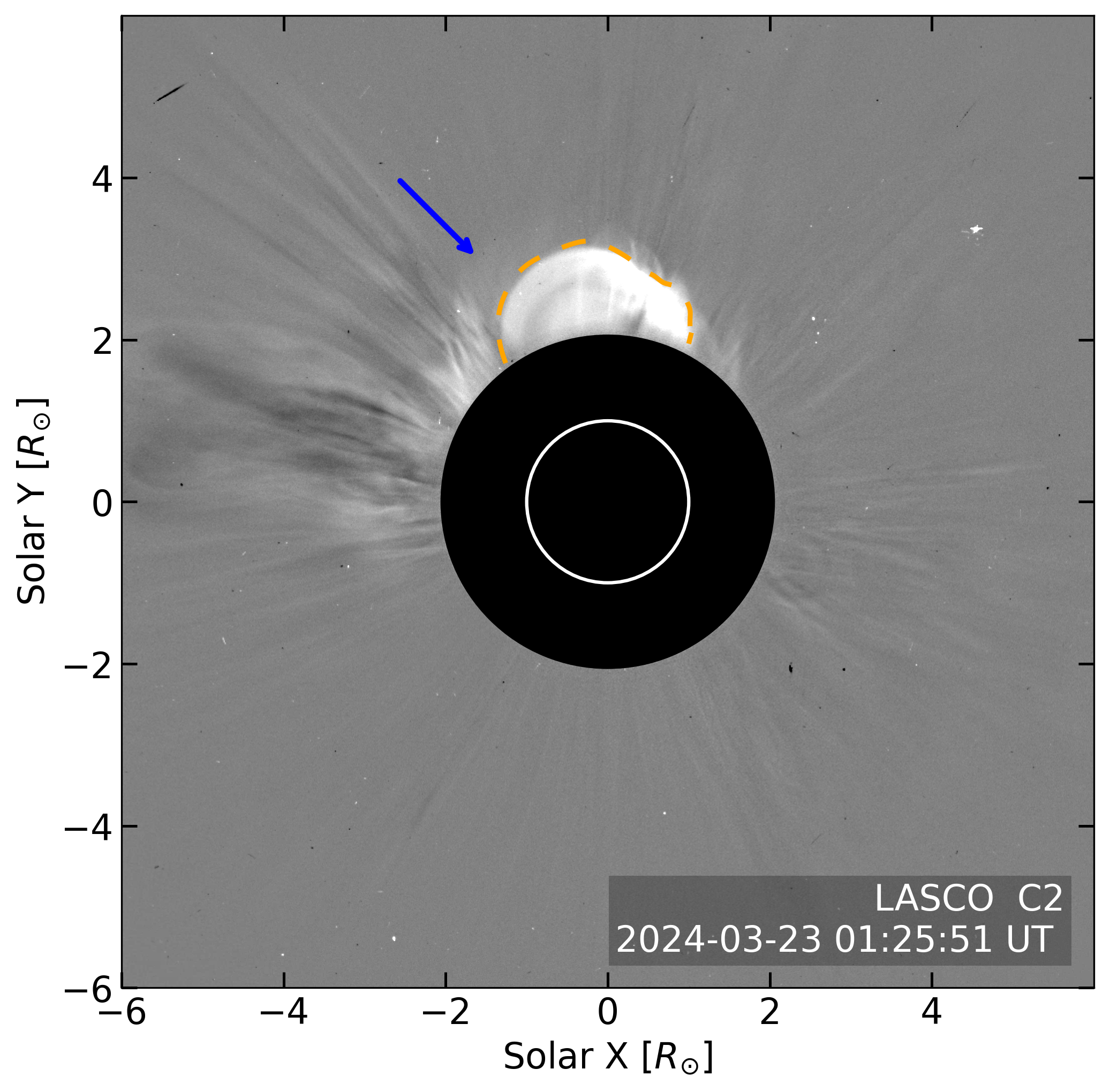}
        \put(15,88){\large\bfseries (a)}
    \end{overpic}
\end{subfigure}
\hfill
\begin{subfigure}{0.33\textwidth}
    \centering
    \begin{overpic}[width=\linewidth]{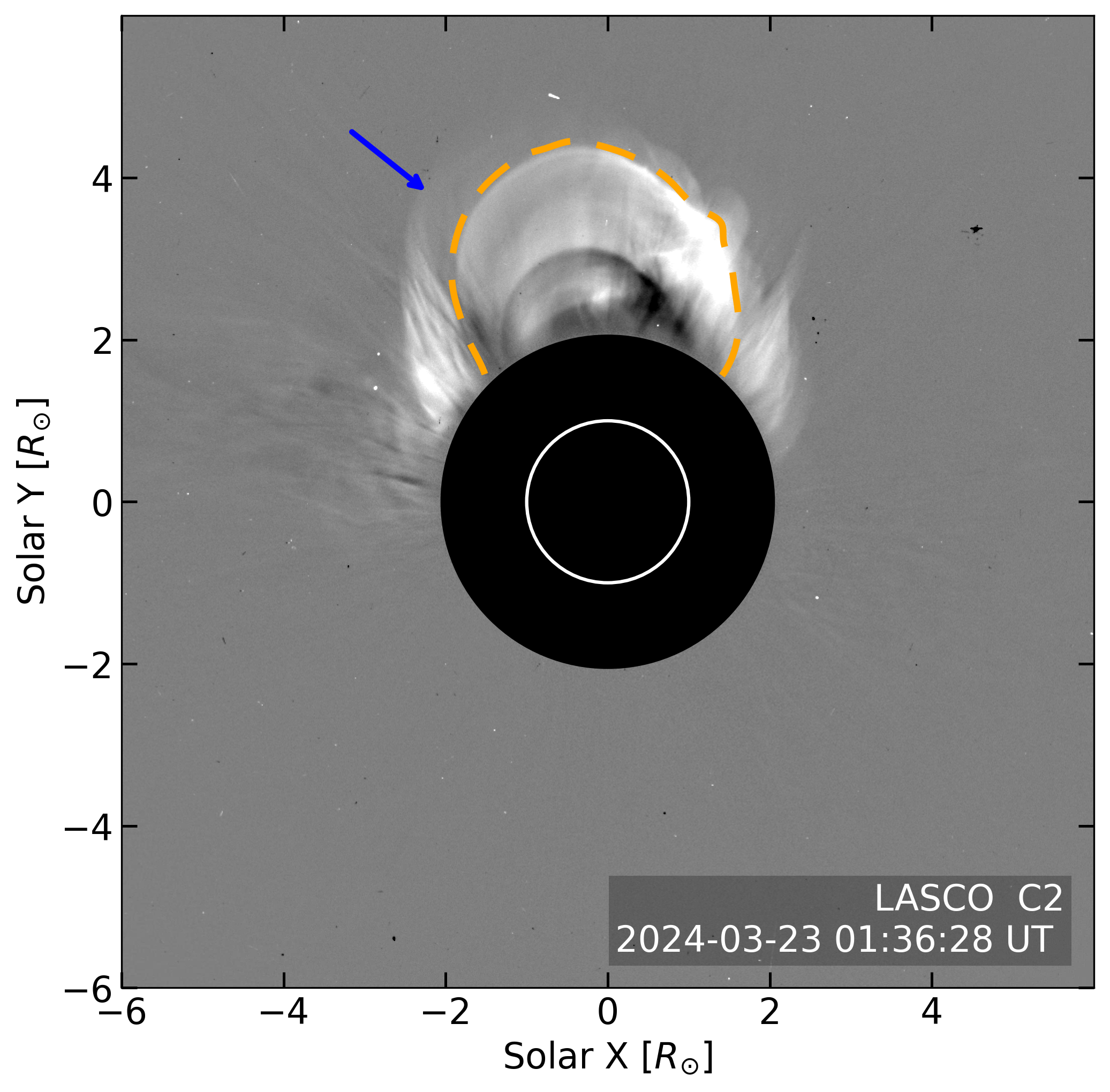}
        \put(15,88){\large\bfseries (b)}
    \end{overpic}
\end{subfigure}
\hfill
\begin{subfigure}{0.33\textwidth}
    \centering
    \begin{overpic}[width=\linewidth]{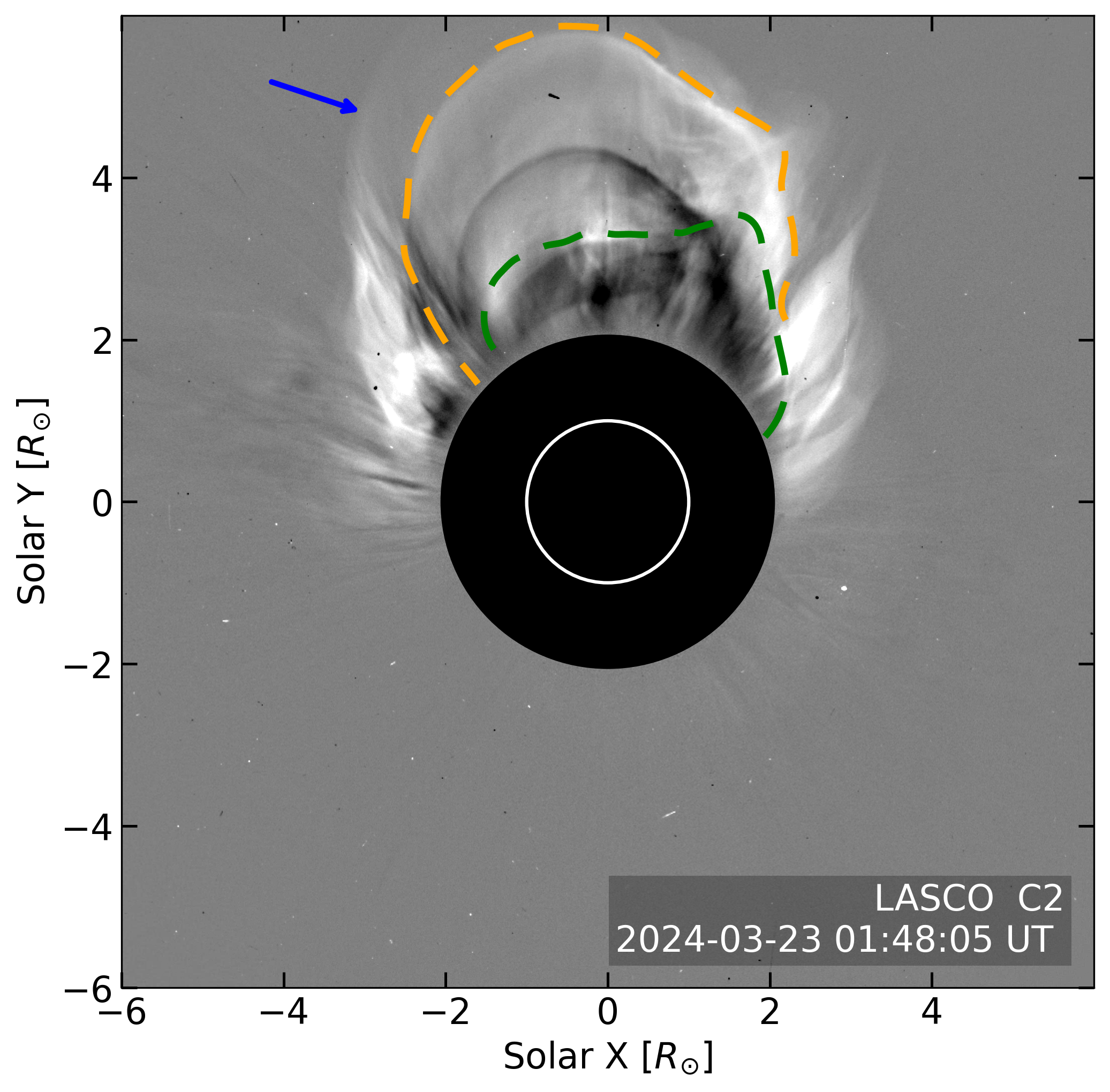}
        \put(15,88){\large\bfseries (c)}
    \end{overpic}
\end{subfigure}

\vspace{0.3cm}

\begin{subfigure}{0.33\textwidth}
    \centering
    \begin{overpic}[width=\linewidth]{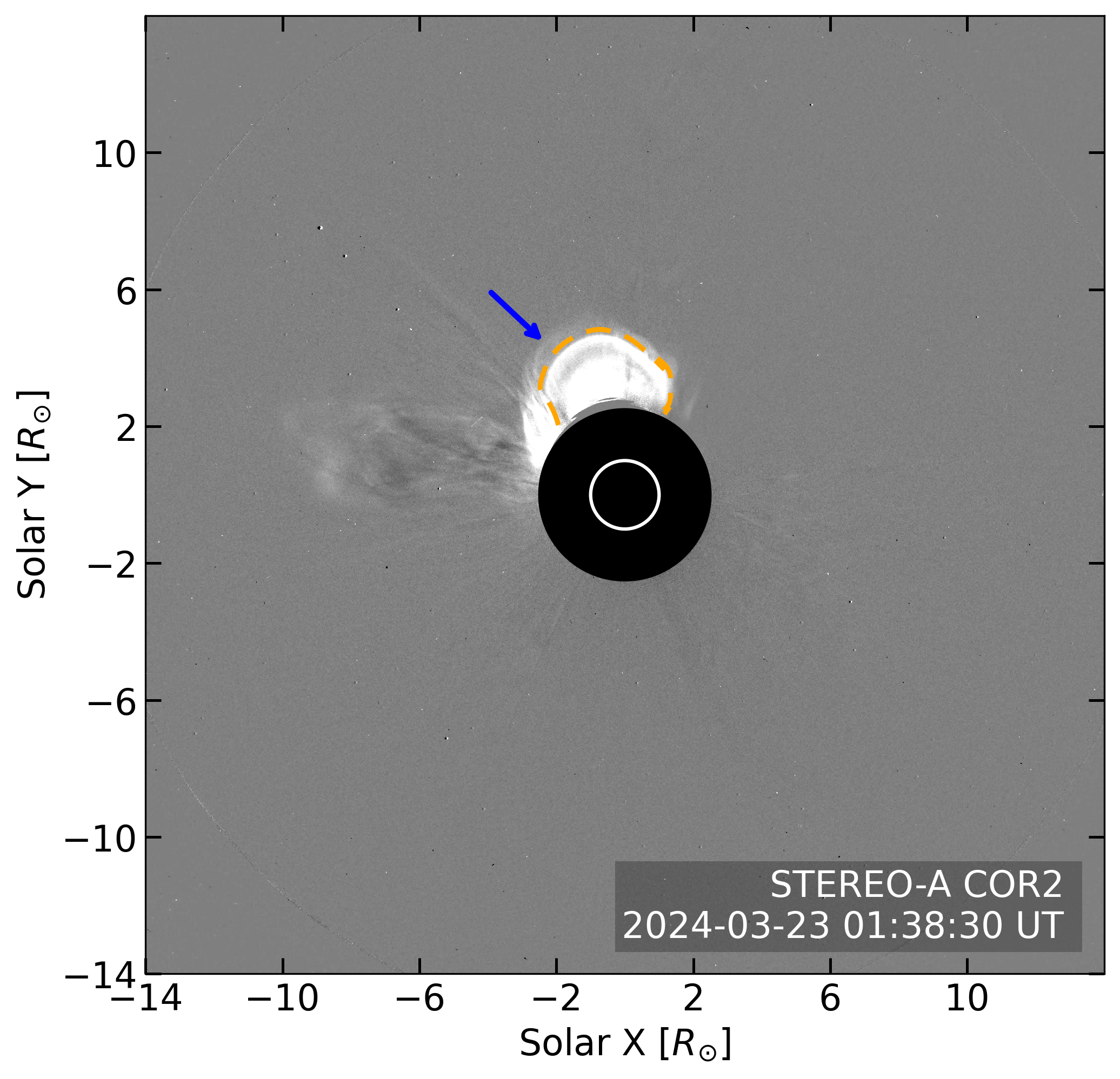}
        \put(15,88){\large\bfseries (d)}
    \end{overpic}
\end{subfigure}
\hfill
\begin{subfigure}{0.33\textwidth}
    \centering
    \begin{overpic}[width=\linewidth]{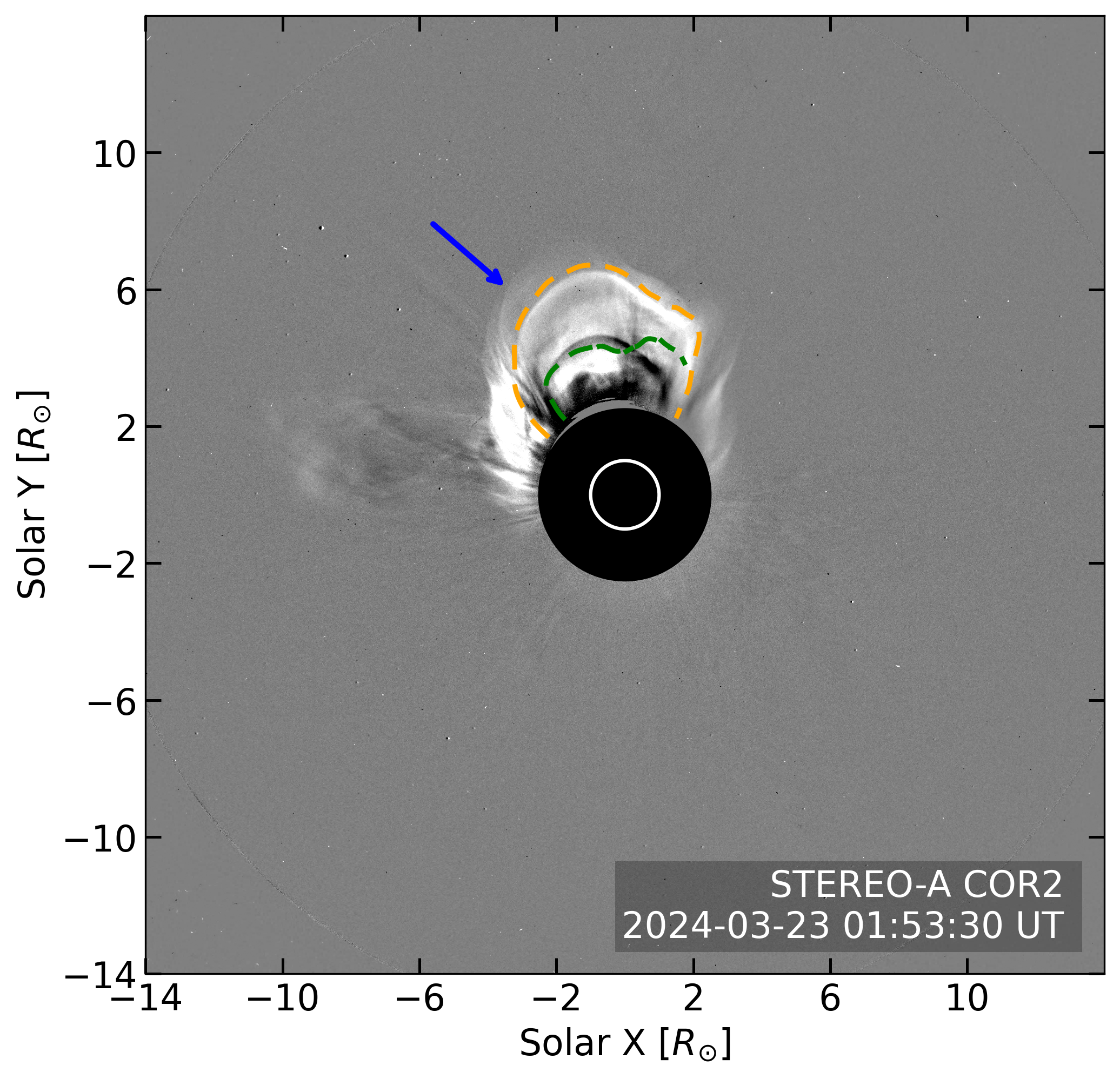}
        \put(15,88){\large\bfseries (e)}
    \end{overpic}
\end{subfigure}
\hfill
\begin{subfigure}{0.33\textwidth}
    \centering
    \begin{overpic}[width=\linewidth]{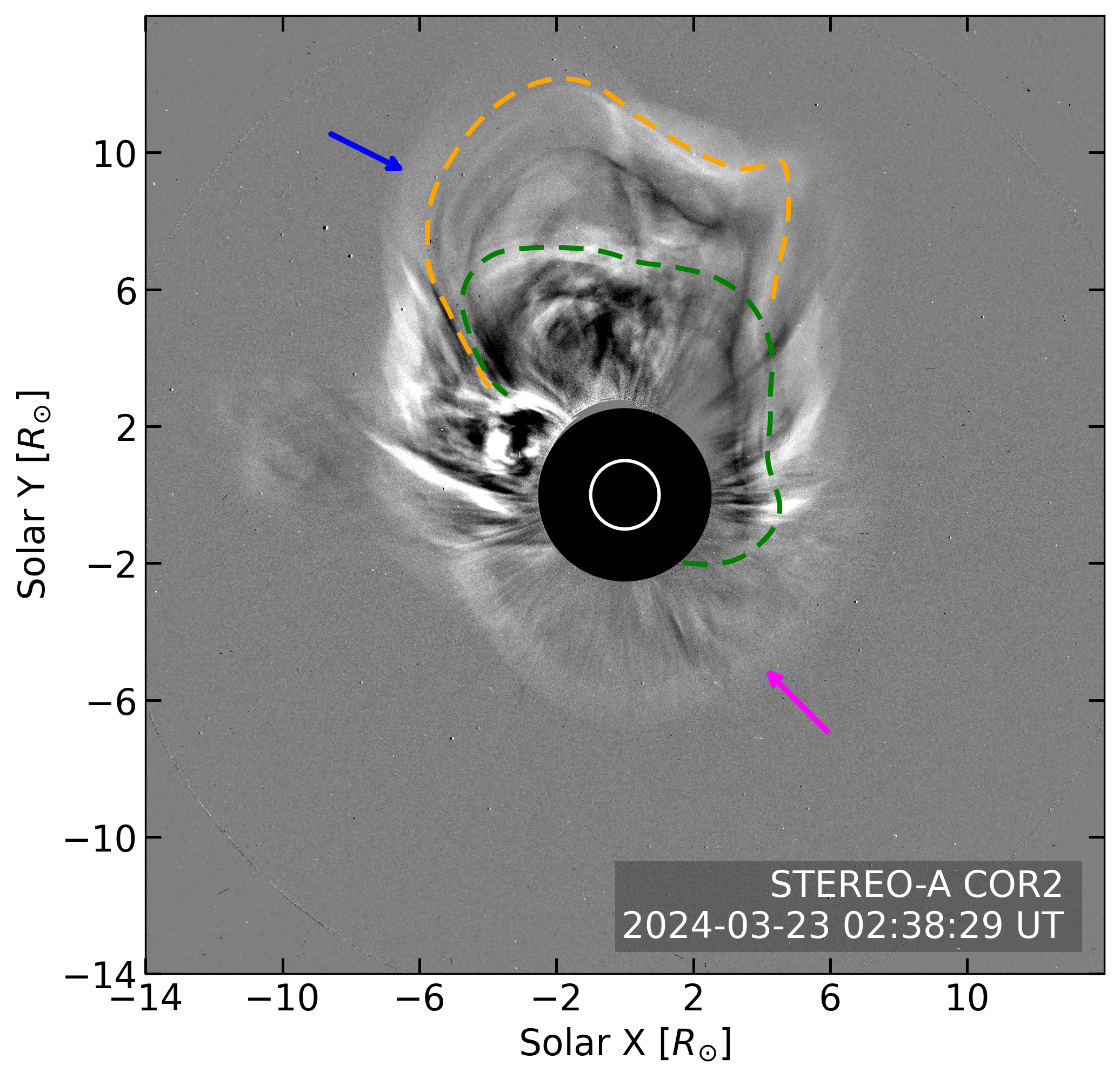}
        \put(15,88){\large\bfseries (f)}
    \end{overpic}
\end{subfigure}

\caption{Overview of the white-light observations of the solar eruption on 2024 March 23. (a)–(c) 12 min running-difference images from the SOHO/LASCO-C2 coronagraph. (d)–(f) 15 min running-difference images from the STEREO-A/COR2 coronagraph. The orange dashed lines outline the leading edge of CME1, while the green dashed lines mark the leading edge of CME2. The blue arrow indicates the faint shock front associated with CME1, and the magenta arrow indicates the shock front associated with CME2.}
\label{fig:wl_obs}
\end{figure*}

\subsection{Near-Sun EUV and White-Light Observations}

Figure~\ref{fig:euv_obs} provides an overview of the low-coronal evolution of the 2024 March 23 eruption. The source region comprises two magnetically connected active regions, AR~3614 and AR~3615, forming a quadrupolar magnetic configuration. The earliest signature of the eruption is the expansion and rise of a faint coronal loop system above AR~3614 (marked by red arrows), which began around 00:50~UT and is best observed in EUVI images, as shown in panels~(a)--(b). Nearly simultaneously, an X1.1 flare occurred in AR~3614 (N25E07). The flare began at 00:58~UT, peaked at 01:33~UT, and ended at 01:52~UT. EUV observations reveal two successive eruptive structures originating from AR~3614: a hot-channel MFR (outlined by the yellow dashed curve in panel~d), identified as the source of the first eruption, and a large filament (outlined by the cyan dashed curve in panel~e), which gives rise to the second eruption. The two eruptions occur in close succession around 01:00~UT, producing two distinct CME components that are subsequently observed in white-light coronagraph images, characteristic of a twin-CME scenario, as also suggested by \citet{2024_Mishev}. Following the eruptions, prominent post-eruptive arcades develop above AR~3614, accompanied by large-scale coronal dimming and a sympathetic eruption from AR~3615, evident in panels~(c) and (f).

Figure~\ref{fig:wl_obs} illustrates the subsequent white-light evolution of the two low-coronal eruptions. In LASCO-C2 observations, the two eruptions appear as successive CMEs propagating primarily northward. The first eruption (hereafter CME1) appears on 23 March at 01:25:53~UT as a bright partial-halo CME (outlined by the orange dashed line), with a faint shock front visible ahead of its leading edge (indicated by the blue arrow), as shown in panels~(a)--(b). The second eruption (hereafter CME2) first appears in the LASCO-C2 field of view at 01:48:07~UT as a bright partial-halo CME (outlined by the green dashed line) propagating northwestward and closely trailing CME1, as shown in panel~(c). Both eruptions are also observed in the STEREO-A/COR2 field of view (panels~(d)--(f)), where CME1 first appears on 23 March at 01:38:43~UT, followed by CME2 at 01:53:43~UT. A second shock front, visible as a faint halo surrounding CME2 (indicated by the magenta arrow), is also identified in the COR2 images.

By the time CME2 first becomes distinguishable in the coronagraph images, its leading edge is already located at a heliocentric distance of approximately $2$--$3~R_{\odot}$ and lies within the central portion of CME1, indicating that the interaction between the two CMEs had already begun at low coronal heights. Moreover, only part of the shock front associated with CME2 can be identified in the COR2 images, while the northern portion of the shock front appears to be embedded within or obscured by the white-light structure of CME1. This morphology is consistent with the interpretation that the northern part of the shock driven by CME2 is already propagating through plasma disturbed by CME1 rather than through the ambient solar wind. The interaction could therefore have begun either when the leading edge of CME2 reached the trailing edge of CME1 or earlier, when the shock driven by CME2 first encountered the trailing edge of CME1 \citep{2013_Lugaz}. However, the exact height at which the interaction commenced cannot be determined from the available observations, as the onset of the interaction may have occurred below the inner boundary of the coronagraph field of view. In addition, data gaps in the STEREO-A/COR2 observations further limit our ability to determine the precise interaction height.

\subsection{Heliospheric Propagation of the 2024 March 23–24 ICME Event Using the ELEvo Model}

The \texttt{HELIO4CAST} LineupCAT is a living catalog that lists CME events measured in situ at more than one spacecraft, i.e., so-called multipoint events. It incorporates information from the Database of Notifications, Knowledge, and Information (DONKI), developed at the Community Coordinated Modeling Center (CCMC). The DONKI catalog provides near-Sun CME kinematics, including speed, propagation direction, and angular width, derived from multi-viewpoint coronagraph observations from SOHO/LASCO C2/C3 and STEREO-A/COR2, together with geometric reconstruction of CME morphology \citep{2015_Mays}.

In order to identify multipoint events, the propagation of the CMEs listed in DONKI is simulated using the ELliptical Evolution \cite[ELEvo;][]{2015_Mostl} model. If a modeled CME passes more than one spacecraft \textit{in situ}, and the magnetic field data from the corresponding spacecraft show signatures of a CME or ME \citep{2006_Zurbuchen,2017_Kilpua}, then the event is added to the LineupCAT. ELEvo is a semi-empirical model that assumes the shock fronts of CMEs to be elliptical. These ellipses are propagated outward through the heliosphere using a simple drag-based model \citep{2012_Vrsnak}. Consequently, the CME kinematics are determined by the interaction between the CME and the ambient solar wind, with slower CMEs being accelerated and faster CMEs being decelerated. However, changes in CME kinematics and propagation direction resulting from interactions with other CMEs or large-scale solar structures are not modeled by ELEvo.

To account for uncertainties in the drag-based model, we vary the input parameters: the initial CME speed, the ambient solar wind speed, and the drag parameter, $\gamma$. To simulate CME propagation, we therefore use an ensemble of 50,000 normally distributed elements, from which the model parameters are randomly selected at each time step. This approach provides estimates of the uncertainties in both the predicted arrival time and arrival speed. The initial values of $\gamma$ and the ambient solar wind speed are $0.1 \times 10^{-7}$~km$^{-1}$ and 400~km,s$^{-1}$, respectively, with associated $1\sigma$ uncertainties of $0.025 \times 10^{-7}$~km$^{-1}$ and 50~km,s$^{-1}$. In addition, the inverse aspect ratio of the ellipse, required by ELEvo, is fixed at 0.7. All other input parameters are adopted from the DONKI catalog.

\begin{figure*}[htbp]
\centering
\includegraphics[width=0.99\textwidth]{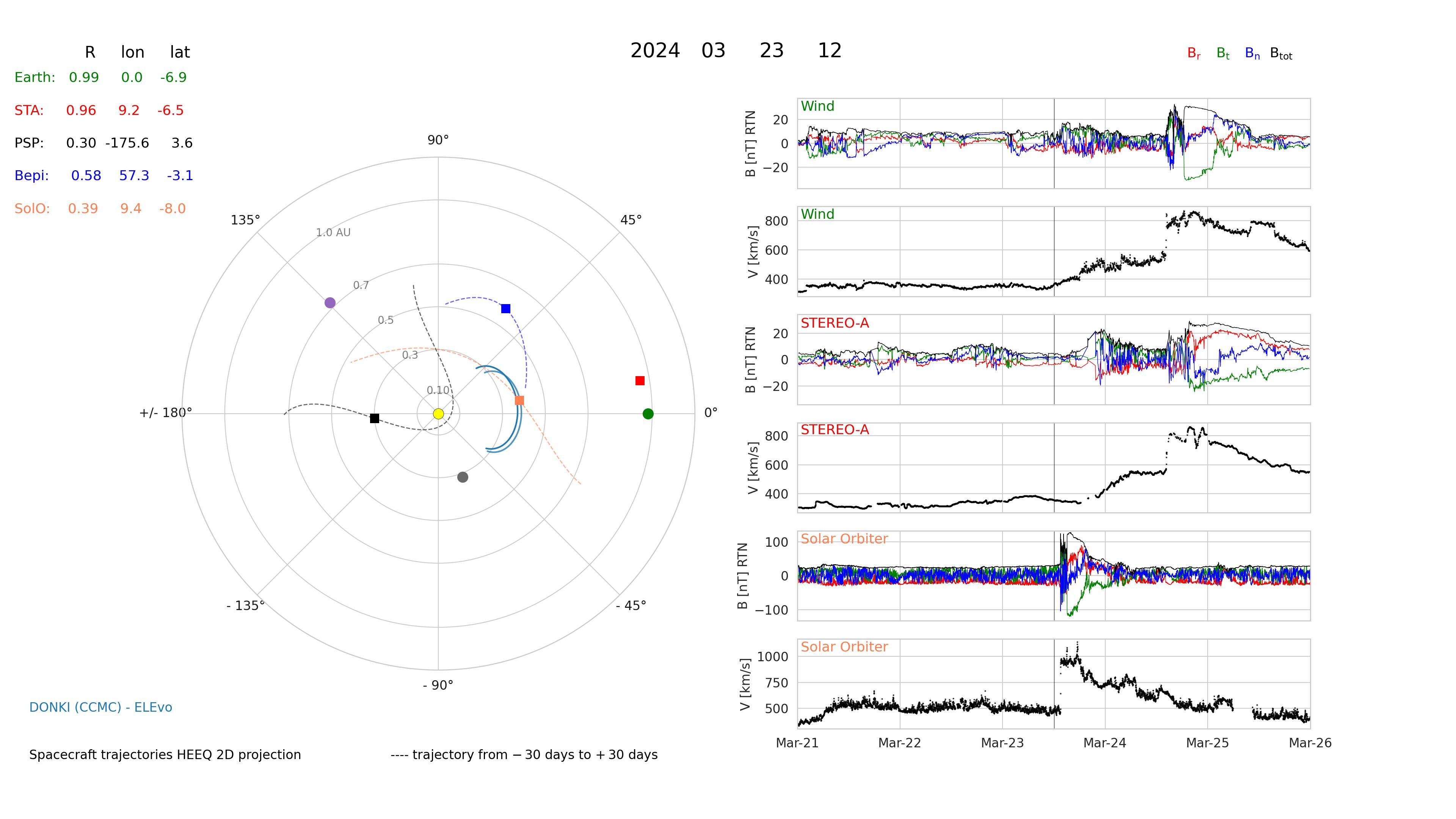}
\caption{Frame from an animation illustrating the 2024 March 23--24 ICME event detected \textit{in situ} by SolO, STA, and Wind. Left: Spacecraft positions in HEEQ coordinates together with the CME fronts (blue ellipses) obtained from the ELEvo model. The green circle at 0$^{\circ}$ represents Earth, the gray and purple circles indicate Mercury and Venus, respectively. Right: Time series of \textit{in situ} measurements from multiple spacecraft. From top to bottom: Magnetic field components in RTN coordinates and solar-wind bulk speed for Wind, STA, and SolO. The animation covers 2024 March 22--27 at 1~hr cadence and illustrates the propagation of the CMEs through the inner heliosphere. Here, a frame from 2024 March 23 12:00~UT is shown, where the vertical black line in the in situ panels corresponds to this time. The animation begins on 2024 March 22 and runs through 2024 March 27 with a cadence of 1~hr, and is available online.}
\label{fig:elevo_icmes}
\end{figure*}

DONKI reports that the 2024 March 23--24 ICME corresponds to the combined arrival of two CMEs launched in close succession (CME1 at 2024 March 23 01:25~UT and CME2 at 2024 March 23 01:48~UT), predicted to arrive at L1 on 2024 March 24 at 14:10~UT. We use the average CME parameters from all measurement types (LE and SH) reported in the DONKI catalog as input to the Ellipse Evolution \cite[ELEvo;][]{2015_Mostl} model to propagate both CMEs and track their heliospheric evolution. The adopted parameters are $(\mathrm{lon},\mathrm{lat})=(2^\circ,22^\circ)$ in HEEQ, a speed of $1613~\mathrm{km~s^{-1}}$, and a half-width of $41^\circ$ for CME1, and $(\mathrm{lon},\mathrm{lat})=(1.5^\circ,7^\circ)$ in HEEQ, a speed of $1571.5~\mathrm{km~s^{-1}}$, and a half-width of $45.5^\circ$ for CME2. Figure~\ref{fig:elevo_icmes} shows a representative frame from the ELEvo simulation. The left panel displays the positions of the spacecraft in HEEQ coordinates, including their trajectories over a time window of $\pm30$ days. The blue ellipses represent the CME1 and CME2 fronts resulting from the ELEvo simulation. The right panel presents the corresponding in situ measurements from SolO, STA, and Wind. An animation of Figure~\ref{fig:elevo_icmes} is available online\footnote{Figure~\ref{fig:elevo_icmes} animation available at \url{https://doi.org/10.6084/m9.figshare.32043171}}.

The magnetic field and plasma signatures observed at SolO, STA, and Wind are consistent with the arrival times predicted by the ELEvo simulation for the two interacting CMEs. The ELEvo simulation further indicates apex encounters for both CMEs at all three spacecraft, consistent with the DONKI-derived CME directions and their large angular widths, which place SolO, STA, and Wind well within the central longitudinal and latitudinal extent of the eruptions. The small longitudinal separations ($0^\circ$--$10^\circ$) and modest latitudinal offsets (approximately $10^\circ$--$30^\circ$) relative to the CME axes further support that each spacecraft intercepted the near-apex region of both CMEs

It is important to note that, as discussed above, ELEvo does not account for CME deflection or rotation during propagation. Significant deflection due to CME–CME interactions or large-scale coronal and heliospheric magnetic structures could alter the spacecraft encounter geometry. Previous studies have shown that most CME deflection occurs within the first few solar radii, where it is governed by the ambient coronal magnetic field and CME properties \citep{2015_kay, 2017_Kay}, while fast CMEs are generally less susceptible to interplanetary deflection than slow CMEs \citep{2004_Wang}. In the present event, however, both CMEs propagated radially in nearly the same direction throughout the coronagraph field of view, with no apparent non-radial deflection. While CME–CME interactions can produce additional deflection, the similar propagation directions, high propagation speeds, and large angular extents of the two CMEs indicate that any such deflection would not be expected to alter the spacecraft encounter geometry from a near-apex to a flank encounter. Nevertheless, the possibility of interplanetary deflection cannot be completely excluded, and its quantitative assessment would require self-consistent modeling of CME–CME interactions within the ambient coronal and heliospheric magnetic field, which is beyond the scope of this study.

\subsection{Interpretation of the 2024 March 23--24 MCL Structure}

The combined remote-sensing, in situ, and ELEvo results suggest that the 2024 March 23--24 event resulted from the interaction and subsequent arrival of two closely launched CMEs and produced MCL signatures at all three in situ spacecraft. The observations and in situ signatures are broadly consistent with the new type of complex ejecta resulting from CME--CME interactions proposed by \citet{2013_Lugaz,2014_Lugaz} based on MHD simulations. In their simulations, a fast CME overtakes a preceding slower CME, and magnetic reconnection between the interacting CMEs modifies the trailing ejecta into an extended ``tail", such that the resulting compound ejecta may be misidentified as a glancing encounter with an isolated CME \citep{2010_Mostl}. A related observational scenario was reported by \citet{Liu_2014}, who investigated the interaction of a fast CME with a preceding slower CME near 1~au. Interestingly, \citet{Liu_2014} found that the interacting ejecta could be reconstructed using the Grad--Shafranov technique despite the ongoing CME interaction and the presence of a penetrating shock. They further anticipated that, once the merging process is complete, the resulting ejecta may resemble an MCL structure with no clear signature of CME--CME interaction. Unlike these previous studies, however, the present event involved two CMEs launched in rapid succession with comparable propagation speeds and nearly identical propagation directions, with the interaction commencing at low coronal heights. The present observations provide observational support for the new type of complex ejecta resulting from CME--CME interactions proposed by \citet{2013_Lugaz, 2014_Lugaz}. The observations further indicate that, when two CMEs are launched nearly simultaneously and interact at low coronal heights, the evolution toward an MCL-like configuration can occur well before 1~au, with MCL signatures already detected at SolO (0.37~au).

The white-light observations further indicate that the shock driven by CME2 propagated through CME1 rather than through the ambient solar wind. Although a second shock front is identified in the coronagraph observations, no distinct second shock is detected in the in situ measurements. A plausible explanation is that the CME2-driven shock weakened as it propagated through the preceding CME1 before reaching the spacecraft \citep{Lugaz_2005, Xiong_2005, Lugaz_2009}. Under this interpretation, the discontinuity observed within the sheath at SolO and Wind may mark the boundary between ambient solar-wind plasma compressed by the CME1-driven shock and CME1 plasma subsequently modified by the CME2-driven shock. Numerical simulations that explicitly model the CME--CME interaction are required to test this interpretation.

\section{Analysis and Results}\label{sec:analysis}

We analyze the radial evolution of the mean magnetic and plasma parameters of the ICME substructures (sheath and ME) for the present event and compare their behavior with results from previous studies of individual ICMEs in the inner heliosphere. Using the substructure boundaries identified in Section~\ref{sec:data}, we obtain the duration ($\Delta t$) of each region and compute the corresponding mean magnetic field $\langle B \rangle$, mean bulk speed $\langle V_p \rangle$, mean proton density $\langle N_p \rangle$, and mean proton temperature $\langle T_p \rangle$ at the locations of SolO, STA, and Wind. The radial size $S$ of each ICME substructure is then estimated as $S = \langle V_p \rangle \times \Delta t$. We further derive the mean magnetic pressure $\langle P_{B} \rangle = \langle B^{2} / (2\mu_{0}) \rangle$, the mean thermal pressure $\langle P_{\mathrm{th}} \rangle = \langle N_{p} k_{B} T_{p} \rangle$, and the mean plasma beta $\langle \beta \rangle = \langle P_{\mathrm{th}} / P_{B} \rangle$. The mean values of these parameters, together with their standard errors for both ICME substructures, are listed in Table~\ref{tab:sheath_event24} and shown in Figure~\ref{fig:icme_par_fit}. 
SolO and STA are approximately radially aligned, with a radial separation of $\sim0.59$~au and an angular separation of only $1.5^\circ$, whereas STA and Wind are separated by only $\sim0.04$~au radially but by $9.2^\circ$ in longitude. The STA--Wind separation falls within the range identified by \citet{Lugaz_2018} as suitable for investigating the mesoscale spatial variability of ICMEs. Therefore, the STA--Wind pair is used to examine mesoscale variations within the ICME substructures along their angular extent,  while the SolO--STA pair is used to investigate the radial evolution of the ICME.

\subsection{Mesoscale Variations in Properties of ICME Substructures}

A comparison of the sheath properties at STA and Wind shows that $\langle B \rangle$, $\langle N_p \rangle$, and $\langle V_p \rangle$ are broadly similar at the two spacecraft. However, the sheath at Wind is thinner by $\sim$19\% and exhibits $\langle T_p \rangle$ approximately 2.5 times greater than at STA. A similar trend is observed within the ME, where the radial size is smaller at Wind by $\sim$46\%, while $\langle V_p \rangle$ exceeds that at STA by $\sim$20\% and $\langle T_p \rangle$ is larger by a factor of $\sim$6.6. In contrast, $\langle B \rangle$ and $\langle N_p \rangle$ within the ME remain broadly comparable between the two spacecraft. Although the quantitative differences should be interpreted in the context of the uncertainties listed in Table~\ref{tab:sheath_event24}, the observed mesoscale variations remain evident. These results demonstrate that the magnetic, plasma, and spatial properties of both the sheath and the ME exhibit measurable variations over a longitudinal separation of only $9.2^\circ$. Similar mesoscale variability has been reported in previous multipoint studies of MCs \citep{KILPUA_2011,Lugaz_2018,Regnault_2024,Agarwal_2025}, whereas \citet{2020_Davies} found no significant variation in the magnetic-field properties of an MC. The observed mesoscale differences may result from the combined effects of non-uniform ambient solar-wind conditions, non-isotropic compression or expansion, intrinsic flux-rope inhomogeneity, and structural distortion of the flux rope \citep{Agarwal_2025,2025_Al-haddad}. In the present event, the interaction between the two closely launched CMEs may have further contributed to this mesoscale variability.

\subsection{Radial Evolution of Properties of ICME Substructures}

\begin{figure*}[htbp]
    \centering
    \includegraphics[width=\linewidth]{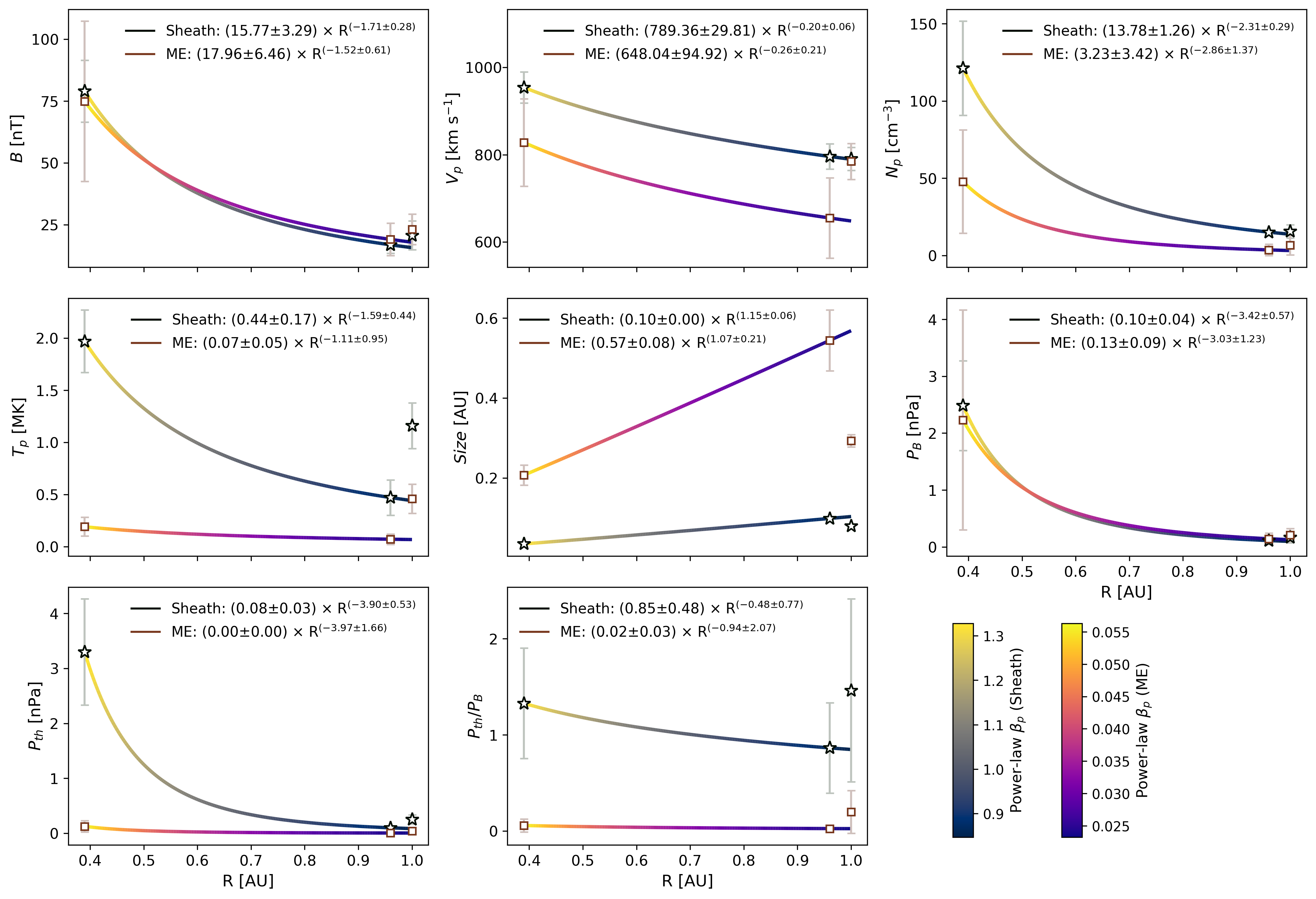} 
    \caption{
Mean plasma properties and the radial size of the ICME sheath (stars) and ME (squares) observed by SolO, STA, and Wind. Error bars represent the standard deviation of the plasma properties within the ICME substructures. Solid curves show power-law fits, $aR^{\alpha}$, to the mean values at SolO and STA; Wind measurements are not included in the fits. The color of the fitted curves indicates the radial variation of plasma $\beta_p$, derived from the fitted exponents $\alpha_{B}$, $\alpha_{T_p}$, and $\alpha_{N_p}$ (see text). Sheath fits use the \textit{cividis} colormap, while ME fits use the \textit{plasma} colormap.
}
    \label{fig:icme_par_fit}
\end{figure*}

To characterize the radial evolution of properties within ICME sheath and ME, we fit power laws of the form $X(R) = a R^{\alpha}$ to $\langle B \rangle$, $\langle V_p \rangle$, $\langle N_{p} \rangle$, $\langle T_{p} \rangle$, and $S$. The resulting best-fit exponents are summarized in Table~\ref{tab:event2_exponents}.
The power-law behavior of the derived quantities, such as magnetic pressure, thermal pressure, and plasma beta (\(\beta \)), follows directly from the fitted exponents. If $B \propto R^{\alpha_B}$, $T_{p} \propto R^{\alpha_{T_p}}$, and $N_{p} \propto R^{\alpha_{N_p}}$, then $P_{B} \propto R^{2\alpha_B}$, $P_{\mathrm{th}} \propto R^{\alpha_{T_p} + \alpha_{N_p}}$, and $\beta \propto R^{\alpha_{T_p} + \alpha_{N_p} - 2\alpha_B }$.
The solid curves in Figure~\ref{fig:icme_par_fit} show the fitted power laws for each parameter, and the color-coded curves show the corresponding radial evolution of $\langle \beta \rangle$. In the following subsections, we compare the derived power-law exponents with those reported for individual ICMEs in the inner heliosphere from statistical studies \citep{Bothmer1998,Leitner2007,Gulisano2010,2015_Winslow,2022_Temmer, 2024_Larrodera,2024_Salman,2026_Mostl} and radially aligned multi-spacecraft observations \citep{2019_Good,2019_Vrvsnak,2020_Salman,2022_Davies,2025b_Zhang}. These comparisons provide an observational context for the radial evolution inferred for the present event. However, the kinematic and thermodynamic evolution of ICMEs depends on several factors, including their initial properties, the ambient solar wind conditions, and CME--CME interactions \citep[and references therein]{2012_Liu,2014_Temmer,2017_Lugaz,10.1093/mnras/stab1721,TEMMER_2023,khuntia2025evolution}. Therefore, the discussion below is limited to a comparison of the derived power-law exponents with previous observations, whereas investigating the physical processes underlying the observed radial evolution requires numerical modeling that accounts for these effects \citep{2004_Schmidt,Lugaz_2005,2023_Palmerio,2024a_Mayank,2024b_Mayank,2025_Manchester}.

\begin{table*}[htbp]
\centering
\caption{Mean values of observed in situ parameters within the ICME sheath and ME regions.}
\renewcommand{\arraystretch}{1.35}   
\setlength{\tabcolsep}{2.8pt}         
\small                             
\begin{tabular}{|c|c|c|c|c|c|c|c|c|c|c|c|}
\hline
Region & Spacecraft & Long, Lat & R & Duration & B & $V_p$ & $N_p$ & $T_p$ & $P_{B}$ & $P_{th}$ & $\beta_p$\\
 &  & ($^{\circ}$) & [AU] & (hours) & [nT] & [$\mathrm{km\,s^{-1}}$] & [$\mathrm{cm^{-3}}$] & [MK] & [nPa] & [nPa] &  \\
\hline
\hline
\multirow{3}{*}{Sheath}
 & SolO & 9.6, -8.0 & 0.39 & 1.53 & 79.02$\pm$12.53 & 954.04$\pm$35.84 & 121.26$\pm$30.42 & 1.97$\pm$0.30 & 2.484$\pm$0.788 & 3.298$\pm$0.968 & 1.53$\pm$1.08 \\
 & STA  & 9.2, -6.5 & 0.96 & 5.17 & 16.91$\pm$3.37 & 795.87$\pm$28.72 & 15.14$\pm$1.31 & 0.47$\pm$0.17 & 0.114$\pm$0.045 & 0.098$\pm$0.037 & 1.58$\pm$1.08 \\
 & Wind & 0.0, -6.9 & 1.00 & 4.22 & 20.74$\pm$5.82 & 790.49$\pm$26.37 & 15.63$\pm$4.21 & 1.16$\pm$0.22 & 0.171$\pm$0.096 & 0.250$\pm$0.082 & 2.05$\pm$1.82 \\
\hline
\hline
\multirow{3}{*}{ME}
 & SolO & 9.6, -8.0 & 0.39 & 11.08 & 74.89$\pm$32.40 & 828.00$\pm$100.22 & 47.89$\pm$33.42 & 0.19$\pm$0.09 & 2.232$\pm$1.931 & 0.126$\pm$0.106 & 0.08$\pm$0.08 \\
 & STA  & 9.2, -6.5 & 0.96 & 34.97 & 19.11$\pm$6.56 & 654.96$\pm$91.71 & 3.63$\pm$3.68 & 0.07$\pm$0.05 & 0.145$\pm$0.100 & 0.004$\pm$0.004 & 0.09$\pm$0.07 \\
 & Wind & 0.0, -6.9 & 1.00 & 15.58 & 23.16$\pm$6.14 & 784.54$\pm$41.12 & 6.66$\pm$6.28 & 0.46$\pm$0.14 & 0.213$\pm$0.113 & 0.042$\pm$0.042 & 0.33$\pm$0.43 \\
\hline
\end{tabular}%
\label{tab:sheath_event24}
\end{table*}

\begin{table}[htbp]
\centering
\caption{Power-law fit exponents ($\alpha$) from the relation $aR^{\alpha}$ applied to the mean values of the observed in situ parameters within the ICME substructures.}
\label{tab:event2_exponents}
\renewcommand{\arraystretch}{1.35} 
\setlength{\tabcolsep}{1.5pt}
\small
\begin{tabular}{c c c c c c}
\hline
Region & \multicolumn{5}{c}{Fit exponents ($\alpha$)} \\
\cline{2-6}
 & Size & B & $V_p$ & $N_p$ & $T_p$ \\
\hline
\hline
Sheath & 1.15$\pm$0.06 & -1.71$\pm$0.28 & -0.20$\pm$0.06 & -2.31$\pm$0.29 &  -1.59$\pm$0.44\\
ME     & 1.07$\pm$0.21 & -1.52$\pm$0.61 & -0.26$\pm$0.20 & -2.86$\pm$1.37 &  -1.11$\pm$0.95 \\
\hline
\end{tabular}
\end{table}

\subsubsection{Radial Evolution of Size of ICME substructures}

The fitted power-law trends for the radial sizes of the ICME substructures are
\[
S_{\mathrm{sheath}} = (0.10 \pm 0.00)\,R^{1.15 \pm 0.06}, \quad
S_{\mathrm{ME}}     = (0.57 \pm 0.08)\,R^{1.07 \pm 0.21}.
\]
These results indicate that both the sheath and the ME broaden with heliocentric distance as the ICME propagates outward from the Sun.  In EUHFORIA simulations of an ICME, \citet{2021_Scolini} reported a sheath-size exponent of 1.88, while a statistical investigation by \citet{2024_Larrodera} found $1.71 \pm 0.74$. The theoretical upper limit derived from Eq.~(8) of \citet{2017_Lee} predicts an exponent of $\sim 0.99$ \citep{2021_Scolini}, indicating more rapid sheath growth in these studies than expected theoretically. In contrast, \citet{2022_Temmer} found a shallower exponent of $\sim$0.48. The exponent derived in the present study (1.15±0.06) lies within the range of previously reported values but exceeds this theoretical upper limit for individual ICME. Previous studies report power-law exponents for the radial size of ICME ejecta ranging from approximately 0.2 to 1.0 \citep{Bothmer1998,Leitner2007,Gulisano2010,2019_Vrvsnak,2025b_Zhang}. Our fitted value of 1.07±0.21 falls within this range and is similar to the exponent predicted for a self-similarly expanding flux rope \citep{2008_Demoulin,2009_Demoulin,Gulisano2010,2019_Vrvsnak}. We note that the radial sizes are estimated as $S=\langle V_p \rangle \times \Delta t$, assuming isotropic self-similar expansion and an approximately constant center-of-mass speed during the spacecraft crossing, and therefore provide a first-order estimate of the local ICME extent. This approximation neglects any evolution of the expansion rate during the crossing itself, which is expected to introduce only a second-order correction to the derived sizes, although the effect may be larger for slow, long-duration MEs observed at greater heliocentric distances. Additional uncertainties may arise from departures from self-similar expansion, including anisotropic expansion or compression, interactions with ambient solar-wind structures, and CME--CME interactions, which may affect both the derived sizes and the fitted power-law exponents.

The fitted exponents further indicate that the sheath exhibits a slightly steeper radial size scaling than the ME, broadly consistent with the trend reported in earlier studies \citep{2019_Janvier, 2021_Scolini}.  Although both ICME substructures expand as they propagate through the solar wind, the faster growth of the sheath is commonly attributed to solar wind pile-up ahead of the ICME ejecta, often described as a “snow-plow” effect. As the ICME propagates outward, upstream solar wind plasma and magnetic field accumulate at the ejecta front, while a portion of the compressed material escapes laterally along the flanks. The resulting sheath thickness, therefore, depends on the balance between plasma pile-up ahead of the CME, lateral escape of compressed material, and intrinsic expansion \citep{SiscoeOdstrcil2008}. MHD simulations further show that lateral expansion of the ejecta prevents the ambient solar wind from fully bypassing the obstacle, promoting plasma pile-up in front of the ICME ejecta \citep{SiscoeOdstrcil2008}. Magnetic reconnection at the sheath--ejecta boundary may also erode or inject flux into the ejecta \citep{2016_Demoulin}, contributing to variations in the sizes of both the sheath and the ME, although such effects likely play a secondary role in determining their radial widths. In the present event, however, magnetic reconnection at the CME--CME interaction interface may have played a more significant role in modifying the ejecta structure \citep{2013_Lugaz, 2014_Lugaz}. Overall, the faster increase in sheath size relative to that of the ME is consistent with solar-wind pile-up and sheath expansion, although reconnection during the CME--CME interaction may also have contributed to the observed radial evolution of the ME.

\subsubsection{Radial Evolution of Mean Parameters within ICME Substructures}

The expansion of ICMEs modifies the average physical conditions within their substructures. Figure~\ref{fig:icme_par_fit} shows the radial evolution of the $\langle B \rangle$, $\langle V_p \rangle$, $\langle N_p \rangle$, and $\langle T_p \rangle$ in both the sheath and the ME, together with the fitted power-law profiles. The corresponding exponents $(\alpha_B, \alpha_{V_p}, \alpha_{N_p}, \alpha_{T_p})$ are listed in Table~\ref{tab:event2_exponents}. 

\paragraph{Radial Evolution within the Sheath.}\label{sec:res_sheath}

As discussed in the previous section, the increase in radial size arises from different physical processes in the sheath and the ejecta; however, in contrast to the ME, the radial evolution of sheath parameters has received limited attention. Because the sheath consists of compressed solar-wind plasma downstream of the CME-driven shock, its radial evolution can be compared with expectations for the ambient solar wind. In the solar wind, Parker’s model \citep{1965_Parker} predicts an approximately constant bulk speed ($\alpha_{V_p} \approx 0$) and, under mass–flux conservation, a density falloff index of $\alpha_{N_p} \approx -2$. For an adiabatically expanding solar wind, conservation of mass and magnetic flux implies a magnetic field decay index between $0$ and $-2$, and an ideal proton temperature decrease of $T_p \propto R^{-4/3}$. These scalings further imply $\alpha_{P_B} \sim -4$ to $0$, $\alpha_{P_{\mathrm{th}}} \sim -10/3$, and $\alpha_{\beta_p} \sim -10/3$ to $2/3$ \citep{2021_Scolini}. These scalings provide a useful reference for interpreting the sheath evolution relative to the ambient solar wind.

The sheath exhibits both similarities to and departures from typical solar wind radial trends. Unlike the ambient solar wind, the ICME sheath is a compressed, dynamically evolving region whose plasma and magnetic properties are continuously modified by solar-wind pile-up, shock compression, and interactions with the driving ejecta. Consequently, mass and magnetic-flux conservation assumptions are not expected to hold strictly within the sheath; departures from ambient solar-wind radial trends are therefore anticipated. However, the magnetic field  within the sheath decreases with $\alpha_B = -1.71 \pm 0.28$ and the density with $\alpha_{N_p} = -2.31 \pm 0.29$; both values are consistent with theoretical expectations and with solar-wind radial scalings derived from the \textit{Helios} database \citep{2011_Hellinger,2018_Venzmer,2019_Perrone}. The fitted speed exponent, $\alpha_{V_p} = -0.20 \pm 0.06$, indicates weak deceleration, likely resulting from the accumulation of upstream solar-wind plasma ahead of the ICME \citep{2018_Benjamin}. As the sheath gathers increasing amounts of slower-moving plasma, the added mass reduces the propagation speed to maintain momentum balance. The radial speed profile may also be influenced by the evolution of the CME-driven shock and the driving ejecta. Meanwhile, sheath temperature decreases $\alpha_{T_p} = -1.59 \pm 0.44$, more rapidly than  an ideal adiabatically expanding solar wind. The mean magnetic and thermal pressures decrease with heliocentric distance, with slopes $\alpha_{P_B} = -3.42 \pm 0.57$ and $\alpha_{P_{\mathrm{th}}} = -3.90 \pm 0.53$, comparable to values reported for fast solar-wind streams \citep{2019_Perrone}. The mean proton plasma beta exhibits only a weak radial dependence, scaling as $\langle \beta_p \rangle \propto R^{-0.48 \pm 0.77}$.

\paragraph{Radial Evolution within the ME}\label{sec:res_ejecta}
The mean magnetic field in the ME decreases with heliocentric distance, with a fitted exponent of $\alpha_B = -1.52 \pm 0.61$. This value lies within the range reported by \citet{Leitner2007, 2019_Good, 2021_Davies_multi, 2022_Davies, 2024_Salman}, although it is lower than the exponents obtained by \citet{Gulisano2010} and \citet{2015_Winslow}, as well as those derived from EUHFORIA simulations \citep{2021_Scolini}. Such discrepancies indicate that a single power-law behavior may not adequately represent all ICMEs, as emphasized by \citet{2020_Salman} and \citet{2022_Davies}. Since the radial evolution of each ICME depends on its initial properties and the surrounding solar-wind environment, while CME--CME interactions can further modify the evolution of interacting events, variations in the derived power-law exponents are expected. Furthermore, as noted by \citet{Gulisano2010}, differences among studies can also arise from variations in event-selection criteria and the range of heliocentric distances sampled. 

The mean proton speed shows a weak radial decrease ($\alpha_{V_p} = -0.26 \pm 0.21$), consistent with gradual deceleration caused by interaction with the ambient solar wind, which transfers momentum from the ejecta to the surrounding plasma. This value is broadly consistent with results from EUHFORIA simulations \citep{2021_Scolini} and with observational estimates reported by \citet{Liu2005} and \citet{2022_Temmer}. The proton density declines more substantially with distance ($\alpha_{N_p} = -2.86 \pm 1.37$), while the mean proton temperature also decreases with heliocentric distance, with a fitted exponent of $\alpha_{T_p} = -1.11 \pm 0.95$. These trends are consistent with both EUHFORIA simulations \citep{2021_Scolini} and observational studies such as \citet{Liu2005}, which likewise report decreasing proton density and temperature with heliocentric distance, as expected for expanding ejecta.

The mean magnetic and proton thermal pressures within the ME decrease with increasing heliocentric distance, with slopes of $\alpha_{P_B} = -3.03 \pm 1.23$ and $\alpha_{P_{\mathrm{th}}} = -3.97 \pm 1.66$. Consequently, the mean proton plasma beta shows a modest decrease, $\alpha_{\beta_p} = -0.94 \pm 2.07$. Direct observational constraints on the radial evolution of pressure terms and plasma beta in CME ejecta remain scarce; however, EUHFORIA simulations \citep{2021_Scolini} report similar behavior, with declining thermal and magnetic pressures together with only weak radial variation in $\beta_p$.

For an ideal, self-similar expansion of a cylindrically symmetric, force-free flux rope, the expected exponents under adiabatic conditions are $\alpha_B = -2$, $\alpha_{N_p} = -3$, and $\alpha_{T_p} = -2$ \citep[e.g.,][]{2009_Demoulin,2011_Chen}. The fitted exponents deviate substantially from these ideal predictions, indicating that the ME does not evolve in a purely self-similar or adiabatic manner. In this event, interaction between two closely spaced CMEs may have altered the internal pressure balance and expansion dynamics of the ejecta while promoting mixing of plasma with different magnetic and thermodynamic properties, potentially contributing to the observed departures from ideal behavior.

\subsection{Determination of the Polytropic Index}

To characterize the thermal state of the ICME, we determine the proton polytropic indices within its substructures (sheath and ME) at each spacecraft location. In logarithmic form, Equation~(\ref{eq:polytropic_relation}) becomes
\begin{equation}
\log_{10} P_{\mathrm{th}} = \Gamma_p \, \log_{10} N_p + \log_{10} F ,
\label{eq:poly_log}
\end{equation}
where \(F\) is a constant. The index \(\Gamma_p\) is obtained by fitting Equation~(\ref{eq:poly_log}) to the data in log--log space for each substructure.
Figure~\ref{fig:gamma_24_m} shows the observed distributions of $\log_{10} P_{\mathrm{th}}$ versus $\log_{10} N_{p}$ for the sheath (panels~a, c, and e) and the ME (panels~b, d, and f) at the SolO, STA, and Wind locations. The data points are color coded by time to illustrate the temporal evolution during each spacecraft crossing. For the SolO observations, however, the $\log_{10} P_{\mathrm{th}}$–$\log_{10} N_{p}$ distributions in both the sheath and the ME exhibit a clear change in slope, indicating that the proton population in these regions cannot be adequately described by a single polytropic index. To quantify this behavior, an optimal transition time ($T_1$) marking the change in slope is determined. First, a single-slope linear least-squares regression is performed over the full duration of the ICME substructure to obtain the residual sum of squares (RSS). A breakpoint is then sought by scanning candidate times between the start ($T_{\mathrm{start}}$) and end ($T_{\mathrm{end}}$) of the ICME substructure interval. For each candidate $T_1$, the data are divided into two sub-intervals, $[T_{\mathrm{start}}, T_1]$ and $[T_1, T_{\mathrm{end}}]$, and linear least-squares fits are performed independently on each segment to obtain the slopes and correlation coefficients ($R_1$, $R_2$).  An F-test is then used to evaluate whether the reduction in RSS achieved by the two-slope model represents a statistically significant improvement over the single-slope fit. A breakpoint is considered statistically significant only when the computed F-statistic exceeds the critical F-value corresponding to a significance level of $\alpha = 0.05$. The optimal $T_1$ is selected as the breakpoint that maximizes the combined goodness of fit, quantified by $R_1^2 + R_2^2$, while satisfying the F-test significance criterion. This procedure provides a robust estimate of the transition between two thermodynamically distinct regimes within the ICME substructure. The same approach is applied to the sheath and ME intervals observed at the STA and Wind locations.

\subsubsection{Dual Polytropic index in ICME substructutes}

\begin{figure*}[htbp]
\centering
\begin{subfigure}[b]{0.33\textwidth}
    \includegraphics[width=\linewidth]{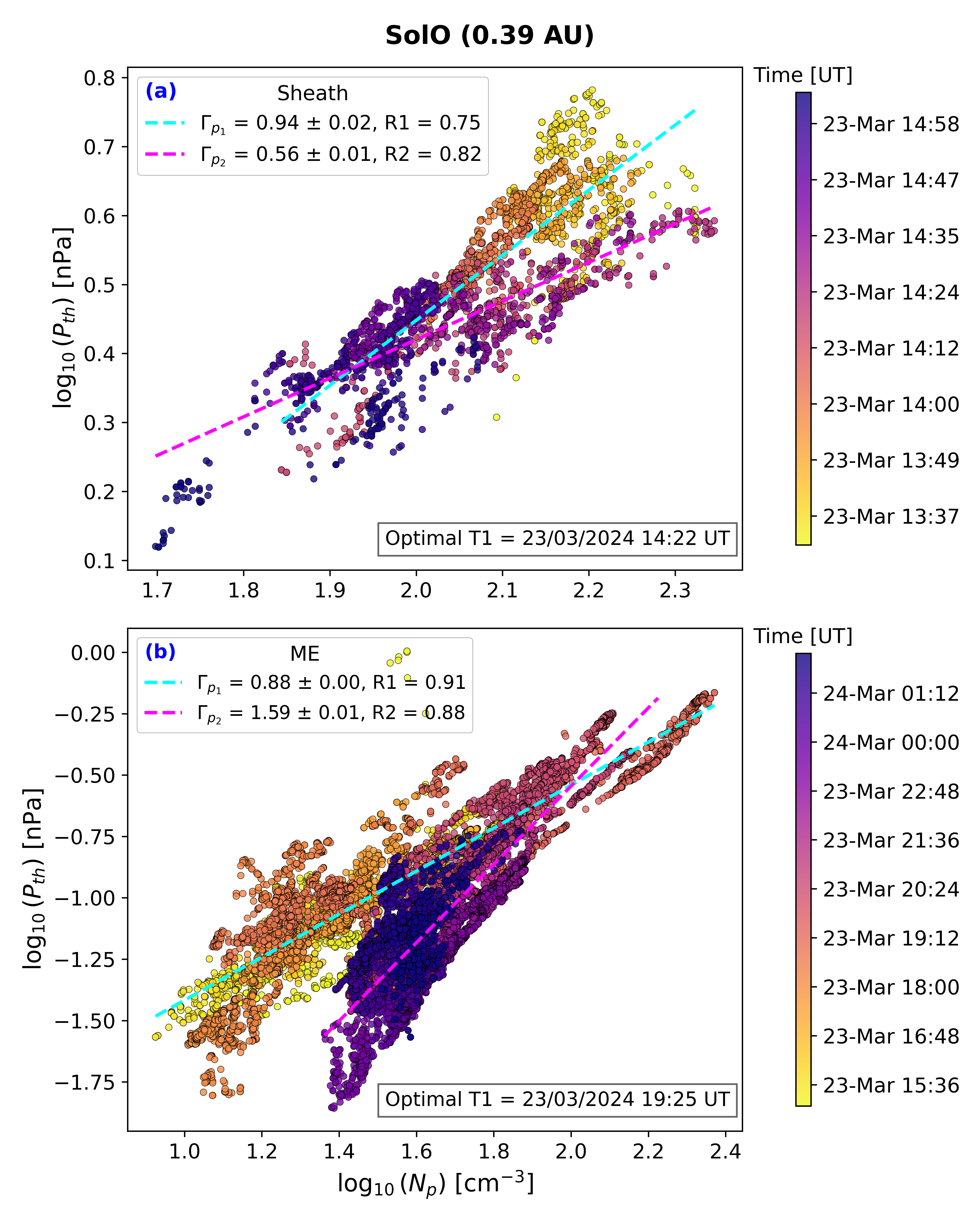}
    \caption{}
\end{subfigure}
\hfill
\begin{subfigure}[b]{0.33\textwidth}
    \includegraphics[width=\linewidth]{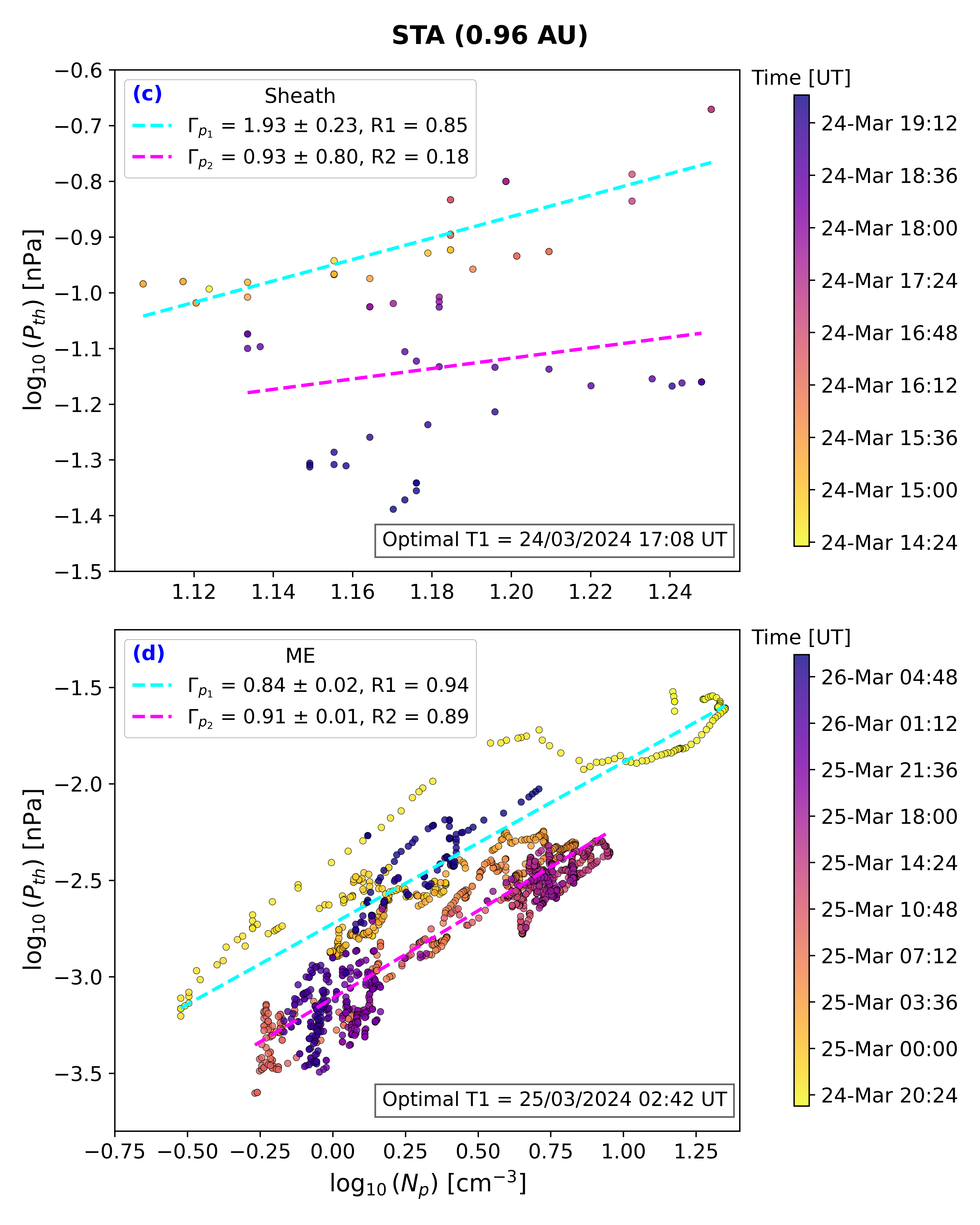}
    \caption{}
\end{subfigure}
\hfill
\begin{subfigure}[b]{0.33\textwidth}
    \includegraphics[width=\linewidth]{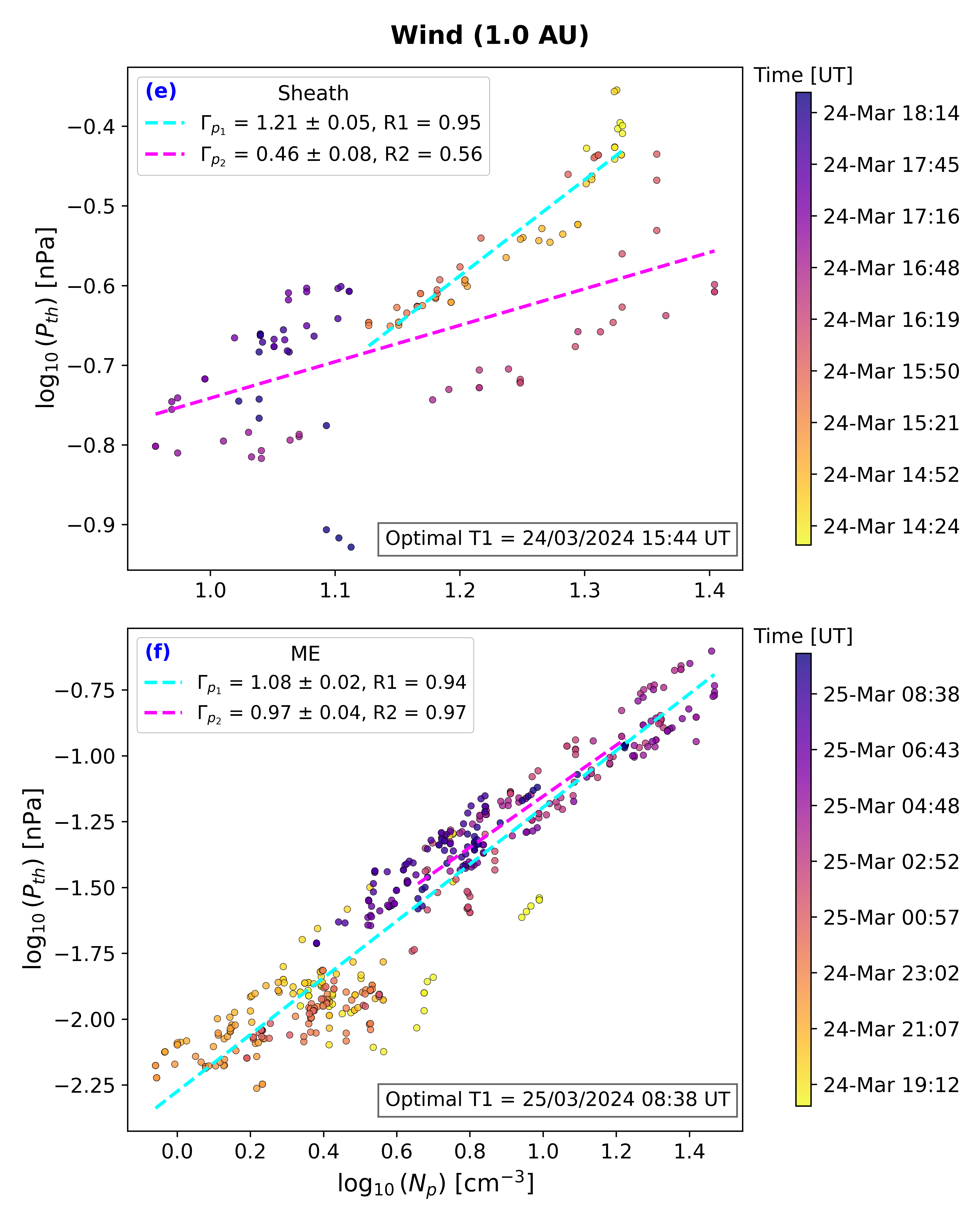}
    \caption{}
\end{subfigure}

\caption{
$\log P_{\mathrm{th}}$--$\log N_p$ distributions for the ICME sheath and ME observed by (i) SolO, (ii) STA, and (iii) Wind. Data points are color-coded by observation time. Dashed line(s) show linear fits (Equation~\ref{eq:poly_log}) used to derive the polytropic index. An F-test is used to evaluate whether the distributions are better represented by single- or two-slope fits; when a two-slope model is favored, the optimal $T_1$
denotes the transition time between thermodynamic regimes.
}

\label{fig:gamma_24_m}
\end{figure*}

The polytropic indices of the ICME substructures (sheath and ME), derived from both single- and double-slope linear fits, are listed in Table~\ref{tab:poly_icme} together with the optimal transition time $T_1$; the corresponding fits are shown in Figure~\ref{fig:gamma_24_m}. The F-test indicates that the $\log P_{\mathrm{th}}$--$\log N_p$ distributions in both the sheath and the ME are generally better represented by a two-slope model, supporting the presence of two thermodynamically distinct regimes at most spacecraft locations. For the ME at Wind, however, the two-slope model does not provide a statistically significant improvement over the single-slope fit. The single-slope polytropic indices obtained here are broadly consistent with previous studies \citep{2006_Liu, 2022_Dayeh, 2023_Khuntia, 2024_Khuntia, 2024_Ghag, 2025_Shaikh}.

The sheath exhibits pronounced dual behavior at SolO and Wind, where a clear change in slope indicates two thermodynamically distinct plasma populations. We further find that the transition time, $T_1$, closely coincides with the arrival time of the discontinuity within the sheath (marked by the purple dashed line in Figure~\ref{fig:raw_data_all}). At STA, the sheath shows only marginal evidence of duality. Although a formal breakpoint is identified, the second interval is weakly correlated, suggesting a less robust separation between the two regimes, possibly due to a plasma data gap within the sheath. The dual polytropic indices within the sheath are consistent with this event involving two closely launched CMEs identified from near-Sun observations and heliospheric modeling. Although no distinct second shock is identified in the in situ measurements, the coronagraph observations indicate that the shock driven by the trailing CME (CME2) propagated through the preceding CME (CME1), suggesting that the sheath contains plasma compressed and accumulated by both disturbances, as also evidenced by the discontinuity within the sheath. The leading portion is likely dominated by ambient solar-wind plasma compressed by the CME1-driven shock, whereas the trailing portion may contain CME1 plasma subsequently processed by the CME2-driven shock, providing a possible explanation for the dual polytropic response observed within the sheath. Furthermore, within the sheath, the magnetic and thermal pressures decrease at comparable rates with increasing heliocentric distance, resulting in only a weak radial dependence of the proton plasma beta (discussed in Section~\ref{sec:res_sheath} and Figure~\ref{fig:icme_par_fit}). This near balance between the magnetic and thermal pressures may help preserve the thermodynamic contrasts from the inner heliosphere to 1~au and support the persistence of the dual polytropic behavior over large distances.

The dual polytropic behavior of the ME is most pronounced at SolO, where the two slopes ($\Gamma_{p_1}=0.88\pm0.01$ and $\Gamma_{p_2}=1.59\pm0.01$) indicate distinct thermodynamic states within the ejecta. At STA and Wind, the ME also exhibits dual behavior, but with only a small separation between the slopes. The indices at STA ($\Gamma_{p_1}=0.84\pm0.02$, $\Gamma_{p_2}=0.91\pm0.01$) and at Wind ($\Gamma_{p_1}=1.08\pm0.02$, $\Gamma_{p_2}=0.97\pm0.04$) differ only modestly and are comparable to the corresponding single-slope indices ($\Gamma_{p}=0.88\pm0.02$ at STA; $\Gamma_{p}=1.08\pm0.02$ at Wind), suggesting a more thermodynamically homogeneous structure at larger heliocentric distances. Moreover, the pronounced dual polytropic behavior observed at SolO is consistent with the ejecta being identified as an MCL structure, characterized by a rotating front region followed by a nearly non-rotating ``back region'' (as discussed in Section~\ref{sec:data}). At SolO, the onset of the MCL ``back region'' closely coincides with the transition time, $T_1$, indicating that the two polytropic regimes correspond to physically distinct portions of the ejecta. Such an interpretation is supported by MHD simulations \citep{2013_Lugaz,2014_Lugaz}, which suggest that the front region may be primarily associated with CME1, whereas the ``back region'' may correspond to CME2. Furthermore, near-Sun observations indicate that the eruptions originated from different source structures, including filament material and hot MFR, thereby introducing plasma with distinct thermal histories into the merged ejecta, as evidenced by the pronounced dual polytropic behavior of the ME. These observations suggest that, although the two CMEs had merged to form a single MCL structure by 0.37~au, they still retained distinct thermodynamic signatures. Our results therefore suggest that polytropic analysis can distinguish the individual ejecta in interacting CME events, particularly at smaller heliocentric distances where their thermodynamic contrast is preserved. In contrast, at STA and Wind, the transition times do not coincide with the onset of the ``back region". This indicates that the correspondence between the two thermodynamic regimes identified by the polytropic analysis and the MCL structure becomes less distinct with increasing heliocentric distance. The small separation between the dual indices at STA and Wind, and their near equivalence to the corresponding single polytropic value, suggests that the thermodynamic distinction between the two ejecta components becomes progressively weaker with heliocentric distance. One possible explanation is that the two ejecta continue to evolve independently toward a near-isothermal thermodynamic state, thereby reducing the contrast between their respective polytropic indices, consistent with previous studies showing that individual ICMEs, despite differing initial thermodynamic states, tend to approach near-isothermal conditions by the time they reach 1 au \citep{LIU20053,2006_Liu,2025b_Khuntia}. Alternatively, the progressive thermodynamic homogenization may result from plasma mixing within the merged MCL structure during heliospheric propagation. Turbulent transport and kinetic instabilities may progressively enhance mixing between plasma originating from the two CME components, thereby reducing their thermodynamic contrast with increasing heliocentric distance. The present observations, however, do not distinguish between these scenarios, and both mechanisms may contribute to the observed evolution.

\begin{table*}
\centering
\caption{Polytropic-index fits for the ICME sheath and magnetic ejecta (ME) at each spacecraft. 
Interval~1 and Interval~2 correspond to the two-segment polytropic fits separated at the optimal transition time $T_1$. 
Single denotes the fit across the full interval. 
$R$ is the Pearson correlation coefficient.}
\label{tab:poly_icme}

\begin{tabular}{@{} l l l  c c  c c  c c  c @{}}
\toprule
Region & Spacecraft & Optimal $T_1$ [UT] 
& \multicolumn{2}{c}{Interval 1} 
& \multicolumn{2}{c}{Interval 2} 
& \multicolumn{2}{c}{Single fit} 
& $F-statistic$ \\
\cmidrule(lr){4-5} \cmidrule(lr){6-7} \cmidrule(lr){8-9}
 &  & 
 & $\Gamma_{p1} \pm \sigma_1$ & $R_1$ 
 & $\Gamma_{p2} \pm \sigma_2$ & $R_2$  
 & $\Gamma_{p} \pm \sigma$ & $R$ 
 &  \\
\midrule

 & SolO & 2024-03-23 14:22:00  
 & $0.94 \pm 0.02$ & $0.75$  
 & $0.56 \pm 0.01$ & $0.82$  
 & $0.82 \pm 0.01$ & $0.81$  
 & $387.79$ \\[3pt]

Sheath & STA   & 2024-03-24 17:08:00  
 & $1.93 \pm 0.23$ & $0.85$  
 & $0.93 \pm 0.80$ & $0.18$  
 & $1.13 \pm 0.55$ & $0.24$  
 & $28.92$ \\[3pt]

 & Wind & 2024-03-24 15:44:00  
 & $1.21 \pm 0.05$ & $0.95$  
 & $0.46 \pm 0.08$ & $0.56$  
 & $0.71 \pm 0.06$ & $0.71$  
 & $28.35$ \\
\midrule

 & SolO & 2024-03-23 19:25:00  
 & $0.88 \pm 0.01$ & $0.91$  
 & $1.59 \pm 0.01$ & $0.88$  
 & $0.89 \pm 0.01$ & $0.74$  
 & $11196.92$ \\[3pt]

ME & STA & 2024-03-25 02:42:00  
 & $0.84 \pm 0.02$ & $0.94$  
 & $0.91 \pm 0.01$ & $0.89$  
 & $0.88 \pm 0.02$ & $0.84$  
 & $562.39$ \\[3pt]

 &Wind  & 2024-03-25 08:38:00  
 & $1.08 \pm 0.02$ & $0.94$  
 & $0.97 \pm 0.04$ & $0.97$  
 & $1.08 \pm 0.02$ & $0.94$  
 & $3.14$ \\
\bottomrule
\end{tabular}

\vspace{2pt}
\footnotesize{
Notes. The optimal transition time $T_1$ marks the boundary between two thermodynamically distinct regimes. 
The $F-statistic$ quantifies the improvement of the two-slope model relative to the single-slope model.
}
\end{table*}

\section{Summary and Conclusions}\label{sec:conclusions}

Using near-radially aligned in situ measurements from SolO, STA, and Wind, we examine the heliospheric evolution of the 2024 March 23--24 ICME from the inner heliosphere to 1~au. At all spacecraft locations, the event exhibits a clear shock, sheath, and magnetic ejecta (ME), with the ejecta displaying an MCL  configuration \citep{2006_Lepping}, consisting of a front region in which the magnetic field rotates and its magnitude declines, followed by a weakly rotating ``back region" with nearly constant field magnitude. Near-Sun EUV and white-light observations reveal that the event resulted from the combined arrival of two closely timed CMEs. MCL signatures are commonly attributed either to magnetic reconnection at the flux-rope leading edge, which produces a tail or back region with minimal field rotation and nearly constant components, or to geometrical effects associated with spacecraft traversal of the ICME flank; both mechanisms may contribute to the observed structure \citep{2006_Dasso, 2007_Dasso, 2007_Marubashi, 2010_Mostl, 2012_Ruffenach, 2024_Rudisser}. By combining near-Sun observations and in situ measurements for this event, we provide direct observational evidence that an MCL-like ejecta can arise from the interaction of two closely spaced CMEs sampled near their apex region, consistent with the new type of complex ejecta resulting from CME–CME interactions proposed by \citet{2013_Lugaz,2014_Lugaz} based on MHD simulations and with the scenario anticipated by \citet{Liu_2014}, in which interacting CMEs may evolve into an MCL-like ejecta with no clear in situ signatures of CME--CME interaction.

Using the SolO--STA pair, we quantify the radial evolution of the magnetic field strength, proton speed, proton density, proton temperature, magnetic and thermal pressures, proton plasma $\beta$, and the radial sizes of the ICME substructures. Both the sheath and the ME broaden with heliocentric distance, with the sheath exhibiting a slightly steeper radial size scaling than the ME, consistent with the expected ``snow-plow'' accumulation of upstream solar wind plasma \citep{2019_Janvier,2020_Lugaz}. In both substructures, the magnetic field strength, proton density, proton temperature, and bulk speed decrease with heliocentric distance. The sheath deceleration agrees with the expected upstream solar-wind pile-up, while its evolution is also influenced by the CME-driven shock and the driving ejecta. The weak deceleration of the ME reflects the combined influence of its initial properties, interaction with the ambient solar wind, and the CME--CME interaction. In both the sheath and the ME, magnetic and thermal pressures decrease at comparable rates, resulting in only a weak radial dependence of the proton plasma $\beta$. Overall, the radial evolution of the interacting ejecta broadly agrees with previous observational and numerical studies of individual ICMEs \citep{Wang2005,Liu2005,Leitner2007,2020_Salman,2021_Scolini,2022_Temmer}; however, the derived power-law exponents deviate from those expected for ideal self-similar adiabatic expansion, suggesting that the interaction between the two closely launched CMEs may have modified the expansion and thermodynamic evolution of the ejecta.

Using the STA--Wind pair, we identify measurable mesoscale variations in the sheath and the ME over a longitudinal separation of only $9.2^\circ$. While the magnetic field strength, proton density, and sheath bulk speed remain broadly comparable, the sheath at Wind is thinner by $\sim$19\% and exhibits a proton temperature approximately 2.5 times higher than that at STA. Within the ME, the radial size is smaller by $\sim$46\%, while the proton bulk speed and temperature are higher by $\sim$20\% and a factor of $\sim$6.6, respectively. These observations reveal appreciable mesoscale variability in the magnetic, plasma, and spatial properties of the ICME and further support the growing evidence that single-spacecraft observations primarily sample the local properties of an ICME rather than its global properties \citep{Moestl_2012,10.1093/mnras/stab1721,Lugaz_2022,Agarwal_2025}. Our findings highlight the importance of coordinated multipoint in situ measurements across the angular extent of ICMEs for understanding the physical processes that drive inhomogeneity at different scales.

The polytropic analysis provides evidence for two thermodynamically distinct regimes in both the sheath and the ME, manifested by dual polytropic indices, consistent with the interaction of two closely timed CMEs that merged to form an MCL-like ejecta. The dual polytropic behavior in the sheath likely reflects thermodynamically distinct plasma populations accumulated by both disturbances, which experienced different compression and heating histories associated with the interaction of the two closely launched CMEs. This thermodynamic distinction within the sheath remains evident from the inner heliosphere to 1~au. Within the ME, the dual behavior is most pronounced at SolO and becomes progressively weaker at STA and Wind, indicating increasing thermodynamic homogenization during heliospheric propagation. At SolO, the two thermodynamic regimes correspond to physically distinct portions of the MCL ejecta---the rotating front region and the weakly rotating ``back region''---each of which is likely associated with a different CME component, as suggested by the MHD simulations of \citet{2013_Lugaz,2014_Lugaz}. The presence of dual thermodynamic regimes within the ME is therefore consistent with the interaction of closely timed CMEs that introduce plasma with different origins and thermal histories into the merged ejecta. In contrast, at STA and Wind, the weaker contrast between the dual polytropic indices makes the correspondence between the thermodynamic regimes and the MCL structure less distinct. This evolution may reflect progressive thermodynamic homogenization during heliospheric propagation, either through plasma mixing within the merged ejecta or through the independent evolution of the ejecta components toward near-isothermal states by the time they reach 1~au \citep{LIU20053,2006_Liu,2025b_Khuntia}. Our results show that, although the magnetic signatures of interacting CMEs may appear merged in situ, distinct thermodynamic signatures can remain identifiable through polytropic analysis, particularly at smaller heliocentric distances. These findings suggest that polytropic analysis can help distinguish individual ejecta within interacting CME events, provided that their thermodynamic contrast is preserved, and can provide additional insight into the origin of MCL ejecta that might otherwise be interpreted as flank encounters of a single CME. Future work will investigate how the radial evolution of CME properties and the observed polytropic indices depend on CME source-region characteristics using coordinated multi-spacecraft observations.

\section{Data Availability}
All data used in our study are publicly accessible via NASA Coordinated Data Analysis Web (CDAWeb; \url{https://cdaweb.gsfc.nasa.gov/}). The ATHARV visualization plots presented in this work were generated using the interactive ATHARV web interface, available at \url{https://vivekmenon42.github.io/ATHARV/}. Supplementary movies and animations for this study are archived at \url{https://doi.org/10.6084/m9.figshare.32043171}.

\begin{acknowledgements}
We acknowledge the use of the ICME LineupCAT hosted at \url{https://helioforecast.space/lineups}, developed by Christian M\"ostl and collaborators as part of the HELIO4CAST project. We acknowledge the use of NASA’s Goddard Space Flight Center (GSFC) Space Physics Data Facility’s Coordinated Data Analysis Web (CDAWeb) to provide access to the data utilized in this study. E.W., C.M., and E.E.D. are funded by the European Union (ERC, HELIO4CAST, 101042188). Views and opinions expressed are however those of the author(s) only and do not necessarily reflect those of the European Union or the European Research Council Executive Agency. Neither the European Union nor the granting authority can be held responsible for them. We acknowledge the Community Coordinated Modeling Center (CCMC) at Goddard Space Flight Center for the use of the Space Weather Database Of Notifications, Knowledge, Information (DONKI), \url{https://kauai.ccmc.gsfc.nasa.gov/DONKI/}.

\end{acknowledgements}


\begin{thebibliography}{}
\expandafter\ifx\csname natexlab\endcsname\relax\def\natexlab#1{#1}\fi
\providecommand{\url}[1]{\href{#1}{#1}}
\providecommand{\dodoi}[1]{doi:~\href{http://doi.org/#1}{\nolinkurl{#1}}}
\providecommand{\doeprint}[1]{\href{http://ascl.net/#1}{\nolinkurl{http://ascl.net/#1}}}
\providecommand{\doarXiv}[1]{\href{https://arxiv.org/abs/#1}{\nolinkurl{https://arxiv.org/abs/#1}}}

\bibitem[{M.~H. {Acu{\~n}a} {et~al.}(2008){Acu{\~n}a}, {Curtis}, {Scheifele}, {et~al.}}]{Acuna2008}
{Acu{\~n}a}, M.~H., {Curtis}, D., {Scheifele}, J., {et~al.} 2008, \bibinfo{title}{{The STEREO/IMPACT Magnetic Field Experiment},} Space Science Reviews, 136, 203, \dodoi{10.1007/s11214-007-9259-2}

\bibitem[{A. Agarwal \& W. Mishra(2025)Agarwal \& Mishra}]{Agarwal_2025}
Agarwal, A., \& Mishra, W. 2025, \bibinfo{title}{Disparities in Magnetic Cloud Observations between Two Spacecraft Having Small Radial and Angular Separations near 1 au,} The Astrophysical Journal, 982, 183, \dodoi{10.3847/1538-4357/adbaeb}

\bibitem[{N. {Al-Haddad} \& N. {Lugaz}(2025){Al-Haddad} \& {Lugaz}}]{2025_Al-haddad}
{Al-Haddad}, N., \& {Lugaz}, N. 2025, \bibinfo{title}{{The Magnetic Field Structure of Coronal Mass Ejections: A More Realistic Representation},} \ssr, 221, 12, \dodoi{10.1007/s11214-025-01138-w}

\bibitem[{V. Bothmer \& R. Schwenn(1998)Bothmer \& Schwenn}]{Bothmer1998}
Bothmer, V., \& Schwenn, R. 1998, \bibinfo{title}{The structure and origin of magnetic clouds in the solar wind,} Annales Geophysicae, 16, 1, \dodoi{10.1007/s00585-997-0001-x}

\bibitem[{G.~E. Brueckner {et~al.}(1995)Brueckner, Howard, Koomen, Korendyke, {et~al.}}]{Brueckner1995}
Brueckner, G.~E., Howard, R.~A., Koomen, M.~J., Korendyke, C.~M., {et~al.} 1995, \bibinfo{title}{The Large Angle Spectroscopic Coronagraph (LASCO),} Solar Physics, 162, 357, \dodoi{10.1007/BF00733434}

\bibitem[{L. {Burlaga} {et~al.}(1981){Burlaga}, {Sittler}, {Mariani}, \& {Schwenn}}]{1981_Burlaga}
{Burlaga}, L., {Sittler}, E., {Mariani}, F., \& {Schwenn}, R. 1981, \bibinfo{title}{{Magnetic loop behind an interplanetary shock: Voyager, Helios, and IMP 8 observations},} \jgr, 86, 6673, \dodoi{10.1029/JA086iA08p06673}

\bibitem[{L.~F. {Burlaga}(1988){Burlaga}}]{1988_Burlaga}
{Burlaga}, L.~F. 1988, \bibinfo{title}{{Magnetic clouds and force-free fields with constant alpha},} \jgr, 93, 7217, \dodoi{10.1029/JA093iA07p07217}

\bibitem[{L.~F. Burlaga \& K.~W. Behannon(1982)Burlaga \& Behannon}]{Burlaga1982}
Burlaga, L.~F., \& Behannon, K.~W. 1982, \bibinfo{title}{Magnetic clouds: Voyager observations between 2 and 4 AU,} Solar Physics, 81, 181, \dodoi{10.1007/BF00151994}

\bibitem[{H.~V. Cane \& I.~G. Richardson(2003)Cane \& Richardson}]{2002_Cane}
Cane, H.~V., \& Richardson, I.~G. 2003, \bibinfo{title}{Interplanetary coronal mass ejections in the near-Earth solar wind during 1996–2002,} Journal of Geophysical Research: Space Physics, 108, \dodoi{https://doi.org/10.1029/2002JA009817}

\bibitem[{S. {Chandrasekhar}(1933){Chandrasekhar}}]{1933_Chandrasekhar}
{Chandrasekhar}, S. 1933, \bibinfo{title}{{The equilibrium of distorted polytropes. I. The rotational problem},} \mnras, 93, 390, \dodoi{10.1093/mnras/93.5.390}

\bibitem[{J. {Chen}(1996){Chen}}]{1996_Chen}
{Chen}, J. 1996, \bibinfo{title}{{Theory of prominence eruption and propagation: Interplanetary consequences},} \jgr, 101, 27499, \dodoi{10.1029/96JA02644}

\bibitem[{P.~F. {Chen}(2011){Chen}}]{2011_Chen}
{Chen}, P.~F. 2011, \bibinfo{title}{{Coronal Mass Ejections: Models and Their Observational Basis},} Living Reviews in Solar Physics, 8, 1, \dodoi{10.12942/lrsp-2011-1}

\bibitem[{S. {Dasso} {et~al.}(2006){Dasso}, {Mandrini}, {D{\'e}moulin}, \& {Luoni}}]{2006_Dasso}
{Dasso}, S., {Mandrini}, C.~H., {D{\'e}moulin}, P., \& {Luoni}, M.~L. 2006, \bibinfo{title}{{A new model-independent method to compute magnetic helicity in magnetic clouds},} \aap, 455, 349, \dodoi{10.1051/0004-6361:20064806}

\bibitem[{S. {Dasso} {et~al.}(2007){Dasso}, {Nakwacki}, {D{\'e}moulin}, \& {Mandrini}}]{2007_Dasso}
{Dasso}, S., {Nakwacki}, M.~S., {D{\'e}moulin}, P., \& {Mandrini}, C.~H. 2007, \bibinfo{title}{{Progressive Transformation of a Flux Rope to an ICME. Comparative Analysis Using the Direct and Fitted Expansion Methods},} \solphys, 244, 115, \dodoi{10.1007/s11207-007-9034-2}

\bibitem[{E.~E. {Davies} {et~al.}(2020){Davies}, {Forsyth}, {Good}, \& {Kilpua}}]{2020_Davies}
{Davies}, E.~E., {Forsyth}, R.~J., {Good}, S.~W., \& {Kilpua}, E. K.~J. 2020, \bibinfo{title}{{On the Radial and Longitudinal Variation of a Magnetic Cloud: ACE, Wind, ARTEMIS and Juno Observations},} \solphys, 295, 157, \dodoi{10.1007/s11207-020-01714-z}

\bibitem[{E.~E. {Davies} {et~al.}(2021{\natexlab{a}}){Davies}, {Forsyth}, {Winslow}, {M{\"o}stl}, \& {Lugaz}}]{2021_Davies}
{Davies}, E.~E., {Forsyth}, R.~J., {Winslow}, R.~M., {M{\"o}stl}, C., \& {Lugaz}, N. 2021{\natexlab{a}}, \bibinfo{title}{{A Catalog of Interplanetary Coronal Mass Ejections Observed by Juno between 1 and 5.4 au},} \apj, 923, 136, \dodoi{10.3847/1538-4357/ac2ccb}

\bibitem[{E.~E. {Davies} {et~al.}(2022){Davies}, {Winslow}, {Scolini}, {Forsyth}, {M{\"o}stl}, {Lugaz}, \& {Galvin}}]{2022_Davies}
{Davies}, E.~E., {Winslow}, R.~M., {Scolini}, C., {et~al.} 2022, \bibinfo{title}{{Multi-spacecraft Observations of the Evolution of Interplanetary Coronal Mass Ejections between 0.3 and 2.2 au: Conjunctions with the Juno Spacecraft},} \apj, 933, 127, \dodoi{10.3847/1538-4357/ac731a}

\bibitem[{E.~E. {Davies} {et~al.}(2021{\natexlab{b}}){Davies}, {M{\"o}stl}, {Owens}, {Weiss}, {Amerstorfer}, {Hinterreiter}, {Bauer}, {Bailey}, {Reiss}, {Forsyth}, {Horbury}, {O'Brien}, {Evans}, {Angelini}, {Heyner}, {Richter}, {Auster}, {Magnes}, {Baumjohann}, {Fischer}, {Barnes}, {Davies}, \& {Harrison}}]{2021_Davies_multi}
{Davies}, E.~E., {M{\"o}stl}, C., {Owens}, M.~J., {et~al.} 2021{\natexlab{b}}, \bibinfo{title}{{In situ multi-spacecraft and remote imaging observations of the first CME detected by Solar Orbiter and BepiColombo},} \aap, 656, A2, \dodoi{10.1051/0004-6361/202040113}

\bibitem[{E.~E. {Davies} {et~al.}(2025){Davies}, {Weiler}, {M{\"o}stl}, {Majumdar}, {R{\"u}disser}, {Horbury}, {O'Brien}, {Morris}, \& {Crabtree}}]{2025arXiv250813892D}
{Davies}, E.~E., {Weiler}, E., {M{\"o}stl}, C., {et~al.} 2025, \bibinfo{title}{{Real-time prediction of geomagnetic storms using Solar Orbiter as a far upstream solar wind monitor},} arXiv e-prints, arXiv:2508.13892, \dodoi{10.48550/arXiv.2508.13892}

\bibitem[{M.~A. {Dayeh} \& G. {Livadiotis}(2022){Dayeh} \& {Livadiotis}}]{2022_Dayeh}
{Dayeh}, M.~A., \& {Livadiotis}, G. 2022, \bibinfo{title}{{Polytropic Behavior in the Structures of Interplanetary Coronal Mass Ejections},} \apjl, 941, L26, \dodoi{10.3847/2041-8213/aca673}

\bibitem[{P. {D{\'e}moulin} \& S. {Dasso}(2009){D{\'e}moulin} \& {Dasso}}]{2009_Demoulin}
{D{\'e}moulin}, P., \& {Dasso}, S. 2009, \bibinfo{title}{{Causes and consequences of magnetic cloud expansion},} \aap, 498, 551, \dodoi{10.1051/0004-6361/200810971}

\bibitem[{P. {D{\'e}moulin} {et~al.}(2016){D{\'e}moulin}, {Janvier}, {Mas{\'\i}as-Meza}, \& {Dasso}}]{2016_Demoulin}
{D{\'e}moulin}, P., {Janvier}, M., {Mas{\'\i}as-Meza}, J.~J., \& {Dasso}, S. 2016, \bibinfo{title}{{Quantitative model for the generic 3D shape of ICMEs at 1 AU},} \aap, 595, A19, \dodoi{10.1051/0004-6361/201628164}

\bibitem[{P. {D{\'e}moulin} {et~al.}(2008){D{\'e}moulin}, {Nakwacki}, {Dasso}, \& {Mandrini}}]{2008_Demoulin}
{D{\'e}moulin}, P., {Nakwacki}, M.~S., {Dasso}, S., \& {Mandrini}, C.~H. 2008, \bibinfo{title}{{Expected in Situ Velocities from a Hierarchical Model for Expanding Interplanetary Coronal Mass Ejections},} \solphys, 250, 347, \dodoi{10.1007/s11207-008-9221-9}

\bibitem[{V. {Domingo} {et~al.}(1995){Domingo}, {Fleck}, \& {Poland}}]{1995_Domingo}
{Domingo}, V., {Fleck}, B., \& {Poland}, A.~I. 1995, \bibinfo{title}{{SOHO: The Solar and Heliospheric Observatory},} \ssr, 72, 81, \dodoi{10.1007/BF00768758}

\bibitem[{R.~J. {Forsyth} {et~al.}(2006){Forsyth}, {Bothmer}, {Cid}, {Crooker}, {Horbury}, {Kecskemety}, {Klecker}, {Linker}, {Odstrcil}, {Reiner}, {Richardson}, {Rodriguez-Pacheco}, {Schmidt}, \& {Wimmer-Schweingruber}}]{2006_Forsyth}
{Forsyth}, R.~J., {Bothmer}, V., {Cid}, C., {et~al.} 2006, \bibinfo{title}{{ICMEs in the Inner Heliosphere: Origin, Evolution and Propagation Effects. Report of Working Group G},} \ssr, 123, 383, \dodoi{10.1007/s11214-006-9022-0}

\bibitem[{A.~B. {Galvin} {et~al.}(2008){Galvin}, {Kistler}, {Popecki}, {et~al.}}]{Galvin2008}
{Galvin}, A.~B., {Kistler}, L.~M., {Popecki}, M.~A., {et~al.} 2008, \bibinfo{title}{{The Plasma and Suprathermal Ion Composition (PLASTIC) Investigation},} Space Science Reviews, 136, 437, \dodoi{10.1007/s11214-007-9296-x}

\bibitem[{K. {Ghag} {et~al.}(2024){Ghag}, {Pathare}, {Raghav}, {Nicolaou}, {Shaikh}, {Dhamane}, {Panchal}, {Kumbhar}, {Tari}, {Sathe}, {Pawaskar}, \& {Hilbert}}]{2024_Ghag}
{Ghag}, K., {Pathare}, P., {Raghav}, A., {et~al.} 2024, \bibinfo{title}{{Studying the polytropic behavior of an ICME using Multi-spacecraft observation by STEREO-A, STEREO-B, and WIND},} Advances in Space Research, 73, 1064, \dodoi{10.1016/j.asr.2023.09.010}

\bibitem[{S.~E. {Gibson} \& B.~C. {Low}(1998){Gibson} \& {Low}}]{1998_Gibson}
{Gibson}, S.~E., \& {Low}, B.~C. 1998, \bibinfo{title}{{A Time-Dependent Three-Dimensional Magnetohydrodynamic Model of the Coronal Mass Ejection},} \apj, 493, 460, \dodoi{10.1086/305107}

\bibitem[{T. {Gold} \& F. {Hoyle}(1960){Gold} \& {Hoyle}}]{1960_Gold}
{Gold}, T., \& {Hoyle}, F. 1960, \bibinfo{title}{{On the origin of solar flares},} \mnras, 120, 89, \dodoi{10.1093/mnras/120.2.89}

\bibitem[{S.~W. {Good} {et~al.}(2015){Good}, {Forsyth}, {Raines}, {Gershman}, {Slavin}, \& {Zurbuchen}}]{2015_Good}
{Good}, S.~W., {Forsyth}, R.~J., {Raines}, J.~M., {et~al.} 2015, \bibinfo{title}{{Radial Evolution of a Magnetic Cloud: MESSENGER, STEREO, and Venus Express Observations},} \apj, 807, 177, \dodoi{10.1088/0004-637X/807/2/177}

\bibitem[{S.~W. {Good} {et~al.}(2019){Good}, {Kilpua}, {LaMoury}, {Forsyth}, {Eastwood}, \& {M{\"o}stl}}]{2019_Good}
{Good}, S.~W., {Kilpua}, E.~K.~J., {LaMoury}, A.~T., {et~al.} 2019, \bibinfo{title}{{Self-Similarity of ICME Flux Ropes: Observations by Radially Aligned Spacecraft in the Inner Heliosphere},} Journal of Geophysical Research (Space Physics), 124, 4960, \dodoi{10.1029/2019JA026475}

\bibitem[{B. {Grison} {et~al.}(2018){Grison}, {Sou{\v{c}}ek}, {Krupar}, {P{\'\i}{\v{s}}a}, {Santol{\'\i}k}, {Taubenschuss}, \& {N{\u{e}}mec}}]{2018_Benjamin}
{Grison}, B., {Sou{\v{c}}ek}, J., {Krupar}, V., {et~al.} 2018, \bibinfo{title}{{Shock deceleration in interplanetary coronal mass ejections (ICMEs) beyond Mercury's orbit until one AU},} Journal of Space Weather and Space Climate, 8, A54, \dodoi{10.1051/swsc/2018043}

\bibitem[{A.~M. {Gulisano} {et~al.}(2012){Gulisano}, {D{\'e}moulin}, {Dasso}, \& {Rodriguez}}]{Gulisano2012}
{Gulisano}, A.~M., {D{\'e}moulin}, P., {Dasso}, S., \& {Rodriguez}, L. 2012, \bibinfo{title}{{Expansion of magnetic clouds in the outer heliosphere},} \aap, 543, A107, \dodoi{10.1051/0004-6361/201118748}

\bibitem[{A.~M. {Gulisano} {et~al.}(2010){Gulisano}, {D{\'e}moulin}, {Dasso}, {Ruiz}, \& {Marsch}}]{Gulisano2010}
{Gulisano}, A.~M., {D{\'e}moulin}, P., {Dasso}, S., {Ruiz}, M.~E., \& {Marsch}, E. 2010, \bibinfo{title}{{Global and local expansion of magnetic clouds in the inner heliosphere},} \aap, 509, A39, \dodoi{10.1051/0004-6361/200912375}

\bibitem[{P. Hellinger \& P.~M. Trávníček(2011)Hellinger \& Trávníček}]{2011_Hellinger}
Hellinger, P., \& Trávníček, P.~M. 2011, \bibinfo{title}{Proton core-beam system in the expanding solar wind: Hybrid simulations,} Journal of Geophysical Research: Space Physics, 116, \dodoi{https://doi.org/10.1029/2011JA016940}

\bibitem[{T.~S. {Horbury} {et~al.}(2020){Horbury}, {O'Brien}, {Carrasco Blazquez}, {Bendyk}, {Brown}, {Hudson}, {Evans}, {Oddy}, {Carr}, {Beek}, {Cupido}, {Bhattacharya}, {Dominguez}, {Matthews}, {Myklebust}, {Whiteside}, {Bale}, {Baumjohann}, {Burgess}, {Carbone}, {Cargill}, {Eastwood}, {Erd{\"o}s}, {Fletcher}, {Forsyth}, {Giacalone}, {Glassmeier}, {Goldstein}, {Hoeksema}, {Lockwood}, {Magnes}, {Maksimovic}, {Marsch}, {Matthaeus}, {Murphy}, {Nakariakov}, {Owen}, {Owens}, {Rodriguez-Pacheco}, {Richter}, {Riley}, {Russell}, {Schwartz}, {Vainio}, {Velli}, {Vennerstrom}, {Walsh}, {Wimmer-Schweingruber}, {Zank}, {M{\"u}ller}, {Zouganelis}, \& {Walsh}}]{2020_Horbury}
{Horbury}, T.~S., {O'Brien}, H., {Carrasco Blazquez}, I., {et~al.} 2020, \bibinfo{title}{{The Solar Orbiter magnetometer},} \aap, 642, A9, \dodoi{10.1051/0004-6361/201937257}

\bibitem[{R.~A. Howard {et~al.}(2008)Howard, Moses, Vourlidas, Newmark, {et~al.}}]{Howard2008}
Howard, R.~A., Moses, J.~D., Vourlidas, A., Newmark, J.~S., {et~al.} 2008, \bibinfo{title}{Sun Earth Connection Coronal and Heliospheric Investigation (SECCHI),} Space Science Reviews, 136, 67, \dodoi{10.1007/s11214-008-9341-4}

\bibitem[{Q. {Hu} \& B.~U.~{\"O}. {Sonnerup}(2002){Hu} \& {Sonnerup}}]{2002_Hu}
{Hu}, Q., \& {Sonnerup}, B. U.~{\"O}. 2002, \bibinfo{title}{{Reconstruction of magnetic clouds in the solar wind: Orientations and configurations},} Journal of Geophysical Research (Space Physics), 107, 1142, \dodoi{10.1029/2001JA000293}

\bibitem[{M. Janvier {et~al.}(2019)Janvier, Dasso, Mas\'ias-Meza, \& Lugaz}]{2019_Janvier}
Janvier, M., Dasso, S., Mas\'ias-Meza, J.~J., \& Lugaz, N. 2019, \bibinfo{title}{Generic Magnetic Field Intensity Profiles of Interplanetary Coronal Mass Ejections,} Journal of Geophysical Research: Space Physics, 124, 812, \dodoi{10.1029/2018JA026007}

\bibitem[{L. {Jian} {et~al.}(2006){Jian}, {Russell}, {Luhmann}, \& {Skoug}}]{2006_Jian}
{Jian}, L., {Russell}, C.~T., {Luhmann}, J.~G., \& {Skoug}, R.~M. 2006, \bibinfo{title}{{Properties of Interplanetary Coronal Mass Ejections at One AU During 1995 2004},} \solphys, 239, 393, \dodoi{10.1007/s11207-006-0133-2}

\bibitem[{C. Katsavrias {et~al.}(2025)Katsavrias, Nicolaou, Livadiotis, Vourlidas, Wilson~Iii, \& Sandberg}]{katsavrias2025polytropic}
Katsavrias, C., Nicolaou, G., Livadiotis, G., {et~al.} 2025, \bibinfo{title}{The Polytropic Index of Interplanetary Coronal Mass Ejections near L1,} Astronomy \& Astrophysics, 695, A146

\bibitem[{C. {Kay} {et~al.}(2017){Kay}, {Gopalswamy}, {Xie}, \& {Yashiro}}]{2017_Kay}
{Kay}, C., {Gopalswamy}, N., {Xie}, H., \& {Yashiro}, S. 2017, \bibinfo{title}{{Deflection and Rotation of CMEs from Active Region 11158},} \solphys, 292, 78, \dodoi{10.1007/s11207-017-1098-z}

\bibitem[{C. Kay {et~al.}(2015)Kay, Opher, \& Evans}]{2015_kay}
Kay, C., Opher, M., \& Evans, R. 2015, \bibinfo{title}{Global trends of CME deflections based on CME and solar parameters,} The Astrophysical Journal, 805, 168

\bibitem[{S. {Khuntia} \& W. {Mishra}(2025){Khuntia} \& {Mishra}}]{2025b_Khuntia}
{Khuntia}, S., \& {Mishra}, W. 2025, \bibinfo{title}{{Thermal and turbulence characteristics of fast and slow coronal mass ejections at 1 AU},} Journal of Astrophysics and Astronomy, 46, 70, \dodoi{10.1007/s12036-025-10085-5}

\bibitem[{S. Khuntia \& W. Mishra(2026)Khuntia \& Mishra}]{khuntia2026thermal}
Khuntia, S., \& Mishra, W. 2026, \bibinfo{title}{Thermal properties of interplanetary coronal mass ejections at 1 au and their connection to geoeffectiveness across solar cycles 23--25,} Monthly Notices of the Royal Astronomical Society, 545, staf2242

\bibitem[{S. Khuntia {et~al.}(2025)Khuntia, Mishra, \& Agarwal}]{khuntia2025evolution}
Khuntia, S., Mishra, W., \& Agarwal, A. 2025, \bibinfo{title}{Evolution of the interacting coronal mass ejections that drove the great geomagnetic storm of 10 May 2024,} Astronomy \& Astrophysics, 698, A79

\bibitem[{S. {Khuntia} {et~al.}(2023){Khuntia}, {Mishra}, {Mishra}, {Wang}, {Zhang}, \& {Lyu}}]{2023_Khuntia}
{Khuntia}, S., {Mishra}, W., {Mishra}, S.~K., {et~al.} 2023, \bibinfo{title}{{Unraveling the Thermodynamic Enigma between Fast and Slow Coronal Mass Ejections},} \apj, 958, 92, \dodoi{10.3847/1538-4357/ad00ba}

\bibitem[{S. {Khuntia} {et~al.}(2024){Khuntia}, {Mishra}, {Wang}, {Mishra}, {Nieves-Chinchilla}, \& {Lyu}}]{2024_Khuntia}
{Khuntia}, S., {Mishra}, W., {Wang}, Y., {et~al.} 2024, \bibinfo{title}{{Deciphering the evolution of thermodynamic properties and their connection to the global kinematics of high-speed coronal mass ejections using FRIS model},} \mnras, 535, 2585, \dodoi{10.1093/mnras/stae2523}

\bibitem[{E. Kilpua {et~al.}(2011)Kilpua, Jian, Li, Luhmann, \& Russell}]{KILPUA_2011}
Kilpua, E., Jian, L., Li, Y., Luhmann, J., \& Russell, C. 2011, \bibinfo{title}{Multipoint ICME encounters: Pre-STEREO and STEREO observations,} Journal of Atmospheric and Solar-Terrestrial Physics, 73, 1228, \dodoi{https://doi.org/10.1016/j.jastp.2010.10.012}

\bibitem[{E. {Kilpua} {et~al.}(2017){Kilpua}, {Koskinen}, \& {Pulkkinen}}]{2017_Kilpua}
{Kilpua}, E., {Koskinen}, H. E.~J., \& {Pulkkinen}, T.~I. 2017, \bibinfo{title}{{Coronal mass ejections and their sheath regions in interplanetary space},} Living Reviews in Solar Physics, 14, 5, \dodoi{10.1007/s41116-017-0009-6}

\bibitem[{J. {Krall} {et~al.}(2000){Krall}, {Chen}, \& {Santoro}}]{2000_Krall}
{Krall}, J., {Chen}, J., \& {Santoro}, R. 2000, \bibinfo{title}{{Drive Mechanisms of Erupting Solar Magnetic Flux Ropes},} \apj, 539, 964, \dodoi{10.1086/309256}

\bibitem[{A. Kumar \& D.~M. Rust(1996)Kumar \& Rust}]{KumarRust1996}
Kumar, A., \& Rust, D.~M. 1996, \bibinfo{title}{Interplanetary Magnetic Clouds, Helicity Conservation, and Current-Core Flux-Rope Models,} Journal of Geophysical Research, 101, 15667, \dodoi{10.1029/96JA00544}

\bibitem[{C. {Larrodera} \& M. {Temmer}(2024){Larrodera} \& {Temmer}}]{2024_Larrodera}
{Larrodera}, C., \& {Temmer}, M. 2024, \bibinfo{title}{{Evolution of coronal mass ejections with and without sheaths from the inner to the outer heliosphere: Statistical investigation for 1975 to 2022},} \aap, 685, A89, \dodoi{10.1051/0004-6361/202348641}

\bibitem[{J.-O. {Lee} {et~al.}(2017){Lee}, {Moon}, {Lee}, {Kim}, \& {Cho}}]{2017_Lee}
{Lee}, J.-O., {Moon}, Y.-J., {Lee}, J.-Y., {Kim}, R.-S., \& {Cho}, K.-S. 2017, \bibinfo{title}{{Which Bow Shock Theory, Gasdynamic or Magnetohydrodynamic, Better Explains CME Stand-off Distance Ratios from LASCO-C2 Observations ?},} \apj, 838, 70, \dodoi{10.3847/1538-4357/aa656f}

\bibitem[{M. Leitner {et~al.}(2007)Leitner, Farrugia, M{\"o}stl, \& et~al.}]{Leitner2007}
Leitner, M., Farrugia, C.~J., M{\"o}stl, C., \& et~al. 2007, \bibinfo{title}{The magnetic field in magnetic clouds and its solar source,} Journal of Geophysical Research, 112, A06113, \dodoi{10.1029/2006JA011940}

\bibitem[{J.~R. Lemen {et~al.}(2012)Lemen, Title, Akin, Boerner, {et~al.}}]{Lemen2012}
Lemen, J.~R., Title, A.~M., Akin, D.~J., Boerner, P.~F., {et~al.} 2012, \bibinfo{title}{The Atmospheric Imaging Assembly (AIA) on the Solar Dynamics Observatory,} Solar Physics, 275, 17, \dodoi{10.1007/s11207-011-9776-8}

\bibitem[{R.~P. {Lepping} {et~al.}(1995){Lepping}, {Ac{\~u}na}, {Burlaga}, {et~al.}}]{Lepping1995}
{Lepping}, R.~P., {Ac{\~u}na}, M.~H., {Burlaga}, L.~F., {et~al.} 1995, \bibinfo{title}{{The Wind Magnetic Field Investigation},} Space Science Reviews, 71, 207, \dodoi{10.1007/BF00751330}

\bibitem[{R.~P. {Lepping} {et~al.}(2006){Lepping}, {Berdichevsky}, {Wu}, {Szabo}, {Narock}, {Mariani}, {Lazarus}, \& {Quivers}}]{2006_Lepping}
{Lepping}, R.~P., {Berdichevsky}, D.~B., {Wu}, C.-C., {et~al.} 2006, \bibinfo{title}{{A summary of WIND magnetic clouds for years 1995-2003: model-fitted parameters, associated errors and classifications},} Annales Geophysicae, 24, 215, \dodoi{10.5194/angeo-24-215-2006}

\bibitem[{R.~P. {Lepping} {et~al.}(1990){Lepping}, {Jones}, \& {Burlaga}}]{1990_Lepping}
{Lepping}, R.~P., {Jones}, J.~A., \& {Burlaga}, L.~F. 1990, \bibinfo{title}{{Magnetic field structure of interplanetary magnetic clouds at 1 AU},} \jgr, 95, 11957, \dodoi{10.1029/JA095iA08p11957}

\bibitem[{Y. Liu {et~al.}(2005{\natexlab{a}})Liu, Richardson, \& Belcher}]{LIU20053}
Liu, Y., Richardson, J., \& Belcher, J. 2005{\natexlab{a}}, \bibinfo{title}{A statistical study of the properties of interplanetary coronal mass ejections from 0.3 to 5.4AU,} Planetary and Space Science, 53, 3, \dodoi{https://doi.org/10.1016/j.pss.2004.09.023}

\bibitem[{Y. Liu {et~al.}(2005{\natexlab{b}})Liu, Richardson, \& Belcher}]{Liu2005}
Liu, Y., Richardson, J.~D., \& Belcher, J.~W. 2005{\natexlab{b}}, \bibinfo{title}{A statistical study of magnetic cloud properties near 1 AU,} Journal of Geophysical Research, 110, A09S03, \dodoi{10.1029/2005JA011195}

\bibitem[{Y. {Liu} {et~al.}(2006){Liu}, {Richardson}, {Belcher}, {Kasper}, \& {Elliott}}]{2006_Liu}
{Liu}, Y., {Richardson}, J.~D., {Belcher}, J.~W., {Kasper}, J.~C., \& {Elliott}, H.~A. 2006, \bibinfo{title}{{Thermodynamic structure of collision-dominated expanding plasma: Heating of interplanetary coronal mass ejections},} Journal of Geophysical Research (Space Physics), 111, A01102, \dodoi{10.1029/2005JA011329}

\bibitem[{Y.~D. Liu {et~al.}(2014)Liu, Yang, Wang, Luhmann, Richardson, \& Lugaz}]{Liu_2014}
Liu, Y.~D., Yang, Z., Wang, R., {et~al.} 2014, \bibinfo{title}{SUN-TO-EARTH CHARACTERISTICS OF TWO CORONAL MASS EJECTIONS INTERACTING NEAR 1 AU: FORMATION OF A COMPLEX EJECTA AND GENERATION OF A TWO-STEP GEOMAGNETIC STORM,} The Astrophysical Journal Letters, 793, L41, \dodoi{10.1088/2041-8205/793/2/L41}

\bibitem[{Y.~D. {Liu} {et~al.}(2012){Liu}, {Luhmann}, {M{\"o}stl}, {Martinez-Oliveros}, {Bale}, {Lin}, {Harrison}, {Temmer}, {Webb}, \& {Odstrcil}}]{2012_Liu}
{Liu}, Y.~D., {Luhmann}, J.~G., {M{\"o}stl}, C., {et~al.} 2012, \bibinfo{title}{{Interactions between Coronal Mass Ejections Viewed in Coordinated Imaging and in situ Observations},} \apjl, 746, L15, \dodoi{10.1088/2041-8205/746/2/L15}

\bibitem[{G. {Livadiotis}(2018{\natexlab{a}}){Livadiotis}}]{2018b_Livadiotis}
{Livadiotis}, G. 2018{\natexlab{a}}, \bibinfo{title}{{Using Kappa Distributions to Identify the Potential Energy},} Journal of Geophysical Research (Space Physics), 123, 1050, \dodoi{10.1002/2017JA024978}

\bibitem[{G. {Livadiotis}(2018{\natexlab{b}}){Livadiotis}}]{2018a_Livadiotis}
{Livadiotis}, G. 2018{\natexlab{b}}, \bibinfo{title}{{Long-Term Independence of Solar Wind Polytropic Index on Plasma Flow Speed},} Entropy, 20, 799, \dodoi{10.3390/e20100799}

\bibitem[{G. {Livadiotis}(2019){Livadiotis}}]{2019_Livadiotis}
{Livadiotis}, G. 2019, \bibinfo{title}{{On the Origin of Polytropic Behavior in Space and Astrophysical Plasmas},} \apj, 874, 10, \dodoi{10.3847/1538-4357/ab05b7}

\bibitem[{R.~E. {Lopez} \& J.~W. {Freeman}(1986){Lopez} \& {Freeman}}]{1986_Lopez}
{Lopez}, R.~E., \& {Freeman}, J.~W. 1986, \bibinfo{title}{{Solar wind proton temperature-velocity relationship},} \jgr, 91, 1701, \dodoi{10.1029/JA091iA02p01701}

\bibitem[{N. {Lugaz} \& C.~J. {Farrugia}(2014){Lugaz} \& {Farrugia}}]{2014_Lugaz}
{Lugaz}, N., \& {Farrugia}, C.~J. 2014, \bibinfo{title}{{A new class of complex ejecta resulting from the interaction of two CMEs and its expected geoeffectiveness},} \grl, 41, 769, \dodoi{10.1002/2013GL058789}

\bibitem[{N. {Lugaz} {et~al.}(2013){Lugaz}, {Farrugia}, {Manchester}, \& {Schwadron}}]{2013_Lugaz}
{Lugaz}, N., {Farrugia}, C.~J., {Manchester}, IV, W.~B., \& {Schwadron}, N. 2013, \bibinfo{title}{{The Interaction of Two Coronal Mass Ejections: Influence of Relative Orientation},} \apj, 778, 20, \dodoi{10.1088/0004-637X/778/1/20}

\bibitem[{N. Lugaz {et~al.}(2018)Lugaz, Farrugia, Winslow, Al-Haddad, Galvin, Nieves-Chinchilla, Lee, \& Janvier}]{Lugaz_2018}
Lugaz, N., Farrugia, C.~J., Winslow, R.~M., {et~al.} 2018, \bibinfo{title}{On the Spatial Coherence of Magnetic Ejecta: Measurements of Coronal Mass Ejections by Multiple Spacecraft Longitudinally Separated by 0.01 au,} The Astrophysical Journal Letters, 864, L7, \dodoi{10.3847/2041-8213/aad9f4}

\bibitem[{N. Lugaz {et~al.}(2005)Lugaz, Manchester~IV, \& Gombosi}]{Lugaz_2005}
Lugaz, N., Manchester~IV, W.~B., \& Gombosi, T.~I. 2005, \bibinfo{title}{Numerical Simulation of the Interaction of Two Coronal Mass Ejections from Sun to Earth,} The Astrophysical Journal, 634, 651, \dodoi{10.1086/491782}

\bibitem[{N. {Lugaz} {et~al.}(2017){Lugaz}, {Temmer}, {Wang}, \& {Farrugia}}]{2017_Lugaz}
{Lugaz}, N., {Temmer}, M., {Wang}, Y., \& {Farrugia}, C.~J. 2017, \bibinfo{title}{{The Interaction of Successive Coronal Mass Ejections: A Review},} \solphys, 292, 64, \dodoi{10.1007/s11207-017-1091-6}

\bibitem[{N. {Lugaz} {et~al.}(2009){Lugaz}, {Vourlidas}, \& {Roussev}}]{Lugaz_2009}
{Lugaz}, N., {Vourlidas}, A., \& {Roussev}, I.~I. 2009, \bibinfo{title}{{Deriving the radial distances of wide coronal mass ejections from elongation measurements in the heliosphere - application to CME-CME interaction},} Annales Geophysicae, 27, 3479, \dodoi{10.5194/angeo-27-3479-2009}

\bibitem[{N. {Lugaz} {et~al.}(2020){Lugaz}, {Winslow}, \& {Farrugia}}]{2020_Lugaz}
{Lugaz}, N., {Winslow}, R.~M., \& {Farrugia}, C.~J. 2020, \bibinfo{title}{{Evolution of a Long-Duration Coronal Mass Ejection and Its Sheath Region Between Mercury and Earth on 9-14 July 2013},} Journal of Geophysical Research (Space Physics), 125, e27213, \dodoi{10.1029/2019JA027213}

\bibitem[{N. Lugaz {et~al.}(2022)Lugaz, Salman, Zhuang, Al-Haddad, Scolini, Farrugia, Yu, Winslow, Möstl, Davies, \& Galvin}]{Lugaz_2022}
Lugaz, N., Salman, T.~M., Zhuang, B., {et~al.} 2022, \bibinfo{title}{A Coronal Mass Ejection and Magnetic Ejecta Observed In Situ by STEREO-A and Wind at 55° Angular Separation,} The Astrophysical Journal, 929, 149, \dodoi{10.3847/1538-4357/ac602f}

\bibitem[{J.~G. {Luhmann} {et~al.}(2008){Luhmann}, {Curtis}, {Schroeder}, {et~al.}}]{Luhmann2008}
{Luhmann}, J.~G., {Curtis}, D.~W., {Schroeder}, P., {et~al.} 2008, \bibinfo{title}{{STEREO IMPACT: Investigation of space weather from the sun to the heliosphere},} Space Science Reviews, 136, 117, \dodoi{10.1007/s11214-007-9170-x}

\bibitem[{S. Lundquist(1950)Lundquist}]{1950_Lundquist}
Lundquist, S. 1950, \bibinfo{title}{Magneto-hydrostatic fields,} Ark. Fys., 2, 361

\bibitem[{W.~B. {Manchester} {et~al.}(2025){Manchester}, {Sachdeva}, {Kilpua}, {Ala-Lahti}, {Soni}, {Huang}, {Chen}, {Jivani}, {van der Holst}, {Szabo}, \& {Akhavan-Tafti}}]{2025_Manchester}
{Manchester}, IV, W.~B., {Sachdeva}, N., {Kilpua}, E., {et~al.} 2025, \bibinfo{title}{{High-resolution Simulation of Coronal Mass Ejection─Corotating Interaction Region Interactions: Mesoscale Solar Wind Structure Formation Observable by the SWIFT Constellation},} \apj, 992, 51, \dodoi{10.3847/1538-4357/adf855}

\bibitem[{K. Marubashi(1986)Marubashi}]{MARUBASHI1986335}
Marubashi, K. 1986, \bibinfo{title}{Structure of the interplanetary magnetic clouds and their solar origins,} Advances in Space Research, 6, 335, \dodoi{https://doi.org/10.1016/0273-1177(86)90172-9}

\bibitem[{K. {Marubashi} \& R.~P. {Lepping}(2007){Marubashi} \& {Lepping}}]{2007_Marubashi}
{Marubashi}, K., \& {Lepping}, R.~P. 2007, \bibinfo{title}{{Long-duration magnetic clouds: a comparison of analyses using torus- and cylinder-shaped flux rope models},} Annales Geophysicae, 25, 2453, \dodoi{10.5194/angeo-25-2453-2007}

\bibitem[{P. {Mayank} {et~al.}(2024{\natexlab{a}}){Mayank}, {Lotz}, {Vaidya}, {Mishra}, \& {Chakrabarty}}]{2024b_Mayank}
{Mayank}, P., {Lotz}, S., {Vaidya}, B., {Mishra}, W., \& {Chakrabarty}, D. 2024{\natexlab{a}}, \bibinfo{title}{{Study of Evolution and Geo-effectiveness of Coronal Mass Ejection─Coronal Mass Ejection Interactions Using Magnetohydrodynamic Simulations with SWASTi Framework},} \apj, 976, 126, \dodoi{10.3847/1538-4357/ad8084}

\bibitem[{P. {Mayank} {et~al.}(2024{\natexlab{b}}){Mayank}, {Vaidya}, {Mishra}, \& {Chakrabarty}}]{2024a_Mayank}
{Mayank}, P., {Vaidya}, B., {Mishra}, W., \& {Chakrabarty}, D. 2024{\natexlab{b}}, \bibinfo{title}{{SWASTi-CME: A Physics-based Model to Study Coronal Mass Ejection Evolution and Its Interaction with Solar Wind},} \apjs, 270, 10, \dodoi{10.3847/1538-4365/ad08c7}

\bibitem[{M.~L. {Mays} {et~al.}(2015){Mays}, {Taktakishvili}, {Pulkkinen}, {MacNeice}, {Rast{\"a}tter}, {Odstrcil}, {Jian}, {Richardson}, {LaSota}, {Zheng}, \& {Kuznetsova}}]{2015_Mays}
{Mays}, M.~L., {Taktakishvili}, A., {Pulkkinen}, A., {et~al.} 2015, \bibinfo{title}{{Ensemble Modeling of CMEs Using the WSA-ENLIL+Cone Model},} \solphys, 290, 1775, \dodoi{10.1007/s11207-015-0692-1}

\bibitem[{A. {Mishev} {et~al.}(2024){Mishev}, {Larsen}, {Asvestari}, {S{\'a}iz}, {Ann Shea}, {Strauss}, {Ruffolo}, {Banglieng}, {Seunarine}, {Duldig}, {Gil}, {Blanco}, {Garc{\'\i}a-Poblaci{\'o}n}, {Cervino-Solana}, {Adams}, \& {Usoskin}}]{2024_Mishev}
{Mishev}, A., {Larsen}, N., {Asvestari}, E., {et~al.} 2024, \bibinfo{title}{{Anisotropic Forbush decrease of 24 March 2024: First look},} Advances in Space Research, 74, 4160, \dodoi{10.1016/j.asr.2024.08.027}

\bibitem[{W. Mishra {et~al.}(2021)Mishra, Dave, Srivastava, \& Teriaca}]{10.1093/mnras/stab1721}
Mishra, W., Dave, K., Srivastava, N., \& Teriaca, L. 2021, \bibinfo{title}{Multipoint remote and in situ observations of interplanetary coronal mass ejection structures during 2011 and associated geomagnetic storms,} Monthly Notices of the Royal Astronomical Society, 506, 1186, \dodoi{10.1093/mnras/stab1721}

\bibitem[{W. {Mishra} \& Y. {Wang}(2018){Mishra} \& {Wang}}]{2018_Mishra}
{Mishra}, W., \& {Wang}, Y. 2018, \bibinfo{title}{{Modeling the Thermodynamic Evolution of Coronal Mass Ejections Using Their Kinematics},} \apj, 865, 50, \dodoi{10.3847/1538-4357/aadb9b}

\bibitem[{W. Mishra {et~al.}(2023)Mishra, Wang, Lyu, \& Khuntia}]{Mishra_2023}
Mishra, W., Wang, Y., Lyu, S., \& Khuntia, S. 2023, \bibinfo{title}{Erratum: “Modeling the Thermodynamic Evolution of Coronal Mass Ejections Using Their Kinematics” (2018, ApJ, 865, 50),} The Astrophysical Journal, 952, 173, \dodoi{10.3847/1538-4357/ace691}

\bibitem[{W. {Mishra} {et~al.}(2020){Mishra}, {Wang}, {Teriaca}, {Zhang}, \& {Chi}}]{2020_Mishra}
{Mishra}, W., {Wang}, Y., {Teriaca}, L., {Zhang}, J., \& {Chi}, Y. 2020, \bibinfo{title}{{Probing the thermodynamic state of a Coronal Mass Ejection (CME) up to 1 AU},} Frontiers in Astronomy and Space Sciences, 7, 1, \dodoi{10.3389/fspas.2020.00001}

\bibitem[{C. {M{\"o}stl} {et~al.}(2010){M{\"o}stl}, {Temmer}, {Rollett}, {Farrugia}, {Liu}, {Veronig}, {Leitner}, {Galvin}, \& {Biernat}}]{2010_Mostl}
{M{\"o}stl}, C., {Temmer}, M., {Rollett}, T., {et~al.} 2010, \bibinfo{title}{{STEREO and Wind observations of a fast ICME flank triggering a prolonged geomagnetic storm on 5-7 April 2010},} \grl, 37, L24103, \dodoi{10.1029/2010GL045175}

\bibitem[{C. {M{\"o}stl} {et~al.}(2015){M{\"o}stl}, {Rollett}, {Frahm}, {Liu}, {Long}, {Colaninno}, {Reiss}, {Temmer}, {Farrugia}, {Posner}, {Dumbovi{\'c}}, {Janvier}, {D{\'e}moulin}, {Boakes}, {Devos}, {Kraaikamp}, {Mays}, \& {Vr{\v{s}}nak}}]{2015_Mostl}
{M{\"o}stl}, C., {Rollett}, T., {Frahm}, R.~A., {et~al.} 2015, \bibinfo{title}{{Strong coronal channelling and interplanetary evolution of a solar storm up to Earth and Mars},} Nature Communications, 6, 7135, \dodoi{10.1038/ncomms8135}

\bibitem[{C. {M{\"o}stl} {et~al.}(2022){M{\"o}stl}, {Weiss}, {Reiss}, {Amerstorfer}, {Bailey}, {Hinterreiter}, {Bauer}, {Barnes}, {Davies}, {Harrison}, {Freiherr von Forstner}, {Davies}, {Heyner}, {Horbury}, \& {Bale}}]{2022_Mostl}
{M{\"o}stl}, C., {Weiss}, A.~J., {Reiss}, M.~A., {et~al.} 2022, \bibinfo{title}{{Multipoint Interplanetary Coronal Mass Ejections Observed with Solar Orbiter, BepiColombo, Parker Solar Probe, Wind, and STEREO-A},} \apjl, 924, L6, \dodoi{10.3847/2041-8213/ac42d0}

\bibitem[{C. {M{\"o}stl} {et~al.}(2026){M{\"o}stl}, {Davies}, {Weiler}, {Amerstorfer}, {Weiss}, {R{\"u}disser}, {Reiss}, {Majumdar}, {Horbury}, {Bale}, \& {Heyner}}]{2026_Mostl}
{M{\"o}stl}, C., {Davies}, E.~E., {Weiler}, E., {et~al.} 2026, \bibinfo{title}{{On the magnetic field evolution of interplanetary coronal mass ejections from 0.07 to 5.4 au},} arXiv e-prints, arXiv:2512.04730, \dodoi{10.48550/arXiv.2512.04730}

\bibitem[{T. Mulligan {et~al.}(2001)Mulligan, Russell, \& Luhmann}]{Mulligan2001}
Mulligan, T., Russell, C.~T., \& Luhmann, J.~G. 2001, \bibinfo{title}{Complex magnetic cloud structure at 1 AU,} Geophysical Research Letters, 28, 4417, \dodoi{10.1029/2001GL013217}

\bibitem[{C. Möstl {et~al.}(2012)Möstl, Farrugia, Kilpua, Jian, Liu, Eastwood, Harrison, Webb, Temmer, Odstrcil, Davies, Rollett, Luhmann, Nitta, Mulligan, Jensen, Forsyth, Lavraud, de~Koning, Veronig, Galvin, Zhang, \& Anderson}]{Moestl_2012}
Möstl, C., Farrugia, C.~J., Kilpua, E. K.~J., {et~al.} 2012, \bibinfo{title}{MULTI-POINT SHOCK AND FLUX ROPE ANALYSIS OF MULTIPLE INTERPLANETARY CORONAL MASS EJECTIONS AROUND 2010 AUGUST 1 IN THE INNER HELIOSPHERE,} The Astrophysical Journal, 758, 10, \dodoi{10.1088/0004-637X/758/1/10}

\bibitem[{G. {Nicolaou} {et~al.}(2023){Nicolaou}, {Livadiotis}, \& {McComas}}]{2023_Nicolaou}
{Nicolaou}, G., {Livadiotis}, G., \& {McComas}, D.~J. 2023, \bibinfo{title}{{The Polytropic Behavior of Solar Wind Protons as Observed by the Ulysses Spacecraft during Solar Minimum},} \apj, 948, 22, \dodoi{10.3847/1538-4357/acbf33}

\bibitem[{G. {Nicolaou} {et~al.}(2014){Nicolaou}, {Livadiotis}, \& {Moussas}}]{2014_Nicolaou}
{Nicolaou}, G., {Livadiotis}, G., \& {Moussas}, X. 2014, \bibinfo{title}{{Long-Term Variability of the Polytropic Index of Solar Wind Protons at 1 AU},} \solphys, 289, 1371, \dodoi{10.1007/s11207-013-0401-x}

\bibitem[{G. {Nicolaou} {et~al.}(2020){Nicolaou}, {Livadiotis}, {Wicks}, {Verscharen}, \& {Maruca}}]{2020_Nicolaou}
{Nicolaou}, G., {Livadiotis}, G., {Wicks}, R.~T., {Verscharen}, D., \& {Maruca}, B.~A. 2020, \bibinfo{title}{{Polytropic Behavior of Solar Wind Protons Observed by Parker Solar Probe},} \apj, 901, 26, \dodoi{10.3847/1538-4357/abaaae}

\bibitem[{T. Nieves-Chinchilla {et~al.}(2019)Nieves-Chinchilla, Jian, Balmaceda, Vourlidas, dos Santos, \& Szabo}]{2019_nieves}
Nieves-Chinchilla, T., Jian, L.~K., Balmaceda, L., {et~al.} 2019, \bibinfo{title}{Unraveling the internal magnetic field structure of the Earth-directed interplanetary coronal mass ejections during 1995--2015,} Solar Physics, 294, 89

\bibitem[{K.~W. {Ogilvie} {et~al.}(1995){Ogilvie}, {Chornay}, {Fritzenreiter}, {et~al.}}]{Ogilvie1995}
{Ogilvie}, K.~W., {Chornay}, D.~J., {Fritzenreiter}, R.~J., {et~al.} 1995, \bibinfo{title}{{SWE: A comprehensive solar wind plasma experiment for the Wind spacecraft},} Space Science Reviews, 71, 55, \dodoi{10.1007/BF00751326}

\bibitem[{V. Osherovich {et~al.}(1993)Osherovich, Farrugia, \& Burlaga}]{1993a_Osherovich}
Osherovich, V., Farrugia, C., \& Burlaga, L. 1993, \bibinfo{title}{Dynamics of aging magnetic clouds,} Advances in Space Research, 13, 57, \dodoi{https://doi.org/10.1016/0273-1177(93)90391-N}

\bibitem[{V.~A. {Osherovich} {et~al.}(1993){Osherovich}, {Farrugia}, {Burlaga}, {Lepping}, {Fainberg}, \& {Stone}}]{1993b_Osherovich}
{Osherovich}, V.~A., {Farrugia}, C.~J., {Burlaga}, L.~F., {et~al.} 1993, \bibinfo{title}{{Polytropic relationship in interplanetary magnetic clouds},} \jgr, 98, 15331, \dodoi{10.1029/93JA01012}

\bibitem[{C.~J. {Owen} {et~al.}(2020){Owen}, {Bruno}, {Livi}, {Louarn}, {Fazakerley}, {Al Janabi}, {et~al.}}]{2020_Owen}
{Owen}, C.~J., {Bruno}, R., {Livi}, S., {et~al.} 2020, \bibinfo{title}{{The Solar Wind Analyser (SWA) suite for the Solar Orbiter mission},} Astronomy and Astrophysics, 642, A16, \dodoi{10.1051/0004-6361/201937259}

\bibitem[{E. {Palmerio} {et~al.}(2023){Palmerio}, {Torok}, \& {Downs}}]{2023_Palmerio}
{Palmerio}, E., {Torok}, T., \& {Downs}, C. 2023, {SHINE: A Numerical Investigation into Coronal Mass Ejections (CME-CME) Interaction Events},, NSF Award Number 2301403. Directorate for Geosciences, Division of Atmospheric and Geospace Sciences. 2023.

\bibitem[{E.~N. {Parker}(1965){Parker}}]{1965_Parker}
{Parker}, E.~N. 1965, \bibinfo{title}{{Dynamical Theory of the Solar Wind},} \ssr, 4, 666, \dodoi{10.1007/BF00216273}

\bibitem[{D. {Perrone} {et~al.}(2019){Perrone}, {Stansby}, {Horbury}, \& {Matteini}}]{2019_Perrone}
{Perrone}, D., {Stansby}, D., {Horbury}, T.~S., \& {Matteini}, L. 2019, \bibinfo{title}{{Radial evolution of the solar wind in pure high-speed streams: HELIOS revised observations},} \mnras, 483, 3730, \dodoi{10.1093/mnras/sty3348}

\bibitem[{W.~D. Pesnell {et~al.}(2012)Pesnell, Thompson, \& Chamberlin}]{Pesnell2012}
Pesnell, W.~D., Thompson, B.~J., \& Chamberlin, P.~C. 2012, \bibinfo{title}{The Solar Dynamics Observatory (SDO),} Solar Physics, 275, 3, \dodoi{10.1007/s11207-011-9841-3}

\bibitem[{F. Regnault {et~al.}(2024)Regnault, Al-Haddad, Lugaz, Farrugia, Yu, Zhuang, \& Davies}]{Regnault_2024}
Regnault, F., Al-Haddad, N., Lugaz, N., {et~al.} 2024, \bibinfo{title}{Discrepancies in the Properties of a Coronal Mass Ejection on Scales of 0.03 au as Revealed by Simultaneous Measurements at Solar Orbiter and Wind: The 2021 November 3–5 Event,} The Astrophysical Journal, 962, 190, \dodoi{10.3847/1538-4357/ad1883}

\bibitem[{I.~G. {Richardson} \& H.~V. {Cane}(2010){Richardson} \& {Cane}}]{2010_Richardson}
{Richardson}, I.~G., \& {Cane}, H.~V. 2010, \bibinfo{title}{{Near-Earth Interplanetary Coronal Mass Ejections During Solar Cycle 23 (1996 - 2009): Catalog and Summary of Properties},} \solphys, 264, 189, \dodoi{10.1007/s11207-010-9568-6}

\bibitem[{A. {Ruffenach} {et~al.}(2012){Ruffenach}, {Lavraud}, {Owens}, {Sauvaud}, {Savani}, {Rouillard}, {D{\'e}moulin}, {Foullon}, {Opitz}, {Fedorov}, {Jacquey}, {G{\'e}not}, {Louarn}, {Luhmann}, {Russell}, {Farrugia}, \& {Galvin}}]{2012_Ruffenach}
{Ruffenach}, A., {Lavraud}, B., {Owens}, M.~J., {et~al.} 2012, \bibinfo{title}{{Multispacecraft observation of magnetic cloud erosion by magnetic reconnection during propagation},} Journal of Geophysical Research (Space Physics), 117, A09101, \dodoi{10.1029/2012JA017624}

\bibitem[{H. Rüdisser {et~al.}(2024)Rüdisser, Weiss, Le~Louëdec, Amerstorfer, Möstl, Davies, \& Lammer}]{2024_Rudisser}
Rüdisser, H., Weiss, A., Le~Louëdec, J., {et~al.} 2024, \bibinfo{title}{Understanding the Effects of Spacecraft Trajectories through Solar Coronal Mass Ejection Flux Ropes Using 3DCOREweb,} The Astrophysical Journal, 973, 150, \dodoi{10.3847/1538-4357/ad660a}

\bibitem[{T.~M. {Salman} {et~al.}(2024){Salman}, {Nieves-Chinchilla}, {Jian}, {Lugaz}, {Carcaboso}, {Davies}, \& {Collado-Vega}}]{2024_Salman}
{Salman}, T.~M., {Nieves-Chinchilla}, T., {Jian}, L.~K., {et~al.} 2024, \bibinfo{title}{{A Survey of Coronal Mass Ejections Measured In Situ by Parker Solar Probe during 2018─2022},} \apj, 966, 118, \dodoi{10.3847/1538-4357/ad320c}

\bibitem[{T.~M. Salman {et~al.}(2020)Salman, Winslow, Lugaz, Farrugia, \& Galvin}]{2020_Salman}
Salman, T.~M., Winslow, R.~M., Lugaz, N., Farrugia, C.~J., \& Galvin, A.~B. 2020, \bibinfo{title}{Radial Evolution of ICMEs between MESSENGER, Venus Express, STEREO, and 1 AU,} Journal of Geophysical Research: Space Physics, 125, e2019JA027084, \dodoi{10.1029/2019JA027084}

\bibitem[{J. {Schmidt} \& P. {Cargill}(2004){Schmidt} \& {Cargill}}]{2004_Schmidt}
{Schmidt}, J., \& {Cargill}, P. 2004, \bibinfo{title}{{A numerical study of two interacting coronal mass ejections},} Annales Geophysicae, 22, 2245, \dodoi{10.5194/angeo-22-2245-2004}

\bibitem[{C. Scolini {et~al.}(2021)Scolini, Dasso, Rodriguez, Poedts, \& Mierla}]{2021_Scolini}
Scolini, C., Dasso, S., Rodriguez, L., Poedts, S., \& Mierla, M. 2021, \bibinfo{title}{Exploring the Radial Evolution of ICMEs Using EUHFORIA,} Astronomy \& Astrophysics, 650, A37, \dodoi{10.1051/0004-6361/202039687}

\bibitem[{Z.~I. {Shaikh} {et~al.}(2025){Shaikh}, {Nicolaou}, {Raghav}, {Ghag}, \& {Dhamane}}]{2025_Shaikh}
{Shaikh}, Z.~I., {Nicolaou}, G., {Raghav}, A.~N., {Ghag}, K., \& {Dhamane}, O. 2025, \bibinfo{title}{{Revealing super-adiabatic features of interplanetary coronal mass ejections at 1 au},} \aap, 693, L12, \dodoi{10.1051/0004-6361/202451372}

\bibitem[{J. {Sheoran} {et~al.}(2023){Sheoran}, {Pant}, {Patel}, \& {Banerjee}}]{2023_Sheoran}
{Sheoran}, J., {Pant}, V., {Patel}, R., \& {Banerjee}, D. 2023, \bibinfo{title}{{Evolution of the Thermodynamic Properties of a Coronal Mass Ejection in the Inner Corona},} Frontiers in Astronomy and Space Sciences, 10, 27, \dodoi{10.3389/fspas.2023.1092881}

\bibitem[{G. Siscoe \& D. Odstr{\v{c}}il(2008)Siscoe \& Odstr{\v{c}}il}]{SiscoeOdstrcil2008}
Siscoe, G., \& Odstr{\v{c}}il, D. 2008, \bibinfo{title}{Ways in Which ICME Sheaths Differ from Magnetic Clouds,} Journal of Geophysical Research, 113, A00B07, \dodoi{10.1029/2007JA012302}

\bibitem[{M. {Temmer} \& V. {Bothmer}(2022){Temmer} \& {Bothmer}}]{2022_Temmer}
{Temmer}, M., \& {Bothmer}, V. 2022, \bibinfo{title}{{Characteristics and evolution of sheath and leading edge structures of interplanetary coronal mass ejections in the inner heliosphere based on Helios and Parker Solar Probe observations},} \aap, 665, A70, \dodoi{10.1051/0004-6361/202243291}

\bibitem[{M. {Temmer} {et~al.}(2014){Temmer}, {Veronig}, {Peinhart}, \& {Vr{\v{s}}nak}}]{2014_Temmer}
{Temmer}, M., {Veronig}, A.~M., {Peinhart}, V., \& {Vr{\v{s}}nak}, B. 2014, \bibinfo{title}{{Asymmetry in the CME-CME Interaction Process for the Events from 2011 February 14-15},} \apj, 785, 85, \dodoi{10.1088/0004-637X/785/2/85}

\bibitem[{M. Temmer {et~al.}(2023)Temmer, Scolini, Richardson, Heinemann, Paouris, Vourlidas, Bisi, Al-Haddad, Amerstorfer, Barnard, Burešová, Hofmeister, Iwai, Jackson, Jarolim, Jian, Linker, Lugaz, Manoharan, Mays, Mishra, Owens, Palmerio, Perri, Pomoell, Pinto, Samara, Singh, Sur, Verbeke, Veronig, \& Zhuang}]{TEMMER_2023}
Temmer, M., Scolini, C., Richardson, I.~G., {et~al.} 2023, \bibinfo{title}{CME propagation through the heliosphere: Status and future of observations and model development,} Advances in Space Research, \dodoi{https://doi.org/10.1016/j.asr.2023.07.003}

\bibitem[{M.~S. {Venzmer} \& V. {Bothmer}(2018){Venzmer} \& {Bothmer}}]{2018_Venzmer}
{Venzmer}, M.~S., \& {Bothmer}, V. 2018, \bibinfo{title}{{Solar-wind predictions for the Parker Solar Probe orbit. Near-Sun extrapolations derived from an empirical solar-wind model based on Helios and OMNI observations},} \aap, 611, A36, \dodoi{10.1051/0004-6361/201731831}

\bibitem[{B. Vrsnak {et~al.}(2012)Vrsnak, Žic, Vrbanec, Temmer, Amerstorfer, Möstl, Veronig, Čalogović, Dumbovic, Lulić, Moon, \& Shanmugaraju}]{2012_Vrsnak}
Vrsnak, B., Žic, T., Vrbanec, D., {et~al.} 2012, \bibinfo{title}{Propagation of Interplanetary Coronal Mass Ejections: The Drag-Based Model,} Solar Physics, 285, \dodoi{10.1007/s11207-012-0035-4}

\bibitem[{B. {Vr{\v{s}}nak} {et~al.}(2019){Vr{\v{s}}nak}, {Amerstorfer}, {Dumbovi{\'c}}, {Leitner}, {Veronig}, {Temmer}, {M{\"o}stl}, {Amerstorfer}, {Farrugia}, \& {Galvin}}]{2019_Vrvsnak}
{Vr{\v{s}}nak}, B., {Amerstorfer}, T., {Dumbovi{\'c}}, M., {et~al.} 2019, \bibinfo{title}{{Heliospheric Evolution of Magnetic Clouds},} \apj, 877, 77, \dodoi{10.3847/1538-4357/ab190a}

\bibitem[{C. Wang {et~al.}(2005)Wang, Du, \& Richardson}]{Wang2005}
Wang, C., Du, D., \& Richardson, J.~D. 2005, \bibinfo{title}{Characteristics of the interplanetary coronal mass ejections in the heliosphere,} Journal of Geophysical Research, 110, A10107, \dodoi{10.1029/2005JA011198}

\bibitem[{Y. {Wang} {et~al.}(2004){Wang}, {Shen}, {Wang}, \& {Ye}}]{2004_Wang}
{Wang}, Y., {Shen}, C., {Wang}, S., \& {Ye}, P. 2004, \bibinfo{title}{{Deflection of coronal mass ejection in the interplanetary medium},} \solphys, 222, 329, \dodoi{10.1023/B:SOLA.0000043576.21942.aa}

\bibitem[{D.~F. {Webb} \& T.~A. {Howard}(2012){Webb} \& {Howard}}]{2012_Webb}
{Webb}, D.~F., \& {Howard}, T.~A. 2012, \bibinfo{title}{{Coronal Mass Ejections: Observations},} Living Reviews in Solar Physics, 9, 3, \dodoi{10.12942/lrsp-2012-3}

\bibitem[{R.~M. {Winslow} {et~al.}(2015){Winslow}, {Lugaz}, {Philpott}, {Schwadron}, {Farrugia}, {Anderson}, \& {Smith}}]{2015_Winslow}
{Winslow}, R.~M., {Lugaz}, N., {Philpott}, L.~C., {et~al.} 2015, \bibinfo{title}{{Interplanetary coronal mass ejections from MESSENGER orbital observations at Mercury},} Journal of Geophysical Research (Space Physics), 120, 6101, \dodoi{10.1002/2015JA021200}

\bibitem[{M. Xiong {et~al.}(2006)Xiong, Zheng, Wang, \& Wang}]{Xiong_2005}
Xiong, M., Zheng, H., Wang, Y., \& Wang, S. 2006, \bibinfo{title}{Magnetohydrodynamic simulation of the interaction between interplanetary strong shock and magnetic cloud and its consequent geoeffectiveness,} Journal of Geophysical Research: Space Physics, 111, \dodoi{https://doi.org/10.1029/2005JA011593}

\bibitem[{Z. {Zhang} {et~al.}(2025){Zhang}, {Shen}, {Chi}, {Mao}, {Liu}, {Xu}, {Zhong}, {Luo}, {Wang}, \& {Wang}}]{2025b_Zhang}
{Zhang}, Z., {Shen}, C., {Chi}, Y., {et~al.} 2025, \bibinfo{title}{{Studying the Evolution of ICMEs in the Heliosphere Through Multipoint Observations},} Journal of Geophysical Research (Space Physics), 130, e2025JA034094, \dodoi{10.1029/2025JA034094}

\bibitem[{B. {Zhuang} {et~al.}(2023){Zhuang}, {Lugaz}, {Al-Haddad}, {Winslow}, {Scolini}, {Farrugia}, \& {Galvin}}]{2023_Zhuang}
{Zhuang}, B., {Lugaz}, N., {Al-Haddad}, N., {et~al.} 2023, \bibinfo{title}{{Evolution of the Radial Size and Expansion of Coronal Mass Ejections Investigated by Combining Remote and In Situ Observations},} \apj, 952, 7, \dodoi{10.3847/1538-4357/acd847}

\bibitem[{T.~H. {Zurbuchen} \& I.~G. {Richardson}(2006){Zurbuchen} \& {Richardson}}]{2006_Zurbuchen}
{Zurbuchen}, T.~H., \& {Richardson}, I.~G. 2006, \bibinfo{title}{{In-Situ Solar Wind and Magnetic Field Signatures of Interplanetary Coronal Mass Ejections},} \ssr, 123, 31, \dodoi{10.1007/s11214-006-9010-4}

\end{thebibliography}

\end{document}